\documentclass{SciPost}

\hypersetup{
    colorlinks,
    linkcolor={red!50!black},
    citecolor={blue!50!black},
    urlcolor={blue!80!black}
}

\usepackage[bitstream-charter]{mathdesign}
\usepackage{tikz}
\usetikzlibrary{3d,shapes.geometric,calc}
\usepackage{listings}
\usepackage{todonotes}
\usepackage{tcolorbox}
\usepackage{physics}
\usepackage{caption}
\usepackage{subcaption}

\definecolor{mblue}  {rgb}{0.349, 0.545, 0.878}
\definecolor{morange}{rgb}{0.880722, 0.611041, 0.142051}
\definecolor{mgreen} {rgb}{0.560181, 0.691569, 0.194885}
\definecolor{mred}   {rgb}{0.922526, 0.385626, 0.209179}
\definecolor{mpurple}{rgb}{0.741, 0.392, 0.698}
\definecolor{mcyan}  {rgb}{0.363898, 0.618501, 0.782349}

\newcommand{\code}[2]{
\begin{codebox}
    \lstinputlisting[language=python, basicstyle=\ttfamily\footnotesize,
    firstline=#1,
    lastline=#2,
    keywordstyle=\color{blue},
    stringstyle=\color{mred},
    commentstyle=\color{mgreen}]{code.py}
\end{codebox}
}

\newtcolorbox[]{codebox}[1][]{%
    colback=gray!15!white,
    colframe=gray!75!black,
    fonttitle=\bfseries,
    #1
}

\newtcolorbox[]{commandline}[1][]{%
    colback=black!75!white,
    colframe=black,
    fonttitle=\bfseries,
    colupper=white,
    #1
}

\newtcolorbox[auto counter,number within=section]{exercise}[2][]{%
    colback=blue!5!white,
    colframe=blue!75!black,
    fonttitle=\bfseries,
    title=Exercise.~\thetcbcounter: #2,
    #1
}

\newcommand{\C}{\mathbb{C}}
\newcommand{\ptn}{PyTreeNet}
\newcommand{\txt}[1]{{\text{#1}}}

\newtheorem{defn}{Definition}

\DeclareSymbolFont{usualmathcal}{OMS}{cmsy}{m}{n}
\DeclareSymbolFontAlphabet{\mathcal}{usualmathcal}

\fancypagestyle{SPstyle}{
\fancyhf{}
\lhead{\colorbox{scipostblue}{\bf \color{white} ~SciPost Physics Codebases }}
\rhead{{\bf \color{scipostdeepblue} ~Submission }}

\fancyfoot[C]{\textbf{\thepage}}
}

\begin{document}

\pagestyle{SPstyle}

\begin{center}{\Large \textbf{\color{scipostdeepblue}{
PyTreeNet: A Python Library for easy Utilisation of Tree Tensor Networks\\
}}}\end{center}

\begin{center}\textbf{
Richard M. Milbradt\textsuperscript{1$\star$},
Qunsheng Huang\textsuperscript{1} and
Christian B. Mendl\textsuperscript{1 2}
}\end{center}

\begin{center}
{\bf 1} School of Computation, Information and Technology, Technical University of Munich, Germany
\\
{\bf 2} Institute for Advanced Studies, Technical University of Munich, Germany
\\[\baselineskip]
$\star$ \href{mailto:email1}{\small r.milbradt@tum.de}
\end{center}

\section*{\color{scipostdeepblue}{Abstract}}
\boldmath\textbf{%
In recent years, tree tensor network methods have proven capable of simulating quantum many-body and other high-dimensional systems. This work is a user guide to our Python library {\ptn}. It includes code examples and exercises to introduce the library's functions and familiarise the reader with the concepts and methods surrounding tree tensor networks. {\ptn} implements all the tools required to implement general tree tensor network methods, such as tensor decompositions and arbitrary tree structures. The main focus is on the time evolution of quantum systems. This includes an introduction to tree tensor network states and operators and the time-evolving block decimation and time-dependent variational principle. The library's capabilities are showcased with the example of a modified transverse field Ising model on tree structures that go far beyond the ability of common state vector methods.
}

\vspace{\baselineskip}

\noindent\textcolor{white!90!black}{%
\fbox{\parbox{0.975\linewidth}{%
\textcolor{white!40!black}{\begin{tabular}{lr}%
  \begin{minipage}{0.6\textwidth}%
    {\small Copyright attribution to authors. \newline
    This work is a submission to SciPost Physics Codebases. \newline
    License information to appear upon publication. \newline
    Publication information to appear upon publication.}
  \end{minipage} & \begin{minipage}{0.4\textwidth}
    {\small Received Date \newline Accepted Date \newline Published Date}%
  \end{minipage}
\end{tabular}}
}}
}


\vspace{10pt}
\noindent\rule{\textwidth}{1pt}
\tableofcontents
\noindent\rule{\textwidth}{1pt}
\vspace{10pt}


\section{Introduction}\label{sec:intro}
Over the last three decades, tensor networks have proven to be a versatile language to theoretically describe and numerically evaluate high-dimensional data and systems. Mainly developed for the simulation of many-body quantum systems, tensor networks were used in a variety of fields in quantum physics such as quantum chemistry \cite{Szalay2015}, open quantum systems \cite{Wood2015, Jaschke2018, Strathearn2018}, condensed matter physics \cite{Schollwock2011, Banuls2023}, and more \cite{Jahn2021, Melnikov2023, Patra2024, Rieser2023}. Tensor networks are also applied in fields unrelated to quantum physics, such as machine learning \cite{Ji2019, Panagakis2021, Sengupta2022, Stoudenmire2016} and databases \cite{AboKhamis2016, Dudek2020, Stoian2023}. Section~\ref{sec:tensor_networks} provides a short introduction to tensor networks, and we refer to other sources for a more extensive exposition \cite{Silvi2019, Schollwock2011, Ran2020Book, Montangero2018, Evenbly2022, Biamonte2017, Biamonte2020, Bridgeman2017} or a complete overview of the field \cite{Orus2019, Ren2022, Cirac2021, Banuls2023}. A major contribution to the success of tensor networks for simulating quantum systems is the matrix product (MP) structure as an efficient description of one-dimensional quantum systems \cite{Schollwock2011}. However, tensor network simulations of inherently two-dimensional structures, such as the projected entangled pair states (PEPS) \cite{Verstraete2004PEPS}, struggle due to the problematic scaling of required computational resources when working with them. For example, evaluating PEPS expectation values is an $\# P$-hard problem \cite{Verstraete2006, Schuch2007} and can generally only be performed approximately \cite{Lubasch2014}. Tree tensor networks (TTN) \cite{Shi2006, Silvi2019} offer a middle ground between one-dimensional and two-dimensional structures. Our Python library {\ptn} \cite{pytreenet} implements a TTN structure and various methods based on it. This user guide aims to facilitate an easy entry into {\ptn} by providing code examples and exercises. The TTN introduced in Section~\ref{sec:TTN} generalises the MP structure that retains many desired properties, such as the canonical form \cite{Bauernfeind2020}. Specifically for the simulation of quantum systems, the TTN can represent a quantum state or operator \cite{Shi2006, Frowis2010} as will be discussed in Sections~\ref{sec:TTNS} and~\ref{sec:TTNO} respectively. These tree tensor networks states and operators were successfully utilised, for example, in the simulation of condensed matter \cite{Bauernfeind2017, Murg2010, Okunishi2023} and quantum chemical systems \cite{Nakatani2013, Larsson2019, Gunst2018, Murg2015}. As we will see in Section~\ref{sec:time_evo}, the main focus of {\ptn} is the simulation of time evolutions of quantum systems. {\ptn} includes two of the most commonly used time evolution methods for tensor networks, the time-evolving block decimation (TEBD) \cite{Verstraete2004, Vidal2004, Daley2004} and time-dependent variational principle (TDVP) \cite{Haegeman2011, Haegeman2016}, that were generalised to TTN structures \cite{Shi2006, Bauernfeind2020}. Their use will be exemplified by a (modified) transverse field Ising model. Extensions to these algorithms and other possible future features to {\ptn} are considered in Section~\ref{sec:developements}.

\subsection{Installation}
As a prerequisite, {\ptn} requires an instance of Python with a version $\geq 3.10$. Then the simplest way to install {\ptn} is by using \texttt{pip}. Once Python and \texttt{pip} are installed, run
\begin{commandline}
\texttt{pip install pytreenet}
\end{commandline}
\noindent to install {\ptn} as a package.

\section{Tensor Networks}\label{sec:tensor_networks}

The fundamental concepts of tensor networks are based on a generalisation of vectors and matrices. To start, note that vectors have one index running over their entries. A matrix has two running indices, one for the rows and one for the columns. Continuing this scheme, we can define a tensor as any object whose entries are enumerated by some indices. Commonly, this is written as $M \in V^{d_1 \times \cdots \times d_k}$, with the entries $M_{i_1 \cdots i_k} \in V$, where $V$ is a field. $k$ is known as the \emph{degree} of a tensor. Accordingly, vectors are degree-$1$ tensors and matrices are degree-$2$ tensors. In the context of {\ptn}, we will work with the vector space $V=\mathbb{C}$, which is commonly used for most current quantum problems. For concrete implementation, {\ptn} uses complex arrays provided by the NumPy library \cite{Harris2020}. However, in general, less restricted sets than vector spaces \cite{Liu2023, Hu2022} and even continuous indices are possible \cite{Verstraete2010, Jennings2015, Tilloy2019}. To work with tensors on paper is convenient due to the graphical depiction of tensors, where the main part is some geometrical shape, and every index is shown as a leg.
\begin{figure}[!ht]
    \centering
        \begin{tikzpicture}[scale=0.5]
        \def\distance{8}
            \filldraw[fill=mblue] (0,1) -- (1.5,0) -- (0,-1) -- cycle;
            \draw[ultra thick] (-1,0) -- (0,0);
            \node at (0.5,0){$\vec{v}$};

            \begin{scope}[shift={(\distance,0)}]
                \draw[ultra thick] (-2,0) -- (2,0);
                \filldraw[fill=mblue] (-1,-1) rectangle (1,1);
                \node at (0,0) {$A$};
            \end{scope}

            \begin{scope}[shift={(2*\distance+0.5,0)}]
                \draw[ultra thick] (-2,0) -- (2,0);
                \draw[ultra thick] (0,0) -- (0,2);
                \filldraw[fill=mblue] (0,0) circle (1);
                \node at (0,0) {$M$};
            \end{scope}
        \end{tikzpicture}
    \caption{The graphical depiction of a vector $\vec{v}$, a matrix $A$, and a degree-$3$ tensor $M$ (from left to right) in the graphical tensor network language.}
    \label{fig:vecmattens}
\end{figure}
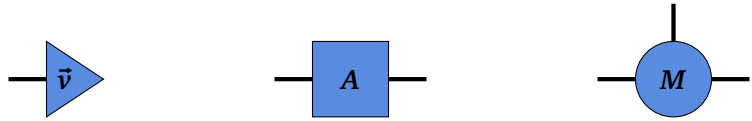
Refer to Figure~\ref{fig:vecmattens} for some examples. In PyTreeNet, a random tensor of any shape is easily generated using the \texttt{crandn} function:

\code{17}{19}

Once tensors are defined, the next step is to have different tensors interact with each other. We can define the combination of two tensors as a \emph{contraction} of tensor legs. A contraction is a generalised matrix multiplication and is achieved by summing over a common index. The elements of the resulting tensor are explicitly defined by the following equation
\begin{equation}\label{eq:tens_contr_symb}
    W_{i_1 \dots i_{L-1} i_{L+1} \dots i_n j_1 \dots j_{K-1} j_{K+1} \dots j_m} = \sum_{\ell=1}^{d} T_{i_1 \dots i_{L-1} \ell i_{L+1} \dots i_n} M_{j_1 \dots j_{K-1} \ell j_{K+1} \dots j_m},
\end{equation}
where $T$ is a degree-$n$ tensor and $M$ a degree-$m$ tensor. Having $d_L = d_K = d$ is a requirement for the sum in \eqref{eq:tens_contr_symb} to be well-defined. Clearly, writing down all the indices is a hassle and error-prone. Instead of the index notation, we can use the graphical version of a tensor contraction. Here, we connect the shared legs of the two tensors. This means \eqref{eq:tens_contr_symb} is graphically represented by:
\begin{equation}
    \begin{tikzpicture}
        \filldraw[fill=mblue] (0,0) rectangle (1,-2.5);
        \node at (0.5,-1.25){$W$};
        \node[anchor=east] (i1) at (-1,0){$i_1$};
        \draw (0,0) -- (i1);
        \node at (-0.5,-0.5){$\vdots$};
        \node[anchor=east] (iLm1) at (-1,-1){$i_{L-1}$};
        \draw (0,-1) -- (iLm1);
        \node[anchor=east] (iLp1) at (-1,-1.5){$i_{L+1}$};
        \draw (0,-1.5) -- (iLp1);
        \node at (-0.5,-2){$\vdots$};
        \node[anchor=east] (in) at (-1,-2.5){$i_1$};
        \draw (0,-2.5) -- (in);
        \node[anchor=west] (i1) at (2,0){$j_1$};
        \draw (1,0) -- (i1);
        \node at (1.5,-0.5){$\vdots$};
        \node[anchor=west] (iLm1) at (2,-1){$j_{K-1}$};
        \draw (1,-1) -- (iLm1);
        \node[anchor=west] (iLp1) at (2,-1.5){$j_{K+1}$};
        \draw (1,-1.5) -- (iLp1);
        \node at (1.5,-2){$\vdots$};
        \node[anchor=west] (in) at (2,-2.5){$j_1$};
        \draw (1,-2.5) -- (in);
    \end{tikzpicture}
    \raisebox{1.4cm}{\, = \,}
    \begin{tikzpicture}
        \filldraw[fill=morange] (0,0) rectangle (1,-2.5);
        \node at (0.5,-1.25){$T$};
        \node[anchor=east] (i1) at (-1,0){$i_1$};
        \draw (0,0) -- (i1);
        \node at (-0.5,-0.5){$\vdots$};
        \node[anchor=east] (iLm1) at (-1,-1){$i_{L-1}$};
        \draw (0,-1) -- (iLm1);
        \node[anchor=east] (iLp1) at (-1,-1.5){$i_{L+1}$};
        \draw (0,-1.5) -- (iLp1);
        \node at (-0.5,-2){$\vdots$};
        \node[anchor=east] (in) at (-1,-2.5){$i_1$};
        \draw (0,-2.5) -- (in);
        \node[anchor=south] at (1.5,-1.25){$\ell$};
        \draw (1,-1.25) -- (2,-1.25);

        \begin{scope}[shift={(2,0)}]
            \draw[fill=mred] (0,0) rectangle (1,-2.5);
            \node at (0.5,-1.25){$M$};
            \node[anchor=west] (i1) at (2,0){$j_1$};
            \draw (1,0) -- (i1);
            \node at (1.5,-0.5){$\vdots$};
            \node[anchor=west] (iLm1) at (2,-1){$j_{K-1}$};
            \draw (1,-1) -- (iLm1);
            \node[anchor=west] (iLp1) at (2,-1.5){$j_{K+1}$};
            \draw (1,-1.5) -- (iLp1);
            \node at (1.5,-2){$\vdots$};
            \node[anchor=west] (in) at (2,-2.5){$j_1$};
            \draw (1,-2.5) -- (in);
        \end{scope}
    \end{tikzpicture}
    .
\end{equation}
A simple, familiar example is the usual multiplication of two matrices
\begin{equation}
\begin{split}
    C_{ij} &= (A \cdot B)_{ij} = \sum_{\ell} A_{i\ell} B_{\ell j} \\
    &=
    \raisebox{-0.15cm}{
    \begin{tikzpicture}
        \draw (0.5,0) -- (2.5,0);
        \node[anchor=south] (i) at (0.5,0){$i$};
        \node[rectangle,fill=morange,draw] (A) at (1,0){$A$};
        \node[anchor=south] (l) at (1.5,0){$\ell$};
        \node[rectangle,fill=mred,draw] (B) at (2,0){$B$};
        \node[anchor=south] (j) at (2.5,0){$j$};
    \end{tikzpicture}
    }
    =
    \raisebox{-0.15cm}{
    \begin{tikzpicture}
        \draw (0.5,0) -- (1.5,0);
        \node[anchor=south] (i) at (0.5,0){$i$};
        \node[rectangle,fill=mblue,draw] (C) at (1,0){$C$};
        \node[anchor=south] (j) at (1.5,0){$j$};
    \end{tikzpicture}
    }.
\end{split}
\end{equation}
To perform a contraction numerically, the NumPy \texttt{tensordot} or \texttt{einsum} functions are used directly on an array representing a tensor.

\subsection{Decomposing Tensors}
\begin{figure}
    \centering
    \begin{tikzpicture}[>=stealth]
        \node[fill=mblue,circle,draw] (A) at (0,0){$A$};
        \draw (-1,0) -- (A) -- (1,0);
        \draw (0,-1) -- (A) -- (0,1);

        \draw[->, ultra thick] (1.3,0) -- (1.8,0);
        \node[anchor=south] at (1.5,0){$1$};

        \begin{scope}[shift={(3,0)}]
            \node[fill=mblue,circle,draw] (A) at (0,0){$A$};
            \draw (-1,0) -- (A) -- (1,0);
            \draw (0,-1) -- (A) -- (0,1);
            \draw[dashed] (-0.75,-0.75) -- (0.75,0.75);
        \end{scope}

        \draw[->, ultra thick] (4.3,0) -- (4.8,0);
        \node[anchor=south] at (4.5,0){$2$};

        \begin{scope}[shift={(6,0)}]
            \node[fill=mblue,circle,draw] (A) at (0,0){$A$};
            \draw (-1,0) -- (A) -- (1,0);
            \draw (A) -- (0,0.5) to[out=90, in=90,] (-0.5,0.5) -- (-0.5,0.1) -- (-1,0.1);
            \draw (A) -- (0,-0.5) to[out=-90, in=-90,] (0.5,-0.5) -- (0.5,-0.1) -- (1,-0.1);
        \end{scope}

        \draw[->, ultra thick] (5.7,-1) -- (5.7,-2);
        \node[anchor=west] at (5.7,-1.5){$3$};

        \begin{scope}[shift={(1,-3)}]
            \draw (-1,0) -- (2,0);
            \node[fill=morange,circle,draw] (Q) at (0,0){$Q$};
            \node[fill=mred,circle,draw] (R) at (1,0){$R$};
            \draw (Q) -- (0,1);
            \draw (R) -- (1,-1);
        \end{scope}

        \draw[->, ultra thick] (3.8,-3) -- (3.2,-3);
        \node[anchor=south] at (3.5,-3){$4$};

        \begin{scope}[shift={(5,-3)}]
            \draw (-1,0) -- (2,0);
            \node[fill=morange,circle,draw] (Q) at (0,0){$Q$};
            \node[fill=mred,circle,draw] (R) at (1,0){$R$};
            \draw (Q) -- (0,0.5) to[out=90, in=90,] (-0.5,0.5) -- (-0.5,0.1) -- (-1,0.1);
            \draw (R) -- (1,-0.5) to[out=-90, in=-90,] (1.5,-0.5) -- (1.5,-0.1) -- (2,-0.1);
        \end{scope}
    \end{tikzpicture}
    \caption{Graphical depiction of the steps involved in a tensor decomposition exemplified by a QR-decomposition. $1)$ The legs of a degree-$4$ tensor are split into two sets. $2)$ The legs in a set are combined into one leg of higher dimension. $3)$ The QR decomposition is performed yielding two matrices $Q$ and $R$. $4)$ The legs of the matrices are separated to yield two degree-$3$ tensors.}
    \label{fig:QR_Decomp}
\end{figure}
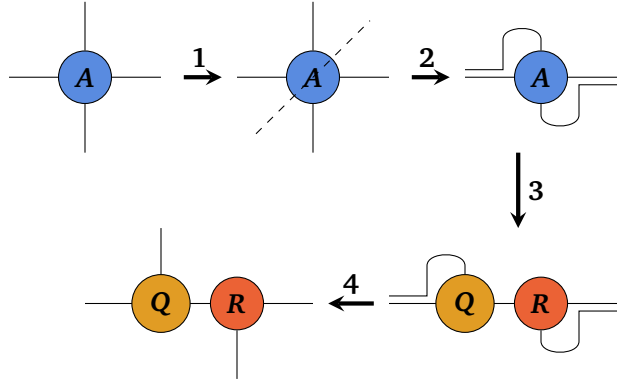

Now, we can combine multiple tensors into a single one. However, we can also decompose a single tensor into two. Such a \emph{tensor decomposition} can be defined using any known matrix decomposition. The reader may refer to \cite{Lu2023} for an overview of matrix decompositions. To apply a tensor decomposition to a degree-$n$ tensor $A \in \C^{d_1 \times \cdots \times d_n}$, one groups the legs $i_1, \cdots , i_n$ of $A$ into two sets $\{ i_1, \dots, i_\ell \}$ and $\{ i_{\ell+1}, \dots, i_n \}$. We choose both sets to be consecutive indices to simplify the notation. However, this is not generally required. Now, we reinterpret each set of indices as one combined index:
\begin{subequations}\label{eq:matricisation_of_indices}
	\begin{align}
		\left\{ i_1, \dots, i_\ell \right\} &\rightarrow \left( i_1, \dots, i_\ell \right) = \textbf{j} \\
		\left\{ i_{\ell+1}, \dots, i_n \right\} &\rightarrow \left( i_{\ell+1}, \dots, i_n \right) = \textbf{k}
	\end{align}
\end{subequations}
Therefore, we can reinterpret the tensor $A$ as a matrix with entries $A_{jk} = A_{i_1 \cdots i_n}$. Thus, the desired matrix decomposition can be applied to the matrix version of $A$, yielding two new matrices $Q \in \C^{(d_1 \times \cdots \times d_\ell) \times D}$ and $R \in \C^{D \times (d_{\ell+1} \times \cdots \times d_n)}$. We can see that a new leg has appeared as one of the matrix legs. We can now reverse our interpretation of the legs \eqref{eq:matricisation_of_indices} to obtain the degree-$(\ell +1)$ tensor $Q$ and the degree-$(n -\ell +1)$ tensor $R$. Note that a contraction of $Q$ and $R$ along their respective $D$-dimensional leg results in the tensor $A$. Figure~\ref{fig:QR_Decomp} provides a graphical depiction of the steps explained above with the QR-decomposition as an example.

One can easily define a custom tensor decomposition corresponding to a given matrix decomposition using the utility functions provided by {\ptn}. The two most commonly required decompositions have already been implemented. The first is the QR decomposition, already used in Figure~\ref{fig:QR_Decomp}. For a matrix $A \in \mathbb{C}^{m\times n}$ with $m\geq n$ the QR-decomposition is given by
\begin{equation}\label{eq:mat_qr}
    A = Q \cdot R
\end{equation}
with a unitary matrix $Q \in \mathbb{C}^{m \times m}$ and an upper triangular matrix $R \in \C^{m \times n}$. As explained before, the QR decomposition can be used to split tensors. An example of this is shown in Figure~\ref{fig:QR_Decomp}. The code to run the process shown Figure~\ref{fig:QR_Decomp} in {\ptn} would be

\code{23}{29}

Let us take a closer look at the graphical depiction of the QR tensor decomposition and the dimensions involved. We can draw the process shown in Figure~\ref{fig:QR_Decomp} in one step as
\begin{equation}\label{eq:QR_dimensions}
    \begin{tikzpicture}
        \draw (1,0) -- (-1,0);
        \draw (0,1) -- (0,-1);
        \node[fill=mblue,draw,circle] at (0,0) {$A$};
        \node[anchor=south] at (-1,0){$d_1$};
        \node[anchor=south] at (1,0){$d_3$};
        \node[anchor=west] at (0,-1){$d_2$};
        \node[anchor=west] at (0,1){$d_4$};
    \end{tikzpicture}
    \raisebox{1.2cm}{=}
    \begin{tikzpicture}
        \draw (-0.5,0) -- (3,0);
        \node[isosceles triangle,
                isosceles triangle apex angle=60,
                draw,fill=morange,
                anchor=west] (Q) at (0,0){$Q$};
        \node[fill=mred,draw,circle] (R) at (2,0){$R$};
        \draw (Q.left corner) -- (0,1);
        \draw (R) -- (2,-1);
        \node[anchor=south] at (-0.5,0){$d_1$};
        \node[anchor=south] at (3,0){$d_3$};
        \node[anchor=west] at (2,-1){$d_2$};
        \node[anchor=west] at (0,1){$d_4$};
        \node[anchor=south] at (1.3,0){$D$};
    \end{tikzpicture}
    ,
\end{equation}
where $d_i$ and $D$ are the dimensions of the respective leg. $Q$ is written as a triangle to emphasise that it is an isometry. This means $Q Q^\dagger = \mathbb{1}$ if the leg starting from the tip is considered the input leg and all other legs are combined into an output leg. Looking at the dimensions and using the naive QR-decomposition as written in \eqref{eq:mat_qr}, we find $D = m = d_1d_4$. However, if $m>n$, some rows in $Q$ are zero. These and the corresponding columns in $R$ can be dropped without changing the equality in \eqref{eq:mat_qr}, but changing $D=n=d_3d_4$. Furthermore, the QR-decomposition function used in {\ptn} can perform such a decomposition for $m<n$, once more yielding $D=m$ or $D=n$, depending on the desired outcome. {\ptn} supplies the \texttt{SplitNode} Enum to differentiate these possibilities according to the desired final dimensions. \texttt{SplitNode.FULL} will always lead to the dimension $D=m$, \texttt{SplitNode.KEEP} to $D=n$, and \texttt{SplitNode.REDUCED} yields $D = \min(n,m)$. The desired Enum has to be supplied to the \texttt{tensor\_qr\_decomposition} as the \texttt{mode} keyword. As one usually desires the lowest possible dimensions in a tensor network, \texttt{mode} defaults to \texttt{SplitMode.REDUCED}. For the $Q$ tensor, the new leg will always be the last leg, and for the $R$ tensor, it will always be the first leg. Accordingly, with the same tensor and shape as in \eqref{eq:QR_dimensions}, we get the following code example for the different modes

\code{31}{50}

The second tensor decomposition implemented in {\ptn} is the singular value decomposition (SVD). The SVD splits a square matrix $A\in \C^{m\times n}$ into three other matrices
\begin{equation}\label{eq:mat_svd}
    A = U \cdot S \cdot V^\dagger,
\end{equation}
where $U \in \C^{m\times m}$ and $V \in \C^{n\times n}$ are matrices with orthonormal rows and columns respectively. $S \in \C^{m \times n}$ is a diagonal matrix. The entries $s_{ii}$ on the diagonal of $S$ are called \emph{singular values} of $A$. The $s_{ii}$ are positive real numbers sorted in descending order. The number of non-zero singular values of a matrix $A$ is called its \emph{rank}. The SVD admits the same modes as the QR decomposition. Note that we can discard zero-valued singular values without affecting the equality in \eqref{eq:mat_svd}. We can choose between the two modes using the same \texttt{SplitNode} Enum. \texttt{SplitNode.REDUCED} will yield the SVD with discarded zero-valued singular values, and the other two modes yield the full SVD. For example, we create a degree-$4$ tensor by diagonalising and reshaping a vector of singular values. Afterwards, we perform an SVD on that tensor to recreate the vector using different modes

\code{54}{65}

While computationally more demanding, the SVD supports an additional feature compared to the QR decomposition: the \emph{truncation} of small singular values. As stated above, we can always discard the zero-valued singular values. If we accept a small error, we can additionally discard non-zero singular values. The effectiveness of this approximation stems from the fact that for a matrix $A$ of rank $r$, the closest matrix $B$ of rank $r' < r$ is the result of \eqref{eq:mat_svd} with the smallest $|r'-r|$ singular values removed \cite{Eckart1936, Mirsky1960}. Therefore, small singular values can be removed while only slightly changing the overall matrix. In practice, these truncations lead to lower-rank tensors and drastically lower bond dimensions in large tensor networks. There are three main ways to define the truncation condition for a vector of singular values:
\begin{enumerate}
    \item Maximum size (\texttt{max\_bond\_dim}): Here a maximum size $L$ for the sorted vector of singular values is given. Only the $L$ largest singular values are kept. This is generally used to put an explicit limit on the memory resources used by tensors, as it strictly limits the dimensions of a bond or leg of tensors.
    \item Relative tolerance (\texttt{rel\_tol}): The relative tolerance $\varepsilon_{\text{rel}}$ compares every singular value to the maximum singular value $s_{\text{max}}$. If for a singular value $s$ the expression
    \begin{equation}
        s < \varepsilon_{\text{rel}} s_{\text{max}}
    \end{equation}
    is true, $s$ will be discarded.
    \item Total tolerance (\texttt{total\_tol}): Every singular values smaller than the total tolerance $\varepsilon_{\text{tot}}$ is discarded.
\end{enumerate}
In {\ptn}, these parameters are handled by the \texttt{SVDParameters} data class. The attribute names are given in the brackets in the above list. While default values are given, the optimal truncation parameters are highly problem-dependent and need to be optimised in relation to each other. One can also decide to renormalise the truncated vector to have the same norm as before the truncation. This is enabled by setting the \texttt{renorm} Attribute to $True$. Using the same tensor as above, we get the following code example

\code{69}{83}

In the first run the singular value $s_4 = 0$ is truncated as it is smaller than  $\varepsilon_{\text{tot}}=10^{-2}$. In the second run $s_3 = 0.05$ is also truncated, since it is smaller than $\varepsilon_{\text{rel}} s_{\text{max}} = 1*0.1 = 0.1$. And in the final execution all singular values apart from $s_1 = 1$ are truncated, as the maximum size of the singular value vector is set to $1$. Note that for both functions, the leg of the $U$ tensor towards the $S$ tensor is always the last leg, while it is always the first leg for the $V$ tensor. Both functions also split a single tensor $A$ into three tensors
\begin{equation}
    \begin{tikzpicture}
        \draw (1,0) -- (-1,0);
        \draw (0,1) -- (0,-1);
        \node[fill=mblue,draw,
                minimum size = 1cm,
                circle] at (0,0) {$A$};
    \end{tikzpicture}
    \raisebox{1cm}{\, = \,}
    \begin{tikzpicture}
        \draw (-0.5,0) -- (4.5,0);
        \node[isosceles triangle,
                isosceles triangle apex angle=60,
                minimum size = 1cm,
                draw,fill=morange,
                anchor=west] (U) at (0,0){$U$};
        \node[isosceles triangle,
                isosceles triangle apex angle=60,
                rotate=180,
                draw,fill=mred,
                minimum size = 1cm,
                anchor=west] (V) at (4,0){};
        \node at (V.center) {$V^*$};
        \node[fill=mgreen,draw,
                minimum size = 1cm,
                diamond] at (2,0){$S$}; 
        \draw (U.left corner) -- (0,1);
        \draw (V.left corner) -- (4,-1);
    \end{tikzpicture}
    ,
\end{equation}
where $S$ is drawn as a diamond to show it is a diagonal matrix. To obtain only two tensors from a tensor-SVD, the $S$ tensor can either be contracted manually after the decomposition was performed or the function \texttt{contr\_truncated\_svd\_splitting} can be used. It performs the SVD, including truncation, and contracts $S$ into either $U$ or $V$ specified by the enum \texttt{ContractionMode}. All the functions and classes discussed in this subsection can be found in the \texttt{util.tensor\_splitting} sub-module, and are useful for working with any kind of tensor network. This first discussion hopefully got you used to the tensor network notation and showed you how the objects introduced in the next couple of sections work under the hood. We will now make good use of the learned and concentrate on the main subclass of tensor networks {\ptn} is concerned with.

\section{Tree Tensor Networks}\label{sec:TTN}
Clearly, tensor networks and graph theory are related. After all, every tensor network $\mathcal{T} = (\mathcal{M}, \mathcal{C})$, where $\mathcal{M}$ is the set of tensors and $\mathcal{C}$ the set of contracted legs, can be mapped directly into a graph $\mathcal{G} = (\mathcal{V}, \mathcal{E})$, where $\mathcal{V}$ and $\mathcal{E}$ are the sets of vertices and edges respectively. One simply maps every tensor to a vertex and every contraction to an edge. As trees have additional properties when compared to general graphs, it should be no surprise that the set of tensor networks that can be mapped to trees have intriguing additional properties compared to general tensor networks. Let us first define some technical terms on the concept of trees.

\subsection{What is a Tree?}
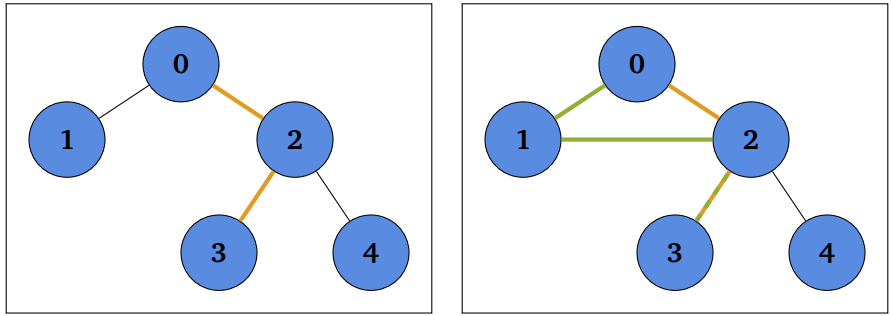
\begin{figure}
    \centering
    \begin{tikzpicture}
        \draw (-2.3,0.8) rectangle (3.3,-3.3);
        \node[fill=mblue,draw,
                circle,
                minimum size = 1cm] (N0) at (0,0){$0$};
        \node[fill=mblue,draw,
                circle,
                minimum size = 1cm] (N1) at (-1.5,-1){$1$};
        \node[fill=mblue,draw,
                circle,
                minimum size = 1cm] (N2) at (1.5,-1){$2$};
        \node[fill=mblue,draw,
                circle,
                minimum size = 1cm] (N3) at (0.5,-2.5){$3$};
        \node[fill=mblue,draw,
                circle,
                minimum size = 1cm] (N4) at (2.5,-2.5){$4$};
        \draw (N0) -- (N1);
        \draw[morange, ultra thick] (N0) -- (N2) --  (N3);
        \draw (N2) -- (N4);

        \begin{scope}[shift={(6,0)}]
            \draw (-2.3,0.8) rectangle (3.3,-3.3);
            \node[fill=mblue,draw,
                    circle,
                    minimum size = 1cm] (N0) at (0,0){$0$};
            \node[fill=mblue,draw,
                    circle,
                    minimum size = 1cm] (N1) at (-1.5,-1){$1$};
            \node[fill=mblue,draw,
                    circle,
                    minimum size = 1cm] (N2) at (1.5,-1){$2$};
            \node[fill=mblue,draw,
                    circle,
                    minimum size = 1cm] (N3) at (0.5,-2.5){$3$};
            \node[fill=mblue,draw,
                    circle,
                    minimum size = 1cm] (N4) at (2.5,-2.5){$4$};
            \draw[morange, ultra thick] (N0) -- (N2) --  (N3);
            \draw (N2) -- (N4);
            \draw[dashed,mgreen,ultra thick] (N3) -- (N2);
            \draw[mgreen, ultra thick] (N2) -- (N1) -- (N0);
        \end{scope}
    \end{tikzpicture}
    \caption{A tree (left) has one unique simple path between two given nodes. For example, between node $3$ and node $0$ (orange). On the other hand, a general graph (right) may have multiple simple paths. For example, the two simple paths between node $3$ and node $0$ (orange, green).}
    \label{fig:tree_vs_graph}
\end{figure}

There are multiple equivalent definitions of a tree \cite{Krumke2005}. We will use the following
\begin{defn}
A \emph{tree} $\mathcal{T}$ is a graph, i.e. a set of vertices $\mathcal{V}$ combined with a set of edges $\mathcal{E}$, each edge $e=(\nu_1,\nu_2)$ connecting two vertices $\nu_1$, $\nu_2$, such that any two vertices are connected by a unique simple path. That is a path which does not contain any vertex twice.
\end{defn}
Figure~\ref{fig:tree_vs_graph} shows a tree and a very similar graph that is not a tree to exemplify the above definition. Due to the trees' properties, the notion of a distance or norm for trees is well-defined.
\begin{defn}\label{def:tree_distance}
The \emph{distance} $\Delta (\nu_1,\nu_2)$ of two nodes $\nu_1$ and $\nu_2$ in a tree $\mathcal{T}$ is the number of edges in the unique simple path between $\nu_1$ and $\nu_2$. If $\gamma (\nu_1,\nu_2)$ denotes this simple path, we can write
\begin{equation}
    \Delta (\nu_1,\nu_2) = \left| \gamma (\nu_1,\nu_2) \right|.
\end{equation}
\end{defn}
The trees, as defined above, are undirected. However, directed trees are easier to use as a data structure and therefore implemented in {\ptn}. This is not a problem, since every undirected tree can be mapped to a directed tree. To do so, one selects a node $r \in \mathcal{T}$ as the so-called \emph{root}. Using $\Delta (r,r) = 0$ as an initial condition, we can recursively define a direction on $\mathcal{T}$. We define the \emph{children} $\mathcal{C}$ of a node $\nu \in \mathcal{T}$ as the set of all nodes $\omega \in \mathcal{T}$ such that
\begin{equation}
    \Delta ( \nu, \omega) = 1 \, \text{and} \, \Delta (r,\omega) = \Delta (r,\nu) + 1.
\end{equation}
This means the children of a node are exactly one edge further away from the root than the node itself. The direction is then defined as from a node $\nu$ to its children. Conversely, $\nu$ is called the \emph{parent} of its children. Clearly, the mapping from a directed graph into an undirected one is simple, so nothing is lost in the implementation.
\begin{figure}
    \centering
    \begin{tikzpicture}
        \draw (-2.3,0.8) rectangle (3.3,-3.3);
        \node[fill=mblue,draw,
                circle,
                minimum size = 1cm] (N0) at (0,0){$0$};
        \node[fill=mblue,draw,
                circle,
                minimum size = 1cm] (N1) at (-1.5,-1){$1$};
        \node[fill=mblue,draw,
                circle,
                minimum size = 1cm] (N2) at (1.5,-1){$2$};
        \node[fill=mblue,draw,
                circle,
                minimum size = 1cm] (N3) at (0.5,-2.5){$3$};
        \node[fill=mblue,draw,
                circle,
                minimum size = 1cm] (N4) at (2.5,-2.5){$4$};
        \draw[-to] (N0) -- (N1);
        \draw[-to] (N0) -- (N2);
        \draw[-to] (N2) -- (N3);
        \draw[-to] (N2) -- (N4);

        \begin{scope}[shift={(6,0)}]
            \draw (-2.3,0.8) rectangle (3.3,-3.3);
            \node[fill=mblue,draw,
                    circle,
                    minimum size = 1cm] (N0) at (-1,-1){$0$};
            \node[fill=mblue,draw,
                    circle,
                    minimum size = 1cm] (N1) at (-1.5,-2.5){$1$};
            \node[fill=mblue,draw,
                    circle,
                    minimum size = 1cm] (N2) at (1.5,0){$2$};
            \node[fill=mblue,draw,
                    circle,
                    minimum size = 1cm] (N3) at (0.5,-1.5){$3$};
            \node[fill=mblue,draw,
                    circle,
                    minimum size = 1cm] (N4) at (2.5,-1.5){$4$};
        \draw[-to] (N0) -- (N1);
        \draw[to-] (N0) -- (N2);
        \draw[-to] (N2) -- (N3);
        \draw[-to] (N2) -- (N4);
        \end{scope}
    \end{tikzpicture}
    \caption{An example mapping of an undirected tree into a directed one. Both directed trees shown are equivalent to the tree in Figure~\ref{fig:tree_vs_graph}. However, in the left one, we chose node $0$ as the root, and node $2$ was chosen for the right one.}
    \label{fig:directed_tree}
\end{figure}
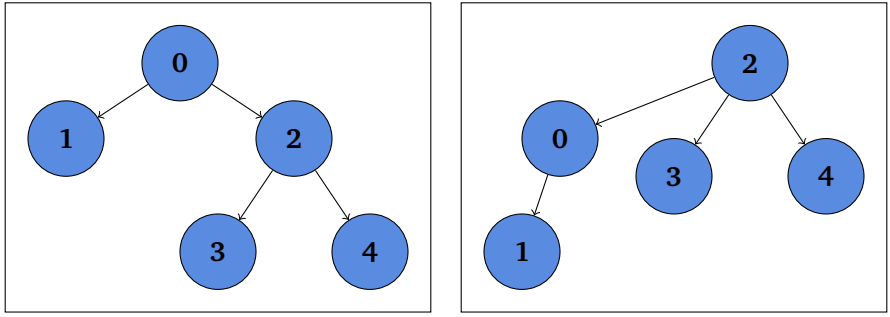
However, there is no unique choice of root for a given undirected tree to turn it into a directed tree. Which node is most suited can depend on the given problem or be completely irrelevant. To showcase this non-uniqueness, two directed trees equivalent to the undirected tree given in Figure~\ref{fig:tree_vs_graph}, but with different root choices, are given in Figure~\ref{fig:directed_tree}. Just as the root, a node without a parent, is named with regard to the biological tree, so are the nodes without children. Such nodes are called \emph{leaves}. There is one last concept we need to define before we continue:
\begin{defn}
A subtree $\mathcal{S}_\nu^{[e]}$ of a tree $\mathcal{T}$ with origin node $\nu$ and with respect to the edge $e$ to which $\nu$ is connected is the set of nodes and edges that can be reached via a path starting at $\nu$ without traversing $e$.
\end{defn}
It should be clear that a subtree is still a tree. As an example, consider the following subtrees of the tree given in Figure~\ref{fig:tree_vs_graph}:
\begin{align}
    \mathcal{S}_2^{[0,2]} &= \{2,3,4,(2,3),(2,4)\} \\
    \mathcal{S}_0^{[0,2]} &= \{0,1,(0,1)\}.
\end{align}
In {\ptn}, everything regarding the tree topology is handled by the \texttt{TreeStructure} class. We can add a root and children, find distances and paths between two given nodes, and replace nodes entirely. We will not go into detail here since the use of the main class representing tree tensor networks is very similar, and that class will be introduced in the next subsection.

\subsection{Building a Tree Tensor Network}
Tee tensor networks (TTN), sometimes known as loop-free tensor networks \cite{Silvi2019}, are tensor networks with an underlying graph topology that is a tree. In {\ptn}, the most general definition of a TTN is used, which allows every site to have an arbitrary number of open legs \cite{Shi2006}. \emph{Open legs} are legs in a tensor network that are not contracted with a different tensor leg. As open legs tend to represent physical Hilbert spaces, they are also referred to as physical legs.
To showcase the use of TTN in {\ptn}, we construct the example TTN
\begin{equation}\label{eq:example_ttn}
    \raisebox{3cm}{$\mathcal{T} = $}
    \begin{tikzpicture}
        \node[draw,fill=mblue,minimum size=1cm,circle] (N0) at (0,0){$0$};
        \node[draw,fill=mblue,minimum size=1cm,circle] (N1) at (-1.5,-1.5){$1$};
        \node[draw,fill=mblue,minimum size=1cm,circle] (N2) at (-2.5,-3){$2$};
        \node[draw,fill=mblue,minimum size=1cm,circle] (N3) at (-1,-3){$3$};
        \node[draw,fill=mblue,minimum size=1cm,circle] (N4) at (0,-1.5){$4$};
        \node[draw,fill=mblue,minimum size=1cm,circle] (N5) at (1.5,-1.5){$5$};
        \node[draw,fill=mblue,minimum size=1cm,circle] (N6) at (2,-3){$6$};

        \draw (N0) -- (N1) -- (N2);
        \draw (N1) -- (N3);
        \draw (N0) -- (N4);
        \draw (N0) -- (N5) -- (N6);

        \draw[mred, ultra thick] (N0) -- (0,1);
        \draw[mred, ultra thick] (N4) -- (0,-2.5);
        \draw[mred, ultra thick] (N4) -- (0.5,-2.3);
        \draw[mred, ultra thick] (N2) -- (-2.7,-4);
        \draw[mred, ultra thick] (N3) -- (-0.8,-4);
        \draw[mred, ultra thick] (N6) -- (2.2,-4);

        \node[anchor=west] at (0,1) {\color{mred} $2$};
        \node[anchor=west] at (0,-2.5) {\color{mred} $2$};
        \node[anchor=west] at (0.5,-2.3) {\color{mred} $3$};
        \node[anchor=west] at (-2.7,-4) {\color{mred} $2$};
        \node[anchor=west] at (-0.8,-4) {\color{mred} $2$};
        \node[anchor=west] at (2.2,-4) {\color{mred} $2$};

        \node[anchor=east] at (-0.75,-0.7) {$4$};
        \node[anchor=east] at (0,-0.75) {$5$};
        \node[anchor=west] at (0.75,-0.7) {$3$};
        \node[anchor=west] at (-2,-2.25) {$2$};
        \node[anchor=west] at (-1.25,-2.25) {$2$};
        \node[anchor=west] at (1.75,-2.25) {$2$};
    \end{tikzpicture},
\end{equation}
where the labels of the legs are the corresponding leg dimensions. The physical legs and their dimensions are coloured in red to highlight them. The tensors will be filled with random complex numbers. Therefore, $\mathcal{T}$ has no deeper physical meaning.\\
As a first step, an empty TTN is generated
\code{87}{87}
A \texttt{TreeTensorNetwork} object holds the nodes and tensors representing a tensor network in separate dictionaries. The node holds all the information about the tree topology and leg connection, while the tensors are purely there to hold the tensor elements as data. To have an example accompanying the explanation, let us fill the TTN we just created. First, we add a root node
\code{89}{91}
leading to the following rather trivial TTN
\begin{equation}
    \begin{tikzpicture}
        \node[fill=mblue,draw,circle,minimum size=1cm] (N0) at (0,0){$0$};
        \draw[mred] (N0.200) -- (-1.25,-0.8);
        \draw (N0.250) -- (-0.5,-1.2);
        \draw (N0.290) -- (0.5,-1.2);
        \draw (N0.330) -- (1.25,-0.8);
        \node[anchor=north] at (-1.25,-0.8){\color{mred} $2$};
        \node[anchor=west] at (-0.5,-1.2){$4$};
        \node[anchor=west] at (0.5,-1.2){$5$};
        \node[anchor=north] at (1.25,-0.8){$3$};
        \node[anchor=east] at (N0.200){\footnotesize \color{mblue} $0$};
        \node[anchor=east] at (N0.250){\footnotesize \color{mblue} $1$};
        \node[anchor=west] at (N0.290){\footnotesize \color{mblue} $2$};
        \node[anchor=west] at (N0.330){\footnotesize \color{mblue} $3$};
    \end{tikzpicture}.
\end{equation}
The two are linked by adding a node and a tensor to a TTN. For one, this results in the node and tensor being saved using the same key in their respective dictionaries, in this case \texttt{"root"}. Additionally, the node object will copy the tensor's shape and record potential leg permutations. This allows us to transpose the actual tensor only if it is called externally. Other permutations of the legs will happen in the node object first. This can avoid unnecessary transpositions of the tensor data. For the same reason, a copy of the shape is kept in the node. We will explain the details of the \texttt{Node} object later. We can attach a node as a child to the root by defining the legs of each node at which the two should be connected. As the tensors are provided in both cases, a check ensures the dimensions are compatible. As a next step, we attach node $1$ to the root
\code{93}{94}
which implies the following graphical representation
\begin{equation}\label{eq:attach_root_to_child}
    \begin{tikzpicture}
        \node[fill=mblue,draw,circle,minimum size=1cm] (N0) at (0,0){$0$};
        \node[fill=mblue,draw,circle,minimum size=1cm] (N1) at (-1.5,-1.5){$1$};
        \draw (N0.200) -- (N1.50);
        \draw[mred] (N0.250) -- (-0.5,-1.2);
        \draw (N0.290) -- (0.5,-1.2);
        \draw (N0.330) -- (1.25,-0.8);
        \node[anchor=north] at (-0.5,-1.2){\color{mred} $2$};
        \node[anchor=east] at (-0.75,-0.5){$4$};
        \node[anchor=west] at (0.5,-1.2){$5$};
        \node[anchor=north] at (1.25,-0.8){$3$};
        \node[anchor=east] at (N0.200){\footnotesize\color{mblue} $0$};
        \node[anchor=east] at (N0.250){\footnotesize\color{mblue} $1$};
        \node[anchor=west] at (N0.290){\footnotesize\color{mblue} $2$};
        \node[anchor=west] at (N0.330){\footnotesize\color{mblue} $3$};
        \node[anchor=south] at (N1.50) {\footnotesize\color{mblue} $0$};
        \draw (N1.220) -- (-2.5,-2.5);
        \draw (N1.310) -- (-1,-2.5);
        \node[anchor=east] at (N1.220){\footnotesize\color{mblue} $1$};
        \node[anchor=west] at (N1.310){\footnotesize\color{mblue} $2$};
        \node[anchor=west] at (-2.5,-2.5){$2$};
        \node[anchor=west] at (-1,-2.5){$2$};
    \end{tikzpicture}.
\end{equation}
Note that this symbolic contraction changed the leg order of the root node. The physical leg now has the index $1$ and the leg of dimension $4$ has index $0$. The reason for this is the convention used for the order of legs of a node. The legs are always sorted such that the first leg points to the node's parent, the next legs point to the node's children in the order the children were added, and all remaining legs are open legs, not connected to any other node. So in short
\begin{equation}\label{eq:leg_convention}
    \texttt{parent} \rightarrow \texttt{children} \rightarrow \texttt{open}
\end{equation}
The above example \eqref{eq:attach_root_to_child} shows this in action. The leg of node $1$ pointing connecting to the parent node $0$ has index $0$. On the other hand, the leg of node $0$ pointing to its first child node $1$ also has the index $0$, as there is no parent to a root. All other legs are open legs for now. The index of a leg to a different node is easily found using the \texttt{neighbour\_index} method of a node:
\code{96}{97}
However, the open legs need to be tracked manually. Therefore it is usually a good practice to generate the tensor with the leg order one desires to add children to it. We can see the open leg of the root being pushed even further if we attach the other children to the root node
\code{99}{102}
The resulting tree tensor network after running this code is
\begin{equation}
    \begin{tikzpicture}
        \node[fill=mblue,draw,circle,minimum size=1cm] (N0) at (0,0){$0$};
        \node[fill=mblue,draw,circle,minimum size=1cm] (N1) at (-1.5,-2){$1$};
        \node[draw,fill=mblue,minimum size=1cm,circle] (N4) at (0,-2){$4$};
        \node[draw,fill=mblue,minimum size=1cm,circle] (N5) at (1.5,-2){$5$};
        \draw (N0.200) -- (N1.50);
        \draw[mred] (N0.90) -- (0,1.2);
        \draw (N0.270) -- (N4.90);
        \draw (N0.330) -- (N5.120);
        \node[anchor=west] at (0,1.2){\color{mred} $2$};
        \node[anchor=east] at (-0.7,-0.7){$4$};
        \node[anchor=west] at (0,-1){$5$};
        \node[anchor=north] at (1.2,-0.7){$3$};
        \node[anchor=east] at (N0.200){\footnotesize\color{mblue} $0$};
        \node[anchor=north east] at (N0.270){\footnotesize\color{mblue} $1$};
        \node[anchor=west] at (N0.330){\footnotesize\color{mblue} $2$};
        \node[anchor=south west] at (N0.90){\footnotesize\color{mblue} $3$};
        \node[anchor=south] at (N1.50) {\footnotesize\color{mblue} $0$};
        \draw (N1.220) -- (-2.5,-3);
        \draw (N1.290) -- (-1.5,-3.5);
        \node[anchor=east] at (N1.220){\footnotesize\color{mblue} $1$};
        \node[anchor=west] at (N1.310){\footnotesize\color{mblue} $2$};
        \node[anchor=west] at (-2.5,-3){$2$};
        \node[anchor=west] at (-1.5,-3.5){$2$};
        \draw[mred] (N4.220) -- (-0.75,-3);
        \draw[mred] (N4.310) -- (0.75,-3);
        \node[anchor=north west] at (-0.75,-3){\color{mred} $2$};
        \node[anchor=north east] at (0.75,-3){\color{mred} $3$};
        \node[anchor=south east] at (N4.90){\footnotesize\color{mblue} $0$};
        \node[anchor=east] at (N4.220){\footnotesize\color{mblue} $1$};
        \node[anchor=west] at (N4.310){\footnotesize\color{mblue} $2$};
        \draw (N5.300) -- (2.2,-3);
        \node[anchor=west] at (2.2,-3){$2$};
        \node[anchor=south] at (N5.120){\footnotesize\color{mblue} $0$};
        \node[anchor=north] at (N5.300){\footnotesize\color{mblue} $1$};
    \end{tikzpicture}.
\end{equation}
As we can see, the first (physical) leg was pushed to have the highest index $3$. We might also note that this convention leaves the order of the two open legs of node $4$ arbitrary. Keeping track of the open leg order is usually not a problem, as the number of open legs on a single node is very small. Now, let us finish our tree tensor network by attaching the leaves
\code{104}{109}
This finally yields the example TTN shown in \eqref{eq:example_ttn}. We can easily access the node and the tensor data by accessing the appropriate attribute dictionary and treating the TTN object as a dictionary.
\code{113}{115}
Remember that solely accessing the node is more efficient if information about the tree topology or leg order is needed, instead of the actual tensor elements. All required information is available from methods of the \texttt{Node} class. Also, since the root is a special node and frequently required, there are two special ways to access it
\code{119}{121}
Now that we have constructed a TTN, what can we do with it?
\begin{exercise}{Building a simple TTN}\label{exc:build_a_ttn}
    Construct the TTN $\mathcal{T}'$ that is depicted on the left in Figure~\ref{fig:node_contr_example}. (Hint: While you can choose all the dimensions yourself, it is a good idea to choose all of them to be different. This way, the implemented checks can spot any mistake.)
\end{exercise}

\subsection{Contraction and Splitting of a TTN}\label{sec:contr_and_split_of_ttn}
Assume we have a TTN, for example, $\mathcal{T}$ as given in \eqref{eq:example_ttn}. Then, the contraction and splitting of tensors, as described in Section~\ref{sec:tensor_networks}, are easily facilitated by methods of the \texttt{TreeTensorNetwork} class. For the contraction, we merely have to specify the identifiers of the involved nodes and a new identifier. If we want to contract the nodes $5$ and $6$, we run
\code{125}{127}
and we obtain the TTN
\begin{equation}\label{eq:ttn_one_node_contr}
    \raisebox{3cm}{$\mathcal{T} = $}
    \begin{tikzpicture}
        \node[draw,fill=mblue,minimum size=1cm,circle] (N0) at (0,0){$0$};
        \node[draw,fill=mblue,minimum size=1cm,circle] (N1) at (-1.5,-1.5){$1$};
        \node[draw,fill=mblue,minimum size=1cm,circle] (N2) at (-2.5,-3){$2$};
        \node[draw,fill=mblue,minimum size=1cm,circle] (N3) at (-1,-3){$3$};
        \node[draw,fill=mblue,minimum size=1cm,circle] (N4) at (0,-1.5){$4$};
        \node[draw,fill=mblue,minimum size=1cm,circle] (N5) at (1.5,-1.5){new};

        \draw (N0) -- (N1) -- (N2);
        \draw (N1) -- (N3);
        \draw (N0) -- (N4);
        \draw (N0) -- (N5);

        \draw[mred, ultra thick] (N0) -- (0,1);
        \draw[mred, ultra thick] (N4) -- (0,-2.5);
        \draw[mred, ultra thick] (N4) -- (0.5,-2.3);
        \draw[mred, ultra thick] (N2) -- (-2.7,-4);
        \draw[mred, ultra thick] (N3) -- (-0.8,-4);
        \draw[mred, ultra thick] (N5) -- (2,-2.7);

        \node[anchor=west] at (0,1) {\color{mred} $2$};
        \node[anchor=west] at (0,-2.5) {\color{mred} $2$};
        \node[anchor=west] at (0.5,-2.3) {\color{mred} $3$};
        \node[anchor=west] at (-2.7,-4) {\color{mred} $2$};
        \node[anchor=west] at (-0.8,-4) {\color{mred} $2$};
        \node[anchor=west] at (2,-2.7) {\color{mred} $2$};

        \node[anchor=east] at (-0.75,-0.7) {$4$};
        \node[anchor=east] at (0,-0.75) {$5$};
        \node[anchor=west] at (0.75,-0.7) {$3$};
        \node[anchor=west] at (-2,-2.25) {$2$};
        \node[anchor=west] at (-1.25,-2.25) {$2$};
    \end{tikzpicture}.
\end{equation}
We can see that nodes $5$ and $6$ were combined into the node \texttt{new}. The general leg order convention \eqref{eq:leg_convention} is kept. However, the other legs are sorted similarly to the convention of NumPy's \texttt{tensordot}. This means the children of the first specified node are the first children legs of the contracted node, the same with the open legs.
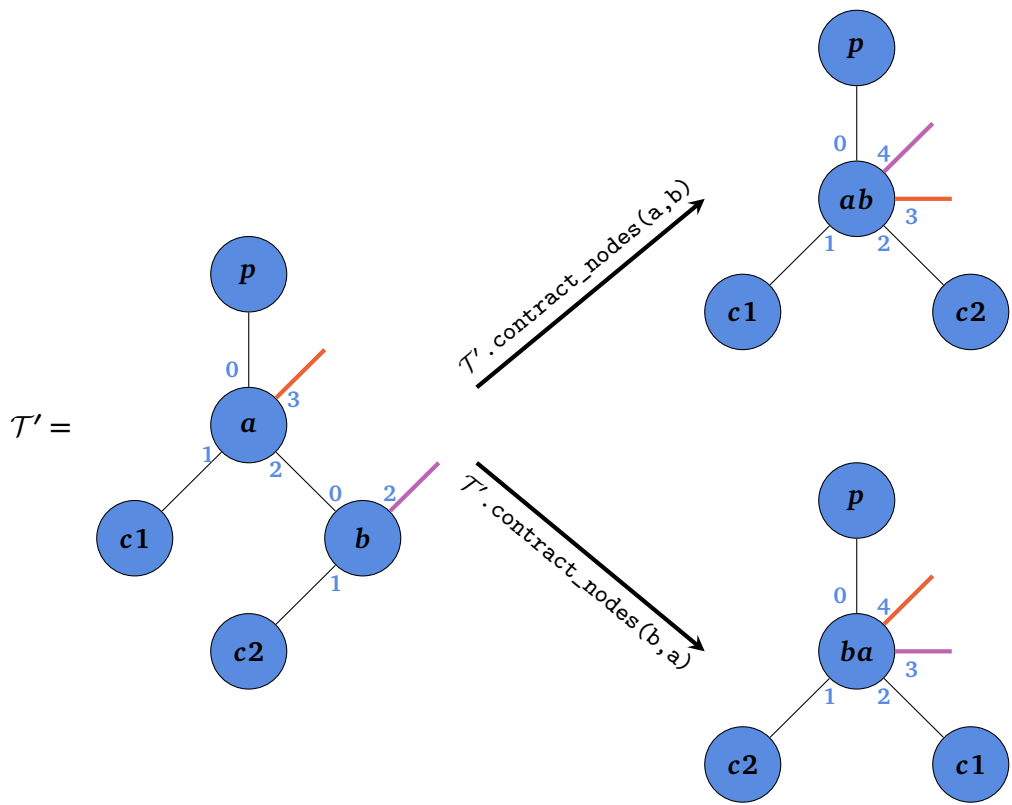
\begin{figure}
    \centering
    \begin{tikzpicture}[>=stealth]
        \node[anchor=east] at (-2.2,0) {$\mathcal{T}'=$};
        \node[fill=mblue,draw,circle,minimum size = 1cm] (a) at (0,0) {$a$};
        \node[fill=mblue,draw,circle,minimum size = 1cm] (p) at (0,2) {$p$};
        \node[fill=mblue,draw,circle,minimum size = 1cm] (c1) at (-1.5,-1.5) {$c1$};
        \node[fill=mblue,draw,circle,minimum size = 1cm] (b) at (1.5,-1.5) {$b$};
        \node[fill=mblue,draw,circle,minimum size = 1cm] (c2) at (0,-3) {$c2$};

        \draw (p) -- (a) -- (b) -- (c2);
        \draw (a) -- (c1);
        \draw[mred, ultra thick] (a) -- (1,1);
        \draw[mpurple, ultra thick] (b) -- (2.5,-0.5);

        \node[anchor=south east] at (a.90){\footnotesize\color{mblue} $0$};
        \node[anchor=east] at (a.230){\footnotesize\color{mblue} $1$};
        \node[anchor=north] at (a.315){\footnotesize\color{mblue} $2$};
        \node[anchor=west] at (a.42){\footnotesize\color{mblue} $3$};

        \node[anchor=south] at (b.135){\footnotesize\color{mblue} $0$};
        \node[anchor=north] at (b.225){\footnotesize\color{mblue} $1$};
        \node[anchor=south] at (b.45){\footnotesize\color{mblue} $2$};

        \draw[->, ultra thick] (3,0.5) -- (6,3) node[midway,above,sloped] () {{\footnotesize $\mathcal{T}'\texttt{.contract\_nodes(a,b)}$}};
        \draw[->, ultra thick] (3,-0.5) -- (6,-3) node[midway,below,sloped] () {{\footnotesize $\mathcal{T}'\texttt{.contract\_nodes(b,a)}$}};

        \begin{scope}[shift={(8,3)}]
            \node[fill=mblue,draw,circle,minimum size = 1cm] (ab) at (0,0) {$ab$};
            \node[fill=mblue,draw,circle,minimum size = 1cm] (p) at (0,2) {$p$};
            \node[fill=mblue,draw,circle,minimum size = 1cm] (c1) at (-1.5,-1.5) {$c1$};
            \node[fill=mblue,draw,circle,minimum size = 1cm] (c2) at (1.5,-1.5) {$c2$};
    
            \draw (p) -- (ab) -- (c2);
            \draw (ab) -- (c1);
            \draw[mpurple, ultra thick] (ab) -- (1,1);
            \draw[mred, ultra thick] (ab) -- (1.25,0);
    
            \node[anchor=south east] at (ab.90){\footnotesize\color{mblue} $0$};
            \node[anchor=north] at (ab.225){\footnotesize\color{mblue} $1$};
            \node[anchor=north] at (ab.315){\footnotesize\color{mblue} $2$};
            \node[anchor=north west] at (ab.0){\footnotesize\color{mblue} $3$};
            \node[anchor=south] at (ab.45){\footnotesize\color{mblue} $4$};
        \end{scope}

        \begin{scope}[shift={(8,-3)}]
            \node[fill=mblue,draw,circle,minimum size = 1cm] (ab) at (0,0) {$ba$};
            \node[fill=mblue,draw,circle,minimum size = 1cm] (p) at (0,2) {$p$};
            \node[fill=mblue,draw,circle,minimum size = 1cm] (c1) at (-1.5,-1.5) {$c2$};
            \node[fill=mblue,draw,circle,minimum size = 1cm] (c2) at (1.5,-1.5) {$c1$};
    
            \draw (p) -- (ab) -- (c2);
            \draw (ab) -- (c1);
            \draw[mred, ultra thick] (ab) -- (1,1);
            \draw[mpurple, ultra thick] (ab) -- (1.25,0);
    
            \node[anchor=south east] at (ab.90){\footnotesize\color{mblue} $0$};
            \node[anchor=north] at (ab.225){\footnotesize\color{mblue} $1$};
            \node[anchor=north] at (ab.315){\footnotesize\color{mblue} $2$};
            \node[anchor=north west] at (ab.0){\footnotesize\color{mblue} $3$};
            \node[anchor=south] at (ab.45){\footnotesize\color{mblue} $4$};
        \end{scope}
    \end{tikzpicture}
    \caption{Exemplifies the leg order of the node yielded by the contraction of two nodes. The leg indices are indicated by the blue numbers. In this example, we contract the nodes $a$ and $b$ of the TTN $\mathcal{T}'$. The different TTN resulting from the two different orders of contraction are shown on the right. We can see that the most significant difference is the leg order of the children and open legs. The latter are shown in a different colour to be more easily identifiable.}
    \label{fig:node_contr_example}
\end{figure}
Figure~\ref{fig:node_contr_example} gives a minimum example of this. Here, the nodes $a$ and $b$ of a TTN $\mathcal{T}'$ are contracted. Each has a child and an open leg. The leg order of the resulting node depends on the position of the node identifiers as arguments in the TTN method \texttt{contract\_nodes}. In both cases, the leg towards the parent node has index $0$, as the leg convention \eqref{eq:leg_convention} intended. While we can also see that the legs towards the children come before the open legs, their specific order changes in the two different cases. In case $a$ is specified as the first node in the contraction, its child leg towards $c1$ and its open leg are legs $1$ and $3$ respectively. However, if $b$ is the first node, these two legs have the indices $2$ and $4$. In turn, for the child leg of $b$ towards the node $c2$ and the open leg of $b$, the leg indices in the first case are $2$ and $3$, respectively, while they are $1$ and $3$ in the other case, where $b$ is the first specified node. However, one must only be careful with regard to the open legs, as the neighbour legs are easily identified by the identifiers of the connected nodes. One can also easily contract a whole TTN by calling the method \texttt{completely\_contract\_tree}. For example for our already partially contracted $\mathcal{T}$ as shown in \eqref{eq:ttn_one_node_contr}
\code{132}{135}
This is useful for testing code using TTNs with few open legs, but can cause significant memory usage even for a medium number of open legs.
\begin{exercise}{Contracting a TTN}
    Contract the TTN you constructed in Exercise~\ref{exc:build_a_ttn} in two different ways:
    \begin{enumerate}
        \item Contract the nodes individually, i.e. using the \texttt{contract\_nodes} method;
        \item Contract the complete TTN using the \texttt{completely\_contract\_tree}.
    \end{enumerate}
    Check that the resulting tensor is the same in both cases. (Hint: Make sure to use the same contraction order in both cases or transpose the leg of either resulting tensor.)
\end{exercise}

As mentioned before, also the splitting of nodes in a TTN can be done using built-in methods. Once we have a node we want to split, we have to define which two groups we want to organise the legs into. For this, we have the \texttt{LegSpecification} class. We require one leg specification for each new node resulting from a split. In the specification, we define which children and open legs should belong to the new node and which new node has the parent leg. If the node to be split does not have a parent, so it is the root, we also need to specify which new node is supposed to be the new root. For example, if we want to split the node $ab$ of the tensor network in the top right of Figure~\ref{fig:node_contr_example}, we would use the following specifications to reobtain $\mathcal{T}'$
\code{158}{161}
As discussed in Section~\ref{sec:tensor_networks}, the QR- and SV-Decomposition are the two most used splitting methods in tensor networks. Both decompositions are directly implemented as methods in the \texttt{TreeTensorNetwork} class as \texttt{split\_node\_qr} and \texttt{split\_node\_svd}. Their signature is similar to the one for the functions introduced in Section~\ref{sec:tensor_networks}. However, instead of supplying the tensor to split, we provide the node identifier, and instead of supplying the leg indices, we provide two \texttt{LegSpecifications}. Furthermore, the identifiers of the new nodes can be specified. For example, if we use the specifications defined above, we can split the $ab$ node with the QR-Decomposition using the code
\code{162}{164}
In this way, we recover the original tree structure of $\mathcal{T}'$ as shown on the left of Figure~\ref{fig:node_contr_example}. However, the actual numeric tensors corresponding to nodes $a$ and $b$ might be different and notably, the tensor of node $a$ is an isometry. If we desire to split a root node, such as node $0$ of $\mathcal{T}$ in \eqref{eq:example_ttn}, like this
\begin{equation}
    \begin{tikzpicture}
        \node[fill=mblue,draw,circle,minimum size=1cm] (N0) at (0,0){$0$};
        \draw (N0) -- (-1,-1);
        \draw (N0) -- (0,-1);
        \draw (N0) -- (1,-1);
        \draw[mred,ultra thick] (N0) -- (0,1);
        \draw[dashed] (-0.5,0.8) -- (0.5,-0.8);
    \end{tikzpicture}
    \raisebox{0.8cm}{\quad $\rightarrow$ \quad}
    \begin{tikzpicture}
        \node[fill=mblue,draw,circle,minimum size=1cm] (R) at (0,0){$01$};
        \draw (R) -- (-1,-1.5);
        \draw (R) -- (0,-1.5);

        \node[fill=mblue,draw,circle,minimum size=1cm] (N0) at (1,-1){$02$};
        \draw (R) -- (N0);
        \draw (N0) -- (2,-1.5);
        \draw[mred,ultra thick] (N0) -- (1,0);
    \end{tikzpicture},
\end{equation}
we need to specify the new root in the code:
\code{168}{176}
Note that it is easy to implement a new splitting method using the general \texttt{split\_nodes} method of the \texttt{TreeTensorNetwork} class. Among other things, the combination of splitting and contracting tensors in a TTN can be used to bring a TTN into a special form, which will be discussed in the following subsection.
\begin{exercise}{Building a TTN the other way}
    Create a TTN with only a root node of shape $(2,3)$. Obtain a TTN with the tree topology of $\mathcal{T}'$ (see Figure~\ref{fig:node_contr_example}) from the single node TTN by using the node splitting methods. The open leg of node $a$ and node $b$ should have dimensions $3$ and $2$, respectively.
\end{exercise}

\subsection{The Canonical Form}\label{sec:canonical_form}
The canonical form of a TTN is a special form of TTN, where all but one node, called the \emph{orthogonality centre}, fulfils an orthogonality condition. More specific \cite{Bauernfeind2020}
\begin{defn}
    A tree tensor network $\mathcal{T}$ is in \emph{canonical form} if there exists a node $N \in \mathcal{T}$ called the orthogonality centre, such that for all other nodes $A \in \mathcal{T}$ the following holds
    \begin{equation}\label{eq:canon_def_indices}
        \sum_{i_1 \dots i_m} A^*_{i_1 \dots i_m q} A_{i_1 \dots i_m q'} = \delta_{qq'},
    \end{equation}
    where the leg denoted by index $q$ is the unique leg of $A$ pointing towards $N$. We assumed that $q$ is the last leg of $A$ to keep the notation simple. In principle, $q$ can be any leg of $A$. $\delta$ denotes the Kronecker/identity tensor, which is $1$ if both indices coincide and $0$ otherwise. We can alternatively write condition \eqref{eq:canon_def_indices} as
    \begin{equation}\label{eq:orth_condition}
        \begin{tikzpicture}
            \node[isosceles triangle,
            	   isosceles triangle apex angle=70,
            	   draw,
            	   fill=mred,
            	   minimum size=1cm] (A) at (0,0){$A$};

            \node[isosceles triangle,
            	   isosceles triangle apex angle=70,
            	   draw,
            	   fill=mred,
            	   minimum size=1cm] (Astar) at (0,2.5){$A^*$};
            \draw (A.right corner) to[out=180,in=180] (Astar.left corner);
            \draw (A.left corner) to[out=180,in=180] (Astar.right corner);
            \draw (A.150) to[out=180,in=180] (Astar.210);
            \node[anchor=east] at (A.210) {$\vdots$};
            \node[anchor=east] at (Astar.130) {$\vdots$};
            \draw (A.apex) -- (1.5,0);
            \draw (Astar.apex) -- (1.5,2.5);
            \node at (2,1.25){$=$};
            \begin{scope}[shift={(3.5,0)}]
                \draw (1.5,0) -- (0,0) to[out=180,in=180] (0,2.5) -- (1.5,2.5);
            \end{scope}
        \end{tikzpicture},
    \end{equation}
    where the line without a tensor represents the identity.
\end{defn}
While slightly different definitions of the canonical form exist \cite{Shi2006}, we will stick to the definition above, as it generalises the mixed/site-canonical form of the matrix product picture \cite{Schollwock2011, Catarina2023}.\\

Can we find a canonical form for every TTN? To answer this question, note that there is a gauge freedom inherent in the contracted legs of a TTN. We can insert $G G^{-1}$ for any invertible matrix $G$ in between every two contracted tensors. For example
\begin{equation}
    \begin{tikzpicture}
        \def\minsize{3}
        \def\nodedist{1}
        \node[fill=mblue,draw,circle,minimum size=\minsize] (A) at (0,0){$A$};
        \node[fill=mblue,draw,circle,minimum size=\minsize] (B) at (\nodedist,0){$B$};
        \draw (-\nodedist,0) -- (A) -- (B);
        \draw (\nodedist+0.7*\nodedist,0.7*\nodedist) -- (B);
        \draw (\nodedist+0.7*\nodedist,-0.7*\nodedist) -- (B);
    \end{tikzpicture}
    \raisebox{0.6cm}{\, = \,}
    \begin{tikzpicture}
        \def\minsize{3}
        \def\minsizesq{4}
        \def\nodedist{1}
        \node[fill=mblue,draw,circle,minimum size=\minsize] (A) at (0,0){$A$};
        \node[fill=mgreen,draw,rectangle,minimum size=\minsizesq] (G) at (\nodedist,0){$G$};
        \node[fill=mgreen,draw,rectangle,minimum size=\minsizesq] (Ginv) at (2*\nodedist,0){$G^{-1}$};
        \node[fill=mblue,draw,circle,minimum size=\minsize] (B) at (3*\nodedist,0){$B$};
        \draw (-\nodedist,0) -- (A) -- (G) -- (Ginv) -- (B);
        \draw (3*\nodedist+0.7*\nodedist,0.7*\nodedist) -- (B);
        \draw (3*\nodedist+0.7*\nodedist,-0.7*\nodedist) -- (B);
    \end{tikzpicture}
    \raisebox{0.6cm}{\, = \,}
    \begin{tikzpicture}
        \def\minsize{3}
        \def\nodedist{1}
        \node[fill=mred,draw,circle,minimum size=\minsize] (A) at (0,0){$\tilde{A}$};
        \node[fill=mred,draw,circle,minimum size=\minsize] (B) at (\nodedist,0){$\tilde{B}$};
        \draw (-\nodedist,0) -- (A) -- (B);
        \draw (\nodedist+0.7*\nodedist,0.7*\nodedist) -- (B);
        \draw (\nodedist+0.7*\nodedist,-0.7*\nodedist) -- (B);
    \end{tikzpicture}.
\end{equation}
Strictly speaking, inserting $G G^{-1}$ into a TTN $\mathcal{T}$ creates a different TTN $\mathcal{T}'$. The tensor elements and possibly the tree underlying $\mathcal{T}$ and $\mathcal{T}'$ are different. However, contracting each TTN fully will yield the same tensor for both. Therefore, $\mathcal{T}$ and $\mathcal{T}'$ can be treated as equivalent for almost all use cases. A thorough exploration of the impacts of the gauge freedom on TTN can be found in \cite{Silvi2019}.\\
This equivalency allows us to bring any TTN into a canonical form. We can choose any node as the orthogonality centre and use QR-tensor-splitting to fulfil \eqref{eq:canon_def_indices} for all other tensors. To do so for any node $A$, that is not the orthogonality center, we choose the leg towards the planned orthogonalisation center to be the only leg of the new $R$ node. $R$ is then absorbed into the other node the leg is connected to, leaving $A$ with a tensor that is an isometry towards the orthogonality centre. Starting from the nodes furthest from the orthogonalisation centre, we can slowly but surely turn the entire TTN into the correct form. This scheme is sometimes called \emph{pulling-through approach} \cite{Evenbly2022}. As an example, consider the TTN $\mathcal{T}$ in \ref{eq:example_ttn}. If we want to bring $\mathcal{T}$ into canonical form with respect to node $0$ as the orthogonalisation center using {\ptn}, we merely need to write
\code{199}{199}
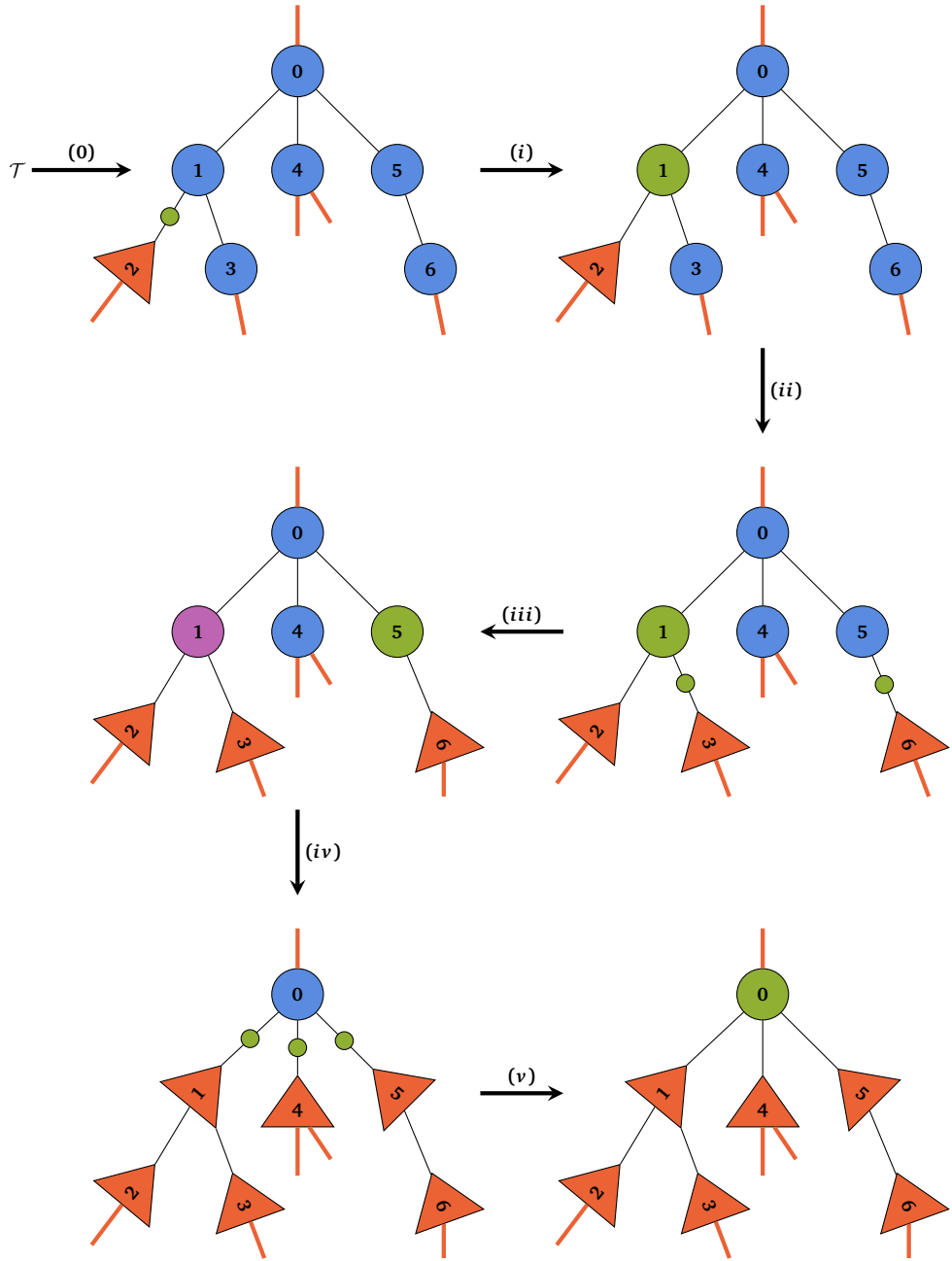
\begin{figure}
    \centering
    \begin{tikzpicture}[>=stealth,scale=0.9,every node/.style={scale=0.7}]
    \node[anchor=east] at (-4,-1.5) {$\mathcal{T}$};
    \draw[ultra thick,->] (-4,-1.5) -- (-2.5,-1.5) node[midway,above] {$(0)$};

    \node[draw,fill=mblue,minimum size=1cm,circle] (N0) at (0,0){$0$};
    \node[draw,fill=mblue,minimum size=1cm,circle] (N1) at (-1.5,-1.5){$1$};
    \node[draw,fill=mred,minimum size=1cm,
        isosceles triangle,
        isosceles triangle apex angle=70,
        rotate=50] (N2) at (-2.5,-3){$2$};
    \node[draw,fill=mblue,minimum size=1cm,circle] (N3) at (-1,-3){$3$};
    \node[draw,fill=mblue,minimum size=1cm,circle] (N4) at (0,-1.5){$4$};
    \node[draw,fill=mblue,minimum size=1cm,circle] (N5) at (1.5,-1.5){$5$};
    \node[draw,fill=mblue,minimum size=1cm,circle] (N6) at (2,-3){$6$};

    \draw (N0) -- (N1) -- (N2.apex) node[draw,fill=mgreen,minimum size=0.1cm,circle,midway] (R2){};
    \draw (N1) -- (N3);
    \draw (N0) -- (N4);
    \draw (N0) -- (N5) -- (N6);

    \draw[mred, ultra thick] (N0) -- (0,1);
    \draw[mred, ultra thick] (N4) -- (0,-2.5);
    \draw[mred, ultra thick] (N4) -- (0.5,-2.3);
    \draw[mred, ultra thick] (N2) -- (-3.1,-3.8);
    \draw[mred, ultra thick] (N3) -- (-0.8,-4);
    \draw[mred, ultra thick] (N6) -- (2.2,-4);

    \draw[ultra thick,->] (2.75,-1.5) -- (4,-1.5) node[midway,above] {$(i)$};
    
    \begin{scope}[shift={(7,0)}]
        \node[draw,fill=mblue,minimum size=1cm,circle] (N0) at (0,0){$0$};
        \node[draw,fill=mgreen,minimum size=1cm,circle] (N1) at (-1.5,-1.5){$1$};
        \node[draw,fill=mred,minimum size=1cm,
            isosceles triangle,
            isosceles triangle apex angle=70,
            rotate=50] (N2) at (-2.5,-3){$2$};
        \node[draw,fill=mblue,minimum size=1cm,circle] (N3) at (-1,-3){$3$};
        \node[draw,fill=mblue,minimum size=1cm,circle] (N4) at (0,-1.5){$4$};
        \node[draw,fill=mblue,minimum size=1cm,circle] (N5) at (1.5,-1.5){$5$};
        \node[draw,fill=mblue,minimum size=1cm,circle] (N6) at (2,-3){$6$};

        \draw (N0) -- (N1) -- (N2.apex);
        \draw (N1) -- (N3);
        \draw (N0) -- (N4);
        \draw (N0) -- (N5) -- (N6);
    
        \draw[mred, ultra thick] (N0) -- (0,1);
        \draw[mred, ultra thick] (N4) -- (0,-2.5);
        \draw[mred, ultra thick] (N4) -- (0.5,-2.3);
        \draw[mred, ultra thick] (N2) -- (-3.1,-3.8);
        \draw[mred, ultra thick] (N3) -- (-0.8,-4);
        \draw[mred, ultra thick] (N6) -- (2.2,-4);
    \end{scope}

    \draw[ultra thick,->] (7,-4.2) -- (7,-5.5) node[midway,right] {$(ii)$};

    \begin{scope}[shift={(7,-7)}]
        \node[draw,fill=mblue,minimum size=1cm,circle] (N0) at (0,0){$0$};
        \node[draw,fill=mgreen,minimum size=1cm,circle] (N1) at (-1.5,-1.5){$1$};
        \node[draw,fill=mred,minimum size=1cm,
            isosceles triangle,
            isosceles triangle apex angle=70,
            rotate=50] (N2) at (-2.5,-3){$2$};
        \node[draw,fill=mred,minimum size=1cm,
            isosceles triangle,
            isosceles triangle apex angle=70,
            rotate=110] (N3) at (-0.8,-3.2){};
        \node[rotate=-70] at (N3){$3$};
        \node[draw,fill=mblue,minimum size=1cm,circle] (N4) at (0,-1.5){$4$};
        \node[draw,fill=mblue,minimum size=1cm,circle] (N5) at (1.5,-1.5){$5$};
        \node[draw,fill=mred,minimum size=1cm,
            isosceles triangle,
            isosceles triangle apex angle=70,
            rotate=110] (N6) at (2.2,-3.2){};
        \node[rotate=-70] at (N6){$6$};
    
        \draw (N0) -- (N1) -- (N2.apex);
        \draw (N1) -- (N3.apex) node[draw,fill=mgreen,minimum size=0.1cm,circle,midway]{};
        \draw (N0) -- (N4);
        \draw (N0) -- (N5) -- (N6) node[draw,fill=mgreen,minimum size=0.1cm,circle,midway]{};
    
        \draw[mred, ultra thick] (N0) -- (0,1);
        \draw[mred, ultra thick] (N4) -- (0,-2.5);
        \draw[mred, ultra thick] (N4) -- (0.5,-2.3);
        \draw[mred, ultra thick] (N2) -- (-3.1,-3.8);
        \draw[mred, ultra thick] (N3) -- (-0.5,-4);
        \draw[mred, ultra thick] (N6) -- (2.5,-4);
    \end{scope}

    \draw[ultra thick,->] (4,-8.5) -- (2.75,-8.5) node[midway,above] {$(iii)$};

    \begin{scope}[shift={(0,-7)}]
        \node[draw,fill=mblue,minimum size=1cm,circle] (N0) at (0,0){$0$};
        \node[draw,fill=mpurple,minimum size=1cm,circle] (N1) at (-1.5,-1.5){$1$};
        \node[draw,fill=mred,minimum size=1cm,
            isosceles triangle,
            isosceles triangle apex angle=70,
            rotate=50] (N2) at (-2.5,-3){$2$};
        \node[draw,fill=mred,minimum size=1cm,
            isosceles triangle,
            isosceles triangle apex angle=70,
            rotate=110] (N3) at (-0.8,-3.2){};
        \node[rotate=-70] at (N3){$3$};
        \node[draw,fill=mblue,minimum size=1cm,circle] (N4) at (0,-1.5){$4$};
        \node[draw,fill=mgreen,minimum size=1cm,circle] (N5) at (1.5,-1.5){$5$};
        \node[draw,fill=mred,minimum size=1cm,
            isosceles triangle,
            isosceles triangle apex angle=70,
            rotate=110] (N6) at (2.2,-3.2){};
        \node[rotate=-70] at (N6){$6$};
    
        \draw (N0) -- (N1) -- (N2.apex);
        \draw (N1) -- (N3.apex);
        \draw (N0) -- (N4);
        \draw (N0) -- (N5) -- (N6);
    
        \draw[mred, ultra thick] (N0) -- (0,1);
        \draw[mred, ultra thick] (N4) -- (0,-2.5);
        \draw[mred, ultra thick] (N4) -- (0.5,-2.3);
        \draw[mred, ultra thick] (N2) -- (-3.1,-3.8);
        \draw[mred, ultra thick] (N3) -- (-0.5,-4);
        \draw[mred, ultra thick] (N6) -- (2.2,-4);
    \end{scope}

    \draw[ultra thick,->] (0,-11.2) -- (0,-12.5) node[midway,right] {$(iv)$};

    \begin{scope}[shift={(0,-14)}]
        \node[draw,fill=mblue,minimum size=1cm,circle] (N0) at (0,0){$0$};
        \node[draw,fill=mred,minimum size=1cm,
            isosceles triangle,
            isosceles triangle apex angle=70,
            rotate=50] (N1) at (-1.5,-1.5){$1$};
        \node[draw,fill=mred,minimum size=1cm,
            isosceles triangle,
            isosceles triangle apex angle=70,
            rotate=50] (N2) at (-2.5,-3){$2$};
        \node[draw,fill=mred,minimum size=1cm,
            isosceles triangle,
            isosceles triangle apex angle=70,
            rotate=110] (N3) at (-0.8,-3.2){};
        \node[rotate=-70] at (N3){$3$};
        \node[draw,fill=mred,minimum size=1cm,
            isosceles triangle,
            isosceles triangle apex angle=70,
            rotate=90] (N4) at (0,-1.75){};
        \node at (N4){$4$};
        \node[draw,fill=mred,minimum size=1cm,
            isosceles triangle,
            isosceles triangle apex angle=70,
            rotate=135] (N5) at (1.5,-1.5){};
        \node[rotate=-40] at (N5){$5$};
        \node[draw,fill=mred,minimum size=1cm,
            isosceles triangle,
            isosceles triangle apex angle=70,
            rotate=110] (N6) at (2.2,-3.2){};
        \node[rotate=-70] at (N6){$6$};
    
        \draw (N0) -- (N1.apex) node[draw,fill=mgreen,minimum size=0.1cm,circle,midway]{};
        \draw (N1) -- (N2.apex);
        \draw (N1.right corner) -- (N3.apex);
        \draw (N0) -- (N4) node[draw,fill=mgreen,minimum size=0.1cm,circle,midway]{};
        \draw (N0) -- (N5) node[draw,fill=mgreen,minimum size=0.1cm,circle,midway]{} -- (N6);
    
        \draw[mred, ultra thick] (N0) -- (0,1);
        \draw[mred, ultra thick] (N4) -- (0,-2.75);
        \draw[mred, ultra thick] (N4) -- (0.5,-2.5);
        \draw[mred, ultra thick] (N2) -- (-3.1,-3.8);
        \draw[mred, ultra thick] (N3) -- (-0.5,-4);
        \draw[mred, ultra thick] (N6) -- (2.2,-4);
    \end{scope}

    \draw[ultra thick,->] (2.75,-15.5) -- (4,-15.5) node[midway,above] {$(v)$};

    \begin{scope}[shift={(7,-14)}]
        \node[draw,fill=mgreen,minimum size=1cm,circle] (N0) at (0,0){$0$};
        \node[draw,fill=mred,minimum size=1cm,
            isosceles triangle,
            isosceles triangle apex angle=70,
            rotate=50] (N1) at (-1.5,-1.5){$1$};
        \node[draw,fill=mred,minimum size=1cm,
            isosceles triangle,
            isosceles triangle apex angle=70,
            rotate=50] (N2) at (-2.5,-3){$2$};
        \node[draw,fill=mred,minimum size=1cm,
            isosceles triangle,
            isosceles triangle apex angle=70,
            rotate=110] (N3) at (-0.8,-3.2){};
        \node[rotate=-70] at (N3){$3$};
        \node[draw,fill=mred,minimum size=1cm,
            isosceles triangle,
            isosceles triangle apex angle=70,
            rotate=90] (N4) at (0,-1.75){};
        \node at (N4){$4$};
        \node[draw,fill=mred,minimum size=1cm,
            isosceles triangle,
            isosceles triangle apex angle=70,
            rotate=135] (N5) at (1.5,-1.5){};
        \node[rotate=-40] at (N5){$5$};
        \node[draw,fill=mred,minimum size=1cm,
            isosceles triangle,
            isosceles triangle apex angle=70,
            rotate=110] (N6) at (2.2,-3.2){};
        \node[rotate=-70] at (N6){$6$};
    
        \draw (N0) -- (N1.apex);
        \draw (N1) -- (N2.apex);
        \draw (N1.right corner) -- (N3.apex);
        \draw (N0) -- (N4);
        \draw (N0) -- (N5) -- (N6);
    
        \draw[mred, ultra thick] (N0) -- (0,1);
        \draw[mred, ultra thick] (N4) -- (0,-2.75);
        \draw[mred, ultra thick] (N4) -- (0.5,-2.5);
        \draw[mred, ultra thick] (N2) -- (-3.1,-3.8);
        \draw[mred, ultra thick] (N3) -- (-0.5,-4);
        \draw[mred, ultra thick] (N6) -- (2.2,-4);
    \end{scope}

\end{tikzpicture}
    \caption{Transforming the example TTN $\mathcal{T}$ given in \eqref{eq:example_ttn} into canonical form with respect to node $0$. In step $(0)$, node $2$ is split into two nodes. At the position of node $2$, an isometric tensor is left. The direction of the isometry is denoted by the triangle. The other node is the $R$-tensor originating from the QR-decomposition \eqref{eq:mat_qr}. The $R$-tensor is then contracted with node $1$. In steps $(ii)$ and $(iii)$, this is repeated for the remaining nodes of the given level. Steps $(iv)$ and $(v)$ perform the canonicalisation for the direct neighbours of node $0$. Now node $0$ is the orthogonalisation center and thus the only node not depicted as a triangle.}
    \label{fig:canonical_form}
\end{figure}
So, all the details are handled using a single method. However, for pedagogical purposes, the detailed process of bringing the example TTN into canonical form is shown in Figure~\ref{fig:canonical_form}. If it exists, the current orthogonalisation centre identifier of a TTN can be accessed using the \texttt{orthogonalisation\_center} attribute. We can also move an existing orthogonality centre. We find the path between the current orthogonality centre and the new one to achieve this. Then, we perform QR-decompositions and contractions with the new $R$ node along that path. This moves the orthogonality centre until it reaches the desired node. Once more, this is covered by a method of the \texttt{TreeTensorNetwork} class in {\ptn}. We merely need to specify the new orthogonalisation centre. For example, if we want to move the centre of $\mathcal{T}$ from node $0$ to node $6$, we write
\code{203}{203}
Having a TTN in canonical form has many advantages, such as causing minimal error during truncation and allowing for quick evaluation of local properties \cite{Schollwock2005, Schollwock2011}. The latter makes a significant difference for single-node expectation values if the TTN represents a quantum state. This exact situation will be illuminated more closely in the next section.
\begin{exercise}{Canonicalise twice}
    In this exercise, we will again consider the TTN $\mathcal{T}'$ shown in Figure~\ref{fig:node_contr_example}.
    \begin{enumerate}
        \item Bring the TTN in canonical form with respect to the root node.
        \item Move the orthogonalisation centre to node $b$.
        \item How can you check that the TTN is actually in the correct form?
        \item What happens, if another ``canonicalisation step'' is performed on node $b$, where the leg chosen to be the $R$ node's leg is the open leg?
    \end{enumerate}
\end{exercise}

\section{Tree Tensor Network States}\label{sec:TTNS}
A \emph{Tree Tensor Network State} (TTNS) is a tree tensor network representing a quantum state $\ket{\psi} \in \C^D$ \cite{Shi2006}. As $\ket{\psi}$ is usually a many-body state, we can reinterpret the Hilbert space $\C^D = \C^{d_1 \times \cdots \times d_L}$, where $L$ is the number of local quantum systems making up the many-body system. Then we write
\begin{equation}
    \ket{\psi} = \sum_{i_1 \cdots i_L} C_{i_1, \cdots , i_L} \ket{i_1 \cdots i_L},
\end{equation}
where ${\ket{i}}_{i=0}^{d_j-1}$ is a basis of the local quantum system $j$. To obtain a desired state as a TTNS, one can, in principle, start with the high degree tensor $C$ representing $\ket{\psi}$. Using the tensor decompositions discussed in Section~\ref{sec:contr_and_split_of_ttn}, $C$ is consecutively brought into the desired tree topology. However, the requirement to have the full tensor defeats the purpose of using tensor networks to reduce memory requirements. Therefore, most initial TTNS are states that can easily be brought into a TTN form, such as product states. It is also noteworthy that the tree topology that minimises the memory requirements to represent a given state is usually not obvious \cite{Hikihara2023, Hikihara2024}. Although in principle a matter of interpretation, in {\ptn} we assume that a TTNS has exactly one open leg for every node. But, this leg can be trivial, i.e. of dimension $1$. So our usual example TTN $\mathcal{T}$ as given in \eqref{eq:example_ttn} is not a TTNS because node $4$ has two open legs. However, the nodes $1$ and $5$ not having one would not be a problem, as we can think of them having a trivial open leg each. But we have to explicitly state this in our construction. Instead of $\mathcal{T}$, we will, from now on, use the following tree tensor network state as an example
\begin{equation}\label{eq:example_ttns}
    \raisebox{1.5cm}{$\ket{T}= \quad$}
    \begin{tikzpicture}
        \def\nodedist{1.3}
        \def\px{0.75};
        \def\py{0.75};
        \node[fill=mblue,draw,circle,minimum size=0.9cm] (R) at (0,0){$0$};
        \node[fill=mblue,draw,circle,minimum size=0.9cm] (N00) at (-\nodedist,0){$00$};
        \node[fill=mblue,draw,circle,minimum size=0.9cm] (N01) at (-2*\nodedist,0){$01$};
        \node[fill=mblue,draw,circle,minimum size=0.9cm] (N10) at (0,-\nodedist){$10$};
        \node[fill=mblue,draw,circle,minimum size=0.9cm] (N11) at (0,-2*\nodedist){$11$};
        \node[fill=mblue,draw,circle,minimum size=0.9cm] (N20) at (\nodedist,0){$20$};
        \node[fill=mblue,draw,circle,minimum size=0.9cm] (N21) at (2*\nodedist,0){$21$};
        \draw (N01) -- (N00) -- (R) -- (N20) -- (N21);
        \draw (R) -- (N10) -- (N11);
        \draw[mred,ultra thick] (R) -- ($(R) + (\px,\py)$);
        \draw[mred,ultra thick] (N00) -- ($(N00) + (\px,\py)$);
        \draw[mred,ultra thick] (N01) -- ($(N01) + (\px,\py)$);
        \draw[mred,ultra thick] (N10) -- ($(N10) + (\px,\py)$);
        \draw[mred,ultra thick] (N11) -- ($(N11) + (\px,\py)$);
        \draw[mred,ultra thick] (N20) -- ($(N20) + (\px,\py)$);
        \draw[mred,ultra thick] (N21) -- ($(N21) + (\px,\py)$);
    \end{tikzpicture}
\end{equation}
In {\ptn}, TTNS can be created as a \texttt{TreeTensorNetworkState} object, a subclass of the \texttt{TreeTensorNetwork}. We could construct $T$ with
\code{207}{219}
While this tree structure is simple, it already causes most of the difficulties occurring during the time evolution of TTNS. The different ways to evolve a TTNS in time using {\ptn} will be explained in later sections.\\

On the other hand, one of the most frequently desired properties of a quantum state $\ket{T}$ is its norm
\begin{equation}
    \norm{T} = \sqrt{\braket{T}{T}}.
\end{equation}
The scalar product for a TTNS can easily be obtained by contracting the state with its complex conjugate
\begin{equation}\label{eq:scalar_prod_ttns}
    \raisebox{1.5cm}{$\braket{T}{T}= \quad$}
    \begin{tikzpicture}
        \def\nodedist{1.3}
        \def\px{0.75};
        \def\py{0.75};
        \node[fill=mpurple,draw,circle,minimum size=0.9cm] (Rs) at (0+\px,0+\py){$0^*$};
        \node[fill=mpurple,draw,circle,minimum size=0.9cm] (N00s) at (-\nodedist+\px,0+\py){$00^*$};
        \node[fill=mpurple,draw,circle,minimum size=0.9cm] (N01s) at (-2*\nodedist+\px,0+\py){$01^*$};
        \node[fill=mpurple,draw,circle,minimum size=0.9cm] (N10s) at (0+\px,-\nodedist+\py){$10^*$};
        \node[fill=mpurple,draw,circle,minimum size=0.9cm] (N11s) at (0+\px,-2*\nodedist+\py){$11^*$};
        \node[fill=mpurple,draw,circle,minimum size=0.9cm] (N20s) at (\nodedist+\px,0+\py){$20^*$};
        \node[fill=mpurple,draw,circle,minimum size=0.9cm] (N21s) at (2*\nodedist+\px,0+\py){$21^*$};
        \draw (N01s) -- (N00s) -- (Rs) -- (N20s) -- (N21s);
        \draw (Rs) -- (N10s) -- (N11s);

        \node[fill=mblue,draw,circle,minimum size=0.9cm] (R) at (0,0){$0$};
        \node[fill=mblue,draw,circle,minimum size=0.9cm] (N00) at (-\nodedist,0){$00$};
        \node[fill=mblue,draw,circle,minimum size=0.9cm] (N01) at (-2*\nodedist,0){$01$};
        \node[fill=mblue,draw,circle,minimum size=0.9cm] (N10) at (0,-\nodedist){$10$};
        \node[fill=mblue,draw,circle,minimum size=0.9cm] (N11) at (0,-2*\nodedist){$11$};
        \node[fill=mblue,draw,circle,minimum size=0.9cm] (N20) at (\nodedist,0){$20$};
        \node[fill=mblue,draw,circle,minimum size=0.9cm] (N21) at (2*\nodedist,0){$21$};
        \draw (N01) -- (N00) -- (R) -- (N20) -- (N21);
        \draw (R) -- (N10) -- (N11);
        \draw[mred,ultra thick] (R) -- ($(R) + (\px,\py)$);
        \draw[mred,ultra thick] (N00) -- ($(N00) + (\px,\py)$);
        \draw[mred,ultra thick] (N01) -- ($(N01) + (\px,\py)$);
        \draw[mred,ultra thick] (N10) -- ($(N10) + (\px,\py)$);
        \draw[mred,ultra thick] (N11) -- ($(N11) + (\px,\py)$);
        \draw[mred,ultra thick] (N20) -- ($(N20) + (\px,\py)$);
        \draw[mred,ultra thick] (N21) -- ($(N21) + (\px,\py)$);.
    \end{tikzpicture}
\end{equation}
In this case, we can find a major advantage in using the canonical form of a TTN. For example if node $20$ is the orthogonalisation center, the diagram in \eqref{eq:scalar_prod_ttns} simplifies to
\begin{equation}\label{eq:simplified_scalar_prod}
    \raisebox{1cm}{$\braket{T}{T}= \quad$}
    \begin{tikzpicture}
        \def\nodedist{1.3}
        \def\px{0.75};
        \def\py{0.75};
        \node[fill=mpurple,draw,circle,minimum size=0.9cm] (Rs) at (0+\px,0+\py){$20^*$};
        \node[fill=mblue,draw,circle,minimum size=0.9cm] (R) at (0,0){$20$};
        \draw[mred,ultra thick] (R) -- ($(R) + (\px,\py)$);
        \draw (R) -- (0.5,0) to[out=0,in=0] (0.5+\px,0+\py) -- (Rs);
        \draw (R) -- (-0.5,0) to[out=180,in=180] (-0.5+\px,\py) -- (Rs);
    \end{tikzpicture}.
\end{equation}
We can achieve this using the following scheme. We start at the leaf tensors of $\ket{T}$, which will cancel with their respective tensors in $\bra{T}$ due to \eqref{eq:orth_condition}. This will connect the virtual legs of their parents in both TTN directly. Using the orthogonality condition \eqref{eq:orth_condition} until only the orthogonality centre nodes are left yields the simplified tensor network \eqref{eq:simplified_scalar_prod}. This scheme can be used for any TTNS in canonical form. In Section \ref{sec:canonical_form}, it was already explained that bringing a TTN into the canonical form does not change the overall tensor represented by the TTN. The same is true for TTNS. Transforming a TTNS into the canonical form does not change the state it represents. Therefore, we do not lose any representation power by bringing a TTNS into canonical form but can avoid costly numerics when determining the norm of a TTNS.\\
We have the same advantage if we want to find the expectation value $\bra{T} O_s \ket{s}$ of a single-site operator $O_{s}$. That is the expectation value of an operator that acts not as the identity only on the site $s$. The operator $O_{s}$ is contracted with the tensors $N^{[s]}$ corresponding to site $s$ in $\ket{T}$. Then we contract $O_{s} \ket{T}$ with $\bra{T}$. If $s$ is the orthogonality centre, all tensors corresponding to other sites cancel due to the orthogonality condition \eqref{eq:orth_condition}. Therefore, only the conjugate tensor $N^{[s]*}$ corresponding to site $s$ in $\bra{T}$ has to be contracted with all legs of $O_s \cdot N^{[s]}$ to find the expectation value. Putting this in pictures for the operator $Z_{20}$ applied to node $20$, which is assumed to be the orthogonality centre, we obtain the already simplified tensor network
\begin{equation}\label{eq:single_site_exp_value}
    \raisebox{1.2cm}{$\bra{T}Z_{20}\ket{T}= \,$}
    \begin{tikzpicture}
        \def\nodedist{1.3}
        \def\px{1.5};
        \def\py{1.5};
        \def\pxhalf{\px / 2};
        \def\pyhalf{\py / 2};
        \node[fill=mpurple,draw,circle,minimum size=0.9cm] (Rs) at (0+\px,0+\py){$20^*$};
        \node[fill=mblue,draw,circle,minimum size=0.9cm] (R) at (0,0){$20$};
        \draw[mred,ultra thick] ($(R) + (\pxhalf,\pyhalf)$) -- ($(R) + (\px,\py)$);
        \node[fill=mgreen,draw,minimum size=0.9cm] (O) at (\pxhalf,\pyhalf) {$Z_{20}$};
        \draw[mred,ultra thick] (R) -- ($(R) + (\pxhalf,\pyhalf)$);
        \draw (R) -- (0.5,0) to[out=0,in=0] (0.5+\px,0+\py) -- (Rs);
        \draw (R) -- (-0.5,0) to[out=180,in=180] (-0.5+\px,\py) -- (Rs);
    \end{tikzpicture}.
\end{equation}
However, if the desired expectation value is of an operator acting on multiple sites or the TTNS is not in the (correct) canonical form, the entire tensor network must be contracted fully. For example, if there is a multi-site operator
\begin{equation}\label{eq:tensor_product}
\Omega=Z_{00} \otimes Z_{10} \otimes Z_{20},
\end{equation}
where $Z_{s}$ is a single-site operator acting on node $s$, the tensor network would be
\begin{equation}\label{eq:tensor_product_exp_value}
    \raisebox{2cm}{$\bra{T}\Omega\ket{T}= \quad$}
    \begin{tikzpicture}[scale=0.9,every node/.style={scale=0.7}]
        \pgfsetxvec{\pgfpoint{1cm}{0cm}}
        \pgfsetyvec{\pgfpoint{0.4cm}{0.3cm}}
        \pgfsetzvec{\pgfpoint{0cm}{1cm}}
        \def\nodedist{1.3}
        \def\disty{3}
        \def\distyhalf{\disty / 2}
        \node[fill=mpurple,draw,circle,minimum size=0.9cm] (Rs) at (0,\disty,0){$0^*$};
        \node[fill=mpurple,draw,circle,minimum size=0.9cm] (N00s) at (-\nodedist,\disty,0){$00^*$};
        \node[fill=mpurple,draw,circle,minimum size=0.9cm] (N01s) at (-2*\nodedist,\disty,0){$01^*$};
        \node[fill=mpurple,draw,circle,minimum size=0.9cm] (N10s) at (0,\disty,-\nodedist){$10^*$};
        \node[fill=mpurple,draw,circle,minimum size=0.9cm] (N11s) at (0,\disty,-2*\nodedist){$11^*$};
        \node[fill=mpurple,draw,circle,minimum size=0.9cm] (N20s) at (\nodedist,\disty,0){$20^*$};
        \node[fill=mpurple,draw,circle,minimum size=0.9cm] (N21s) at (2*\nodedist,\disty,0){$21^*$};
        \draw (N01s) -- (N00s) -- (Rs) -- (N20s) -- (N21s);
        \draw (Rs) -- (N10s) -- (N11s);

        \node[fill=mgreen,draw,minimum size=0.9cm] (O00) at (-\nodedist,\distyhalf,0){$Z_{00}$};
        \node[fill=mgreen,draw,minimum size=0.9cm] (O20) at (\nodedist,\distyhalf,0){$Z_{20}$};
        \node[fill=mgreen,draw,minimum size=0.9cm] (O10) at (0,\distyhalf,-\nodedist){$Z_{10}$};

        \node[fill=mblue,draw,circle,minimum size=0.9cm] (R) at (0,0,0){$0$};
        \node[fill=mblue,draw,circle,minimum size=0.9cm] (N00) at (-\nodedist,0,0){$00$};
        \node[fill=mblue,draw,circle,minimum size=0.9cm] (N01) at (-2*\nodedist,0,0){$01$};
        \node[fill=mblue,draw,circle,minimum size=0.9cm] (N10) at (0,0,-\nodedist){$10$};
        \node[fill=mblue,draw,circle,minimum size=0.9cm] (N11) at (0,0,-2*\nodedist){$11$};
        \node[fill=mblue,draw,circle,minimum size=0.9cm] (N20) at (\nodedist,0,0){$20$};
        \node[fill=mblue,draw,circle,minimum size=0.9cm] (N21) at (2*\nodedist,0,0){$21$};
        \draw (N01) -- (N00) -- (R) -- (N20) -- (N21);
        \draw (R) -- (N10) -- (N11);

        \draw[mred,thick] (R) -- (Rs.center);
        \draw[mred,thick] (N00) -- (O00.center);
        \draw[mred,thick] (O00) -- (N00s.center);
        \draw[mred,thick] (N01) -- (N01s.center);
        \draw[mred,thick] (N10) -- (O10.center);
        \draw[mred,thick] (N11) -- (N11s.center);
        \draw[mred,thick] (N20) -- (O20.center);
        \draw[mred,thick] (O20) -- (N20s.center);
        \draw[mred,thick] (N21) -- (N21s.center);
    \end{tikzpicture}.
\end{equation}
In {\ptn}, operators given as tensor products have their own representation. The conveniently named \texttt{TensorProduct} class records one numeric or symbolic operator associated with a given node. In principle, only non-identity operators need to be recorded. This allows a single instance of a tensor product to be reused for different tree structures. This follows the same convention as ignoring all identities in the tensor product $\Omega$ of \eqref{eq:tensor_product}. The following code recreates $\Omega$ in {\ptn}
\code{223}{224}
The \texttt{TensorProduct} class is a dictionary and can be used as such. However, it has some additional features. For a given tree structure, the \texttt{pad\_with\_identities} method adds all missing identities explicitly to a new tensor product. We will see that this can be required in some of the methods introduced in the following sections. Other utility functions include the exponentiation of a tensor product and turning it into a full matrix or high-degree tensor. The exact form of the tensor product allows us to quickly differentiate between the three cases \eqref{eq:scalar_prod_ttns}, \eqref{eq:single_site_exp_value}, and \eqref{eq:tensor_product_exp_value}. Thus, we can use the same method for all three to find the desired quantities
\code{228}{231}
With the concepts introduced in this section, we can only use static TTNS and determine expectation values of operators given as a tensor product. However, the next sections will unlock their power in simulating quantum models defined by Hamiltonian operators.

\begin{exercise}{The total magnetisation}\label{exc:the_tot_magn}
The total magnetisation of a many-qubit system $\mathcal{Q}$ can be defined as the expectation value of the operator
\begin{equation}
	M = \bigotimes_{s\in \mathcal{Q}} Z_s, 
\end{equation}
where $Z_s$ is the Pauli-$Z$ operator acting on qubit $s$.
\begin{enumerate}
	\item Create a TTNS with the same tree structure as given in \eqref{eq:example_ttns}, but with all open legs of dimension $2$, except for the root node $0$. It should represent a product state $\ket{P}$, where the value of every site is up to your choosing. (Hint: The virtual legs will be trivial but need to be accounted for.)
	\item Create the \texttt{TensorProduct} representing the operator $M$.
	\item Check, that the norm of $\ket{P}$ is $1$ and find the total magnetisation of $\ket{P}$. 
\end{enumerate}
\end{exercise}

\section{Tree Tensor Network Operators}\label{sec:TTNO}
As we learned in the last Section~\ref{sec:TTNS}, we can reduce the exponential memory requirement of many quantum states by representing them as TTNS. The logical next step is to consider a way to achieve the same memory reduction for operators acting on a high-dimensional quantum system using a TTN. This concept is represented by the \emph{Tree Tensor Network Operator} (TTNO). A TTNO is considered to be a TTN with two open legs at each node. To be compatible with the convention commonly used for matrices, the input leg of a TTNO node is the second open leg. Conversely, the output leg of a TTNO node is the first open leg. A simple example of TTNOs are the tensor products we saw in the last section. Just as product states are represented by TTNSs with all virtual legs being trivial, tensor products can be represented by TTNOs with all virtual legs being trivial. More generally a TTNO $A$ with the same underlying tree structure as $\ket{T}$ in \eqref{eq:example_ttns} would be drawn as
\begin{equation}\label{eq:example_ttno}
    \raisebox{2cm}{$A= \quad$}
    \begin{tikzpicture}
        \pgfsetxvec{\pgfpoint{1cm}{0cm}}
        \pgfsetyvec{\pgfpoint{0.4cm}{0.3cm}}
        \pgfsetzvec{\pgfpoint{0cm}{1cm}}
        \def\nodedist{1.3}
        \def\disty{4}
        \def\distyhalf{\disty / 2}
        \node[fill=mgreen,draw,circle,minimum size=0.9cm] (R) at (0,0,0){$A_0$};
        \node[fill=mgreen,draw,circle,minimum size=0.9cm] (N00) at (-\nodedist,0,0){$A_{00}$};
        \node[fill=mgreen,draw,circle,minimum size=0.9cm] (N01) at (-2*\nodedist,0,0){$A_{01}$};
        \node[fill=mgreen,draw,circle,minimum size=0.9cm] (N10) at (0,0,-\nodedist){$A_{10}$};
        \node[fill=mgreen,draw,circle,minimum size=0.9cm] (N11) at (0,0,-2*\nodedist){$A_{11}$};
        \node[fill=mgreen,draw,circle,minimum size=0.9cm] (N20) at (\nodedist,0,0){$A_{20}$};
        \node[fill=mgreen,draw,circle,minimum size=0.9cm] (N21) at (2*\nodedist,0,0){$A_{21}$};
        \draw (N01) -- (N00) -- (R) -- (N20) -- (N21);
        \draw (R) -- (N10) -- (N11);

        \draw[mred,ultra thick] (R) -- ($(R) + (0,\distyhalf,0 )$);
        \draw[mred,ultra thick] (N00) -- ($(N00) + (0,\distyhalf,0 )$);
        \draw[mred,ultra thick] (N01) -- ($(N01) + (0,\distyhalf,0 )$);
        \draw[mred,ultra thick] (N10) -- ($(N10) + (0,\distyhalf,0 )$);
        \draw[mred,ultra thick] (N11) -- ($(N11) + (0,\distyhalf,0 )$);
        \draw[mred,ultra thick] (N20) -- ($(N20) + (0,\distyhalf,0 )$);
        \draw[mred,ultra thick] (N21) -- ($(N21) + (0,\distyhalf,0 )$);

        \draw[mred,ultra thick] (R.center) -- ($(R) + (0,-\distyhalf,0 )$);
        \draw[mred,ultra thick] (N00.center) -- ($(N00) + (0,-\distyhalf,0 )$);
        \draw[mred,ultra thick] (N01.center) -- ($(N01) + (0,-\distyhalf,0 )$);
        \draw[mred,ultra thick] (N10.center) -- ($(N10) + (0,-\distyhalf,0 )$);
        \draw[mred,ultra thick] (N11.center) -- ($(N11) + (0,-\distyhalf,0 )$);
        \draw[mred,ultra thick] (N20.center) -- ($(N20) + (0,-\distyhalf,0 )$);
        \draw[mred,ultra thick] (N21.center) -- ($(N21) + (0,-\distyhalf,0 )$);
    \end{tikzpicture},
\end{equation}
Using a TTNO allows us to find the expectation value of more complex operators than mere tensor products. For example, to generate a random operator with the structure $A$, we use the code
\code{235}{247}
Unsurprisingly, this piece of code is similar to the code we used to randomly generate the TTNS $\ket{T}$ shown in \eqref{eq:example_ttns}. The equivalent structure of state and operator in TTN form allows for a sandwich-like tensor network representing the expectation value of a TTNO with respect to a given TTNS. For the state $\ket{T}$ \eqref{eq:example_ttns} and the TTNO $A$ \eqref{eq:example_ttno} we can draw this as
\begin{equation}\label{eq:ttno_ttns_contraction}
    \raisebox{2cm}{$\bra{T} A \ket{T} = \quad$}
    \begin{tikzpicture}[scale=0.8,every node/.style={scale=0.6}]
        \pgfsetxvec{\pgfpoint{1cm}{0cm}}
        \pgfsetyvec{\pgfpoint{0.4cm}{0.3cm}}
        \pgfsetzvec{\pgfpoint{0cm}{1cm}}
        \def\nodedist{1.6}
        \def\disty{2}
        \def\distyhalf{\disty / 2}
        \def\minsize{1.1cm}

        \node[fill=mpurple,draw,circle,minimum size=\minsize] (Rs) at (0,\disty,0){$0^*$};
        \node[fill=mpurple,draw,circle,minimum size=\minsize] (N00s) at (-\nodedist,\disty,0){$00^*$};
        \node[fill=mpurple,draw,circle,minimum size=\minsize] (N01s) at (-2*\nodedist,\disty,0){$01^*$};
        \node[fill=mpurple,draw,circle,minimum size=\minsize] (N10s) at (0,\disty,-\nodedist){$10^*$};
        \node[fill=mpurple,draw,circle,minimum size=\minsize] (N11s) at (0,\disty,-2*\nodedist){$11^*$};
        \node[fill=mpurple,draw,circle,minimum size=\minsize] (N20s) at (\nodedist,\disty,0){$20^*$};
        \node[fill=mpurple,draw,circle,minimum size=\minsize] (N21s) at (2*\nodedist,\disty,0){$21^*$};
        \draw (N01s) -- (N00s) -- (Rs) -- (N20s) -- (N21s);
        \draw (Rs) -- (N10s) -- (N11s);

        \node[fill=mgreen,draw,circle,minimum size=\minsize] (OR) at (0,0,0){$A_0$};
        \node[fill=mgreen,draw,circle,minimum size=\minsize] (O00) at (-\nodedist,0,0){$A_{00}$};
        \node[fill=mgreen,draw,circle,minimum size=\minsize] (O01) at (-2*\nodedist,0,0){$A_{01}$};
        \node[fill=mgreen,draw,circle,minimum size=\minsize] (O10) at (0,0,-\nodedist){$A_{10}$};
        \node[fill=mgreen,draw,circle,minimum size=\minsize] (O11) at (0,0,-2*\nodedist){$A_{11}$};
        \node[fill=mgreen,draw,circle,minimum size=\minsize] (O20) at (\nodedist,0,0){$A_{20}$};
        \node[fill=mgreen,draw,circle,minimum size=\minsize] (O21) at (2*\nodedist,0,0){$A_{21}$};
        \draw (O01) -- (O00) -- (OR) -- (O20) -- (O21);
        \draw (OR) -- (O10) -- (O11);

        \draw[mred,thick] (Rs.center) -- (OR);
        \draw[mred,thick] (N00s.center) -- (O00);
        \draw[mred,thick] (N01s.center) -- (O01);
        \draw[mred,thick] (N10s.center) -- (O10);
        \draw[mred,thick] (N11s.center) -- (O11);
        \draw[mred,thick] (N20s.center) -- (O20);
        \draw[mred,thick] (N21s.center) -- (O21);

        \node[fill=mblue,draw,circle,minimum size=\minsize] (R) at (0,-\disty,0){$0$};
        \node[fill=mblue,draw,circle,minimum size=\minsize] (N00) at (-\nodedist,-\disty,0){$00$};
        \node[fill=mblue,draw,circle,minimum size=\minsize] (N01) at (-2*\nodedist,-\disty,0){$01$};
        \node[fill=mblue,draw,circle,minimum size=\minsize] (N10) at (0,-\disty,-\nodedist){$10$};
        \node[fill=mblue,draw,circle,minimum size=\minsize] (N11) at (0,-\disty,-2*\nodedist){$11$};
        \node[fill=mblue,draw,circle,minimum size=\minsize] (N20) at (\nodedist,-\disty,0){$20$};
        \node[fill=mblue,draw,circle,minimum size=\minsize] (N21) at (2*\nodedist,-\disty,0){$21$};
        \draw (N01) -- (N00) -- (R) -- (N20) -- (N21);
        \draw (R) -- (N10) -- (N11);

        \draw[mred,thick] (R) -- (OR.center);
        \draw[mred,thick] (N00) -- (O00.center);
        \draw[mred,thick] (N01) -- (O01.center);
        \draw[mred,thick] (N10) -- (O10.center);
        \draw[mred,thick] (N11) -- (O11.center);
        \draw[mred,thick] (N20) -- (O20.center);
        \draw[mred,thick] (N21) -- (O21.center);
    \end{tikzpicture}.
\end{equation}
In {\ptn}, this contraction is achieved with the same method as the other expectation values that were introduced in Section~\ref{sec:TTNS}
\code{251}{251}
Note that behind this method, the contraction is performed from the leaves to the root. For every leaf $\ell$ we contract the tensor $N^{[\ell]}$ in $\ket{T}$ with node $A_\ell$ in $A$. Then the resulting tensor is contracted with $N^{[\ell]*}$ in $\bra{T}$. Then we choose a site $s$ for which all subtrees $\mathcal{S}^{[(s,c)]}_c$ originating from a child $c$ of $s$ are already fully contracted. The tensor representing a fully contracted subtree is a degree-$3$ tensor $B_c$. The legs correspond to the virtual legs towards the parent node. For site $s$ the tensor $N^{[s]}$ of $\ket{T}$ is contracted with all the tensors $B_c$. The same is then done for $A_s$ in $A$ and $N^{[s]*}$ in $\bra{T}$. We repeat this process until the entire network is fully contracted. This contraction order has a better scaling than the alternative of contracting the three nodes corresponding to a site and then fully contracting these.

Now let us move to a concrete example of an operator that  can not be represented as a tensor product. A non-trivial, non-random TTNO ix the following operator inspired by the quantum game of life \cite{Bleh2012, Ney2022} 
\begin{equation}\label{eq:example_non_tp_operator}
\mathcal{N}^{(1)}_{0} = P_1^{[00]}P_0^{[10]}P_0^{[20]} + P_0^{[00]}P_1^{[10]}P_0^{[20]} + P_0^{[00]}P_0^{[10]}P_1^{[20]},
\end{equation}
where $P_i^{[mn]}$ is the projector $\ket{i}\bra{i}$ acting on node $mn$. The expectation value of $\mathcal{N}^{(1)}_{0}$ can be interpreted as the probability of finding exactly one neighbour of node $0$ in state $\ket{1}$. Representing $\mathcal{N}$ as a single tensor product is impossible. However, it can be written as a TTNO with a rather low bond dimension, whose tensors are defined by
\begin{subequations}\label{eq:example_ttno_tensor_elements}
	\begin{align}
		(A_0)_{ijk} &= \mathbb{1} \delta_{(i,j,k) \in \{ (1,0,0), (0,1,0), (0,0,1) \} } \\
		(A_{i0})_j &= P_j^{[i0]} \\
		A_{i1} &= \mathbb{1},
	\end{align}
\end{subequations}
where $\delta_{x \in O}$ is $1$ if $x$ is an element of the set $O$ and $0$ otherwise. Already, for such a small system, we can see a significant reduction in memory requirement. Only $68$ elements need to be saved for the tensors defined in \eqref{eq:example_ttno_tensor_elements}. On the other hand, a full matrix representation of $\mathcal{N}^{(1)}_{0}$ would require $(2^7)^2 = 16 \, 384$ matrix elements to be saved. To reach even higher savings, we can go to larger systems. However, large systems can contain more complex operators, for example, Hamiltonians defined on a complex tree structure. This makes it difficult to construct an equivalent TTNO manually. Therefore, it is useful to automatically convert a symbolically defined operator to a TTNO. A method to do so will be explored in the next section.

\begin{exercise}{The Quantum Game of Life TTNO}\label{exc:quantum_game_of_life_ttno}
\begin{enumerate}
	\item Using the definition of the TTNO tensor in \eqref{eq:example_ttno_tensor_elements}, initialise a TTNO $A$ representing $\mathcal{N}^{(1)}_0$. (Hint: Some virtual bonds will be trivial)
	\item Find the expectation value of $A$ with respect to the TTNS in used in Exercise~\ref{exc:the_tot_magn}. Does the result fit with your intuition? How about other TTNS product states?
	\item Construct the complete matrix representation of $\mathcal{N}^{(1)}_0$ using \eqref{eq:example_non_tp_operator}. Does it coincide with the contraction of $A$?
\end{enumerate}
\end{exercise}

\subsection{Hamiltonians and State Diagrams}\label{sec:ham_and_sd}
Finding a TTNO corresponding to a given operator is not a trivial task. The simplest case of a tree structure is the one-dimensional chain, where the TTNO is called the matrix product operator (MPO). A lot of research went into the automatic construction and optimisation of MPOs on one-dimensional chains \cite{McCulloch2007, Crosswhite2008, Frowis2010, Pirvu2010, Keller2015, Chan2016, Hubig2017, Ren2020, Wall2020, Nusseler2021}. Automatically constructing the TTNO for a general tree structure is more complicated.\cite{Cakir_to_be_publ} However, it was found that a special data structure, called state diagrams, can be used \cite{Milbradt2024} to represent a symbolic operator on a tree structure and allow for a simple read-off of a TTNO.
\begin{defn}
    A \emph{state diagram} $\mathcal{D}$ corresponding to a tree structure $\mathcal{T} = (V, E)$ is a labelled hypergraph $(\Omega,\mathcal{E})$, i.e. a graph in which the edges connect more than two nodes, that contains a set of hyperedges $\mathcal{E}_s \subset \mathcal{E}$ for every node $s\in V$ and a set of vertices $\Omega_e \subset \Omega$ for every edge $e \in E$, such that a vertex $\omega \in \Omega_{(s,w)}$ can only connect to hyperedges in $\mathcal{E}_s \cup \mathcal{E}_w$.
\end{defn}
Technically, a state diagram can represent any TTN (and every TN by using a slightly different definition) \cite{Crosswhite2008}, but we focus on their application in the construction of TTNO. In a state diagram, every hyperedge $\varepsilon \in \mathcal{E}_s$ has a label associated with it. This label corresponds to an operator $O^{[s]}$ that can be applied to $s$. Furthermore, the vertices connected to $\varepsilon$ are assigned an integer value. The integer of a vertex corresponds to the indices of the virtual leg of the TTNO tensor $A_s$ at site $s$ at which $O^{[s]}$ is placed. This way, a TTNO can be read off of a state diagram. The condition that a vertex $\omega \in \Omega_{(s,w)}$ can only connect to hyperedges in $\mathcal{E}_s \cup \mathcal{E}_w$ allows the state diagram to be mapped to its corresponding tree structure $\mathcal{T} = (V, E)$.
\begin{figure}
    \centering
    \begin{tikzpicture}[scale=0.8,every node/.style={scale=0.7}]

    \def\midshift{1}
    \def\border{0.7}
    \def\firstvertdist{2.6}
    \def\intervertdist{1}
    \def\vertsize{0.01}
    \def\vertscale{0.6}
    \def\othervertdist{2}
    \filldraw[dashed,color=mpurple,fill=mpurple!50] (-\midshift-\border,\midshift+\border) rectangle (\midshift+\border,-\midshift-\border);
    \filldraw[dashed,color=mpurple,fill=mpurple!50] (-2*\othervertdist-\border,\intervertdist / 2+\border) rectangle (-2*\othervertdist+\border,-\intervertdist / 2-\border);
    \filldraw[dashed,color=mpurple,fill=mpurple!50] (-3.3*\othervertdist-\border,-\border) rectangle (-3.3*\othervertdist+\border,\border);
    \filldraw[dashed,color=mpurple,fill=mpurple!50] (2*\othervertdist-\border,\intervertdist / 2+\border) rectangle (2*\othervertdist+\border,-\intervertdist / 2-\border);
    \filldraw[dashed,color=mpurple,fill=mpurple!50] (3.3*\othervertdist-\border,-\border) rectangle (3.3*\othervertdist+\border,\border);
    \filldraw[dashed,color=mpurple,fill=mpurple!50] (\intervertdist / 2+\border,-2*\othervertdist-\border) rectangle (-\intervertdist / 2-\border,-2*\othervertdist+\border);
    \filldraw[dashed,color=mpurple,fill=mpurple!50] (-\border,-3.3*\othervertdist-\border) rectangle (\border,-3.3*\othervertdist+\border);

    \node[fill=mgreen,draw,minimum size=0.9cm] (Mid0) at (0,0){$\mathbb{1}$};
    \node[fill=mgreen,draw,minimum size=0.9cm] (Left0) at (-\midshift,\midshift){$\mathbb{1}$};
    \node[fill=mgreen,draw,minimum size=0.9cm] (Right0) at (\midshift,-\midshift){$\mathbb{1}$};
    
    \node[fill=black,minimum size=\vertsize,scale=\vertscale,circle] (E0000) at (-\firstvertdist,\intervertdist / 2){};
    \node[fill=black,minimum size=\vertsize,scale=\vertscale,circle] (E0001) at (-\firstvertdist,-\intervertdist / 2){};
    \node[fill=mgreen,draw,minimum size=0.9cm] (Left001) at (-2*\othervertdist,\intervertdist / 2){$P_1$};
    \node[fill=mgreen,draw,minimum size=0.9cm] (Left000) at (-2*\othervertdist,-\intervertdist / 2){$P_0$};
    \node[fill=black,minimum size=\vertsize,scale=\vertscale,circle] (E00010) at (-2.7*\othervertdist,0){};
    \node[fill=mgreen,draw,minimum size=0.9cm] (Left01) at (-3.3*\othervertdist,0){$\mathbb{1}$};
    \draw (Left0) to[out=180,in=0] (E0000) -- (Left001) to[out=180,in=0] (E00010) -- (Left01);
    \draw (Mid0) to[out=180,in=0] (E0001) -- (Left000) to[out=180,in=0] (E00010);
    \draw (Right0) to[out=180,in=0] (E0001);
    \node[fill=black,minimum size=\vertsize,scale=\vertscale,circle] (E2000) at (\firstvertdist,\intervertdist / 2){};
    \node[fill=black,minimum size=\vertsize,scale=\vertscale,circle] (E2001) at (\firstvertdist,-\intervertdist / 2){};
    \node[fill=mgreen,draw,minimum size=0.9cm] (R001) at (2*\othervertdist,\intervertdist / 2){$P_1$};
    \node[fill=mgreen,draw,minimum size=0.9cm] (R000) at (2*\othervertdist,-\intervertdist / 2){$P_0$};
    \node[fill=black,minimum size=\vertsize,scale=\vertscale,circle] (E20210) at (2.7*\othervertdist,0){};
    \node[fill=mgreen,draw,minimum size=0.9cm] (R21) at (3.3*\othervertdist,0){$\mathbb{1}$};
    \draw (Left0) to[out=0,in=180] (E2001) -- (R000) to[out=0,in=180] (E20210) -- (R21);
    \draw (Mid0) to[out=0,in=180] (E2000) -- (R001) to[out=0,in=180] (E20210);
    \draw (Right0) to[out=0,in=180] (E2001);
    \node[fill=black,minimum size=\vertsize,scale=\vertscale,circle] (E1000) at (\intervertdist / 2,-\firstvertdist){};
    \node[fill=black,minimum size=\vertsize,scale=\vertscale,circle] (E1001) at (-\intervertdist / 2,-\firstvertdist){};
    \node[fill=mgreen,draw,minimum size=0.9cm] (Lo001) at (\intervertdist / 2,-2*\othervertdist){$P_1$};
    \node[fill=mgreen,draw,minimum size=0.9cm] (Lo000) at (-\intervertdist / 2,-2*\othervertdist){$P_0$};
    \node[fill=black,minimum size=\vertsize,scale=\vertscale,circle] (E10110) at (0,-2.7*\othervertdist){};
    \node[fill=mgreen,draw,minimum size=0.9cm] (Lo11) at (0,-3.3*\othervertdist){$\mathbb{1}$};
    \draw (Left0) to[out=-90,in=90] (E1001) -- (Lo000) to[out=-90,in=90] (E10110) -- (Lo11);
    \draw (Mid0) to[out=-90,in=90] (E1001) -- (Lo000);
    \draw (Right0) to[out=-90,in=90] (E1000) -- (Lo001) to[out=-90,in=90] (E10110) -- (Lo11);

    \node[anchor=south] at (E0000){{\color{mred} $1$}};
    \node[anchor=south] at (E0001){{\color{mred} $0$}};
    \node[anchor=south] at (E00010){{\color{mred} $0$}};
    \node[anchor=south] at (E2000){{\color{mred} $1$}};
    \node[anchor=south] at (E2001){{\color{mred} $0$}};
    \node[anchor=south] at (E20210){{\color{mred} $0$}};
    \node[anchor=west] at (E1000){{\color{mred} $1$}};
    \node[anchor=west] at (E1001){{\color{mred} $0$}};
    \node[anchor=west] at (E10110){{\color{mred} $0$}};
    
\end{tikzpicture}
    \caption{The state diagram corresponding to the operator $\mathcal{N}_0^{[1]}$ given in \eqref{eq:example_non_tp_operator}. The green squares represent the labels of the hyperedges, while the black dots are the vertices of the state diagram. The areas shaded in purple each correspond to a set of hyperedges $\mathcal{E}_s$. For example, the big square in the middle is $\mathcal{E}_0$ corresponding to the node $0$ in \eqref{eq:example_ttno}. On the other hand, the sets of vertices $\Omega_e$ between two hyperedge sets correspond to an edge $e$ in \eqref{eq:example_ttno}. We can assign an index value to every vertex. Doing it as we did with the red indices, the resulting TTNO tensors would exactly be the ones defined in \eqref{eq:example_ttno_tensor_elements}.}
    \label{fig:example_state_diagram}
\end{figure}
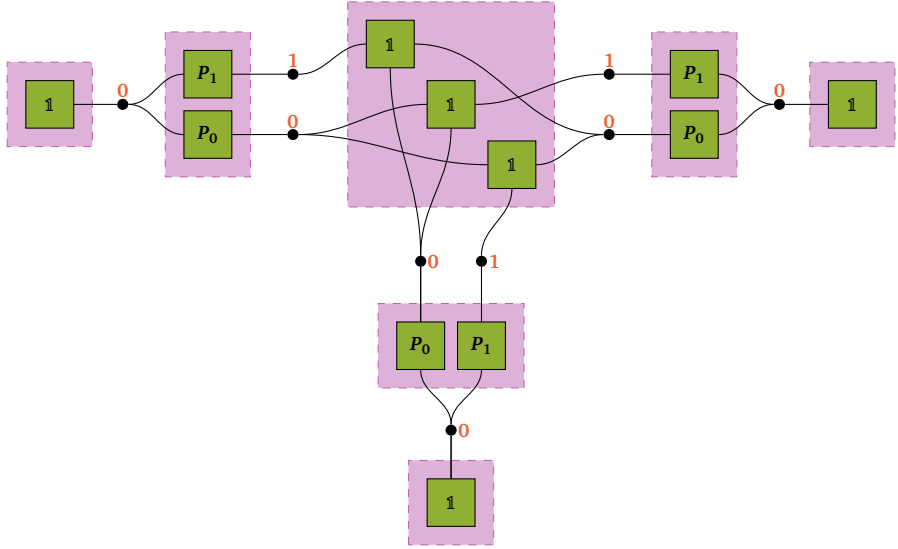
Figure~\ref{fig:example_state_diagram} gives an example of a state diagram. The indices are already assigned to the vertices, such that the resulting TTNO tensors are the tensors in \eqref{eq:example_ttno_tensor_elements}. We can clearly see that the tree structure is kept.

How can we use the state diagram to generate a TTNO from a given operator? As a first step, we assume that any operator we desire to represent is a sum of tensor products
\begin{equation}\label{eq:general_operator}
    \mathcal{N} = \sum_{i=1}^N \bigotimes_{s \in \mathcal{Q}} O^{[s]}_j,
\end{equation}
where $\mathcal{Q}$ is a set of sites in a quantum system and $O^{[s]}$ an operator applied to site $s$. This condition is not very restrictive, as long-range or multi-site operators tend to be given in the format \eqref{eq:general_operator}. On the other hand, short-range or few-site operators can be brought into the desired form by using an operator basis made from tensor products. Once we have $\mathcal{N}$, we need to generate the equivalent state diagram. Note that the state diagram of a single tensor product is rather trivial. Every site in the underlying tree structure has exactly one hyperedge corresponding to it, and every edge in the underlying tree structure has one vertex corresponding to it. Therefore, we generate a trivial diagram for every term in $\mathcal{N}$. We combine all these trivial diagrams into one state diagram without connections between them. This big state diagram is already equivalent to $\mathcal{N}$. However, it contains an unnecessary number of vertices. These would lead to a larger than necessary bond dimension in the TTNO, read from it. However, we can compare the different trivial diagrams and drop redundancies. In turn, the state diagram becomes more and more interconnected. For details on the compression algorithms, see \cite{Milbradt2024, Cakir2024}.

\begin{figure}
     \centering
     \begin{subfigure}[b]{0.45\textwidth}
         \centering
         \includegraphics[width=\textwidth]{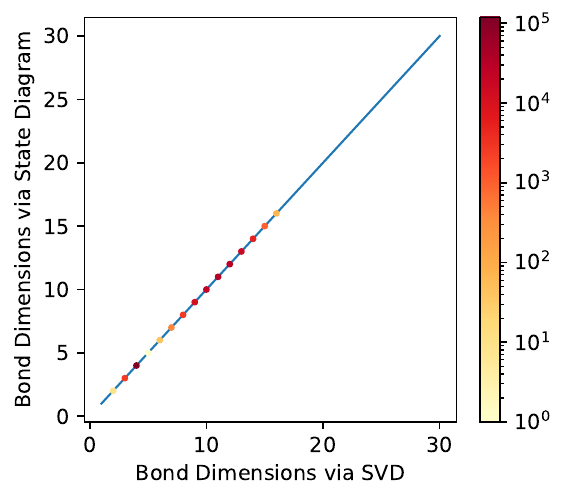}
         \caption{The bond dimensions obtained for the tree structure \eqref{eq:example_ttns}, where node $0$ also has a physical dimension of $2$.}
         \label{fig:bond_dims_T_tree}
     \end{subfigure}
     \hfill
     \begin{subfigure}[b]{0.45\textwidth}
         \centering
         \includegraphics[width=\textwidth]{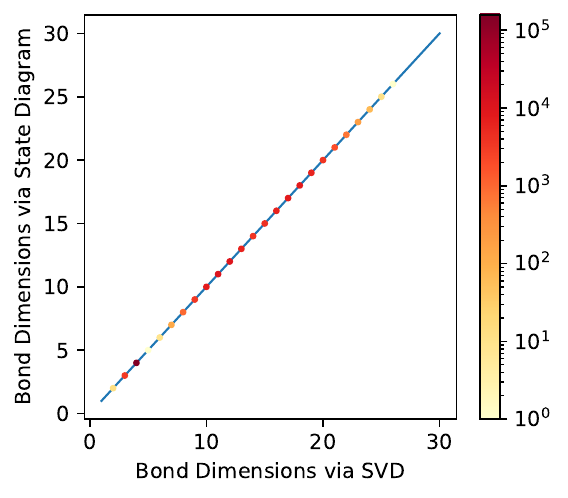}
         \caption{The bond dimensions obtained for the tree structure \eqref{eq:example_ttn}, where all nodes only have a single physical leg of dimension $2$.}
         \label{fig:bond_dims_random_example}
     \end{subfigure}
     \hfill
        \caption{The bond dimensions obtained from generating a TTNO from the full tensor via the SVD against the bond dimension obtained from the state diagram formalism. The colour of each point denotes the number of bonds for which the specific bond dimension combination was obtained. For each configuration, a sample of $40\cdot 10^3$ Hamiltonians was run. Every Hamiltonian was randomly constructed from the set $\{X,Y,Z, \mathbb{1}\}$ with $30$ terms. The blue line corresponds to $x=y$ and runs up to the number of terms, which would be the maximum bond dimension possible.}
        \label{fig:bond_dim_generated_ttno}
\end{figure}

While being a useful representation in {\ptn}, there is no need to consider the details of state diagrams. The construction algorithms work out of the box if we use the \texttt{Hamiltonian} and \texttt{TreeTensorNetworkOperator} classes. The \texttt{Hamiltonian} class is a list of tensor products and can thus represent any operator in the form \eqref{eq:general_operator}. However, the main set of operators one desires to find a TTNO representation of are Hamiltonians $H$ describing a many-body quantum system model. Therefore, the class in {\ptn} is named \texttt{Hamiltonian}. It offers the possibility to map a symbolic representation of \eqref{eq:general_operator} to a numeric one. Using a symbolic representation is advantageous, as it makes it easier to build Hamiltonians and allows for faster and more accurate comparisons during state diagram construction. While we will see more complicated Hamiltonians in later sections, we will stick to $\mathcal{N}_0^{[1]}$ from \eqref{eq:example_non_tp_operator} for demonstration purposes. The \texttt{Hamiltonian} instance corresponding to $\mathcal{N}_0^{[1]}$ is generated by the code
\code{255}{262}
Once we have defined a Hamiltonian, we can simply auto-generate a corresponding TTNO by specifying the desired tree topology the new TTNO should have. The rest is taken care of in one simple line of code
\code{266}{266}
However, note that the provided mapping of symbols to arrays has to include all relevant identity operators. Even though they do not appear explicitly in every Hamiltonian, they are assumed to be there when providing the tree structure. According to numeric evidence, our method produces the optimal bond dimension in the generated TTNO compared to the explicit construction of a TTNO from a high-degree tensor using SVD. For example, we ran some calculations for the already introduced tree structure \eqref{eq:example_ttns} and a modified version of the tree structure \eqref{eq:example_ttn}. We generated $40 \cdot 10^3$ random Hamiltonians for each tree structure from which we constructed the equivalent TTNO. We produced each TTNO once via the state diagram method and once via SVDs from the full matrix representation of the Hamiltonian. The latter gives us the minimum number of bond dimensions required to accurately represent the given Hamiltonian as a TTNO. We found that both methods produce the exact same number of bond dimensions. The results of this are plotted in Figure~\ref{fig:bond_dim_generated_ttno}.

Now that we know how to work with TTNS and how we can automatically generate TTNO from relevant operators, we can consider a main focus of {\ptn}: The time evolution of TTNS governed by a given Hamiltonian.

\begin{exercise}{Generating a TTNO}
    Consider the tree structure \eqref{eq:example_ttns}. A nearest neighbour Hamiltonian is given by
    \begin{equation}
        H_{\text{nn}} = X_{0}X_{00} + X_{0}X_{10} + X_{0}X_{10} + X_{00}X_{01} + X_{10}X_{11} +X_{20}X_{21},
    \end{equation}
    where $X$ is some single-site operator (for example, the Pauli-X operator).
    \begin{enumerate}
        \item Construct the Hamiltonian for $H_{\text{nn}}$ (Hint: Don't forget to include the identity $\mathbb{1}_2$ in the dictionary.)
        \item Generate a TTNO representing $H_{\text{nn}}$. What are its bond dimensions?
        \item Just as in question \ref{exc:quantum_game_of_life_ttno}, find the complete matrix representation of $H_{\text{nn}}$ and compare it to the full contraction of your TTNO.
    \end{enumerate}
\end{exercise}

\section{Time Evolution}\label{sec:time_evo}
Simulating the time evolution of quantum systems allows the exploration of a large variety of phenomena. Examples include quantum quench dynamics \cite{Mitra2018, Smith2019} and the simulation of quantum circuits \cite{Zhou2020, Pan2021}. The time evolution of a quantum system with state $\ket{\psi(t)}$ is described by the Schrödinger equation
\begin{equation}\label{eq:schroedinger}
    \frac{\partial}{\partial t} \ket{\psi(t)} = -iH \ket{\psi(t)},
\end{equation}
where $t$ denotes the time and $H$ the Hamiltonian of the quantum system. For finite-dimensional quantum systems, the Schrödinger equation is solved by the unitary dynamics
\begin{equation}\label{eq:exact_unitary}
    U (t) = e^{-iHt}
\end{equation}
being applied to $\ket{\psi(0)}$. However, many-body quantum systems have such high physical Hilbert space dimensions that it is already too memory-intensive to save the Hamiltonian $H$ as a full matrix. Determining $U(t)$ for a given $H$ is even harder and, therefore, equally impossible for many-body quantum systems. On the other hand, in Section~\ref{sec:ham_and_sd}, we saw that such a $H$ can be saved efficiently as a TTNO. There is, however, no general way to obtain $U(t)$ as a TTNO from $H$. Therefore, we have to use more intricate methods to simulate the dynamics of a many-body quantum system.

\subsection{Trotterisation}\label{sec:trotterisation}
A way to make the dynamics tractable is by splitting $U(t)$ in \eqref{eq:exact_unitary} into smaller parts that can be handled more easily. To introduce the following concepts, we will stick to a simple Hamiltonian
\begin{equation}\label{eq:simple_trotter_hamiltonian}
    H_{\text{simple}} = A + B = A_1 \otimes A_2 + B_1 \otimes B_2
\end{equation}
acting on the still simulatable small quantum system $\C^{d\times d}$. We can split the resulting unitary time evolution up to a final time $T$ into a series of smaller time steps
\begin{equation}\label{eq:unitary_time_step_split}
    U(T) = e^{-iH_\text{simple}T} = \left(e^{-iH_\text{simple}\Delta t} \right)^{\frac{T}{\Delta t}} = U^{\frac{T}{\Delta t}} (\Delta t).
\end{equation}
If one accepts an error, each time step can be split further. The simplest way to split \eqref{eq:unitary_time_step_split} further is using the first order \emph{Suzuki-Trotter splitting}\cite{Trotter1959, Suzuki1976}
\begin{equation}\label{eq:SuzTrot_decomp}
    U(\Delta t) = e^{-iA \Delta t} e^{-iB \Delta t} + \mathcal{O}\left(\Delta t^2\right).
\end{equation}
Inserting \eqref{eq:SuzTrot_decomp} into $U(T)$ of \eqref{eq:unitary_time_step_split} yields total error scaling of $\mathcal{O}(\Delta t)$, due to number of time steps required scaling as $\mathcal{O}(\Delta t ^{-1})$. This explains the initially confusing name of the splitting. A different commonly used splitting method is the \emph{Strang splitting} \cite{Strang1968}
\begin{equation}\label{eq:strang_splitting}
    U(\Delta t) = e^{-iA \frac{\Delta t}{2}} e^{-iB \Delta t} e^{-iA \frac{\Delta t}{2}} + \mathcal{O}\left(\Delta t^3\right).
\end{equation}
This symmetric splitting requires only modestly more operators compared to the first-order Suzuki-Trotter splitting. At the same time, the Strang splitting grants an additional power in the error scaling. A different name for the Strang splitting is second-order Suzuki-Trotter decomposition. Any kind of splitting an exponential operator will be called \emph{Trotterisation} from now on. In the literature, a whole zoo of splitting methods exists \cite{Ostmeyer2023}. Improved schemes aim to achieve a better error scaling while keeping the number of additional operators required low compared to the first order splitting \eqref{eq:SuzTrot_decomp}. However, since the two shown splittings are sufficient for most use cases, we will stick to them in our further explanation. 

\begin{figure}
    \centering
    \includegraphics[width=0.75\textwidth]{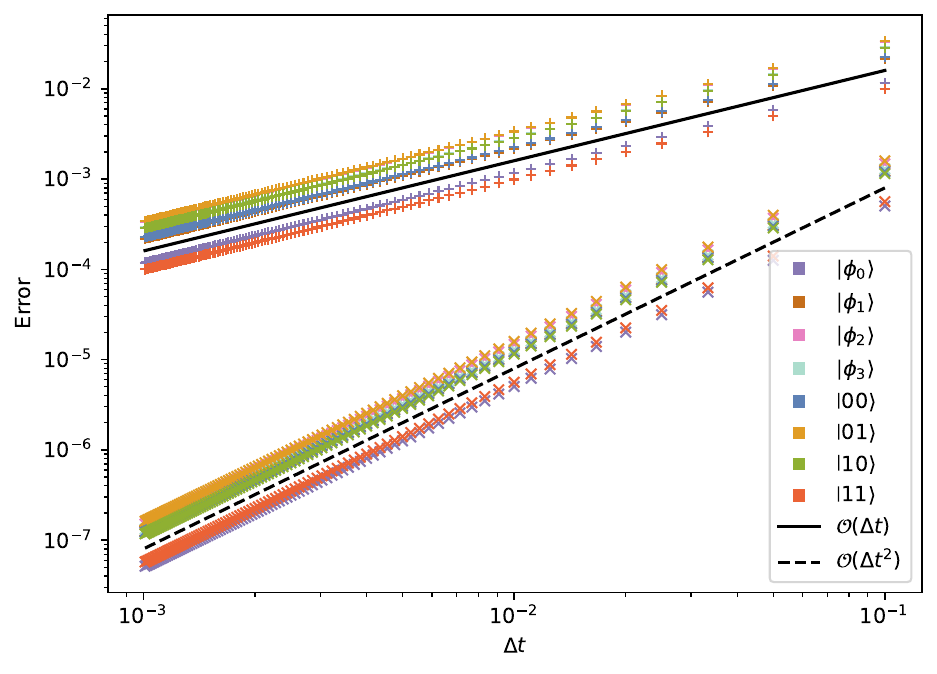}
    \caption{The error of a state accumulated in a Trotterised time evolution. The time evolution was run to a constant final time $T=1$ for different time step sizes $\Delta t$. The plus $+$ denotes the result of the first-order Suzuki-Trotter splitting, and the times $\times$ denotes the results of the Strang splitting. The functions $0.16 \Delta t$ and $0.08 \Delta t^2$ are plotted as the solid black line and dashed black line, respectively.}
    \label{fig:numerics_easy_ham_splitting}
\end{figure}
In {\ptn}, two implemented classes allow the easy implementation of splittings introduced above. The \texttt{TrotterStep} represents a single exponential operator. In addition to a tensor product representing the exponentiated operator, we also need to specify a factor to be multiplied with the tensor. The different steps are then collected in a list and supplied to the \texttt{TrotterSplitting} class. Using the supplied steps, this class will then compute all unitary operators required for a single time step. For example for random operators $A$ and $B$ in \eqref{eq:simple_trotter_hamiltonian} we can generate the first order Suzuki-Trotter splitting \eqref{eq:SuzTrot_decomp} via
\code{270}{281}
For a time step size $\Delta t = 0.01$. Furthermore, we generate the Strang splitting \eqref{eq:strang_splitting} with
\code{285}{289}
To demonstrate the performance of these splittings, we used the above code to generate splittings \eqref{eq:SuzTrot_decomp} and \eqref{eq:strang_splitting} for variable time step sizes $\Delta t$. The system in question is a two-qubit system $\C^{2\times2}$. To have a reference, we evolved an initial state $\ket{\psi(t=0)}$ using the complete time evolution \eqref{eq:unitary_time_step_split}. Thus, the reference state is $\ket{\psi_\txt{ref}} = U(T)\ket{\psi(0)}$. Then the time evolution was performed using both kinds of splitting and a range of time step sizes. At the final time, $T$ the error was evaluated as
\begin{equation}
    \text{Error} = \norm{ \ket{\psi_\txt{ref}(T)} - \ket{\psi_\txt{split}(T)} },
\end{equation}
where $\ket{\psi_\txt{split}(T)}$ is the final state obtained by the split dynamics. The computation was done for every state of both the standard computational basis $\{ \ket{ij} \}_{i=0,j=0}^1$ and the Bell basis $\{ \phi_i \}_{i=0}^3$. The results of the simulations are recorded in Figure~\ref{fig:numerics_easy_ham_splitting}. In the graph, we can clearly see the different scaling for the two different splittings. The Suzuki-Trotter splitting corresponds to the upper points $+$ in the plot. We see that the error scales as desired for a fixed initial state according to \eqref{eq:SuzTrot_decomp}. The same is true for the results of the Strang splitting shown by the lower points $\times$. The scaling follows the prediction in \eqref{eq:strang_splitting}. We can also see that even though the results are dependent on the initial state, the difference for a given splitting is far less than an order of magnitude.

Let us now consider a less trivial example of a Trotterisation. Note that both splitting methods \eqref{eq:SuzTrot_decomp} and \eqref{eq:strang_splitting} can easily be extended to an arbitrary number of terms. Thus, we once more consider a TTNS $\ket{T}$ with the topology \eqref{eq:example_ttns}. On this quantum system $\mathcal{Q}$, we define the almost nearest neighbour Hamiltonian
\begin{equation}\label{eq:almost_nn_ham}
    H_\txt{ann} = \sum_{\Delta(i,j)=1} X_i X_J + X_{00}X_{20},
\end{equation}
where $\Delta$ is the tree distance defined in \ref{def:tree_distance}. The sum runs over all pairs of sites $(i,j) \in \mathcal{Q}\times \mathcal{Q}$ such that $\Delta (i,j) = 1$. In this example, the utility of splitting a Hamiltonian during time evolution becomes more apparent. For any nearest neighbour interaction, we merely have to find the exponent of a two-site operator, so a $2^2$ dimensional matrix, rather than a full system operator with dimension $2^7$. We generate the trotter steps for the nearest neighbour interaction using the code snippet
\code{307}{314}
But what about the term $X_{00}X_{20}$? The nodes $00$ and $20$ are not neighbours, but we still have a two-point interaction. We can use SWAP gates on the TTNS to deal with distant two-point interactions \cite{Stoudenmire2010, Shi2006}. To do so, assume we have the two-point interaction $A = A_i \otimes A_j$, where sites $i$ and $j$ are not nearest neighbours. We will now SWAP $i$ and $j$ with their nearest neighbour sites until they are nearest neighbours. Then, we may exponentiate $A$ and apply the resulting officer to the two sites as usual. Finally, we use SWAP gates to return $i$ and $j$ to their original position. For example, to make $00$ and $20$ neighbours, we merely need to exchange the nodes $00$ and $0$ and swap them back afterwards.
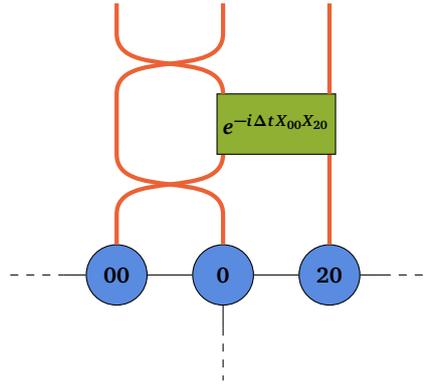
\begin{figure}
    \centering
    \begin{tikzpicture}[scale=0.8,every node/.style={scale=0.8}]
    \def\nodedist{1.75}
    \def\physdist{1}
    \def\minsize{1cm}
    \node[fill=mblue,draw,circle,minimum size=\minsize] (R) at (0,0){$0$};
    \node[fill=mblue,draw,circle,minimum size=\minsize] (N00) at (\nodedist,0){$20$};
    \node[fill=mblue,draw,circle,minimum size=\minsize] (N20) at (-\nodedist,0){$00$};

    \draw[dashed] (-2*\nodedist,0) -- (-1.5*\nodedist,0);
    \draw (-1.5*\nodedist,0) -- (N20) -- (R) -- (N00) -- (1.5*\nodedist,0);
    \draw[dashed] (1.5*\nodedist,0) -- (2*\nodedist,0);
    \draw (R) -- (0,-0.5*\nodedist);
    \draw[dashed] (0,-0.5*\nodedist) -- (0,-1*\nodedist);

    \draw[color=mred,ultra thick] (R) -- (0,\physdist) to[out=90,in=-90] (-\nodedist,2*\physdist) -- (-\nodedist,3*\physdist) to[out=90,in=-90] (0,4*\physdist) -- (0,4.5*\physdist);
    \draw[color=mred,ultra thick] (N00) -- (\nodedist,2*\physdist);
    \draw[color=mred,ultra thick] (N20) -- (-\nodedist,\physdist) to[out=90,in=-90] (0,2*\physdist);

    \filldraw[fill=mgreen] (-0.1,2*\physdist) rectangle (\nodedist+0.1,3*\physdist);
    \node at (0.5*\nodedist,2.5*\physdist) {$e^{-i\Delta t X_{00} X_{20}}$};

    \draw[color=mred,ultra thick] (0,3*\physdist) to[out=90,in=-90] (-\nodedist,4*\physdist) -- (-\nodedist,4.5*\physdist);
    \draw[color=mred,ultra thick] (\nodedist,3*\physdist) -- (\nodedist, 4.5*\physdist);
\end{tikzpicture}
    \caption{A graphical depiction of the SWAP gates required to create a nearest neighbour interaction between site $00$ and $20$. Most of the TTNS $\ket{T}$ \eqref{eq:example_ttns} was left out for convenience.}
    \label{fig:trotter_swapping}
\end{figure}
A graphical depiction of this process is given in Figure~\ref{fig:trotter_swapping}. In {\ptn}, this process is easily facilitated using the \texttt{SWAPList} class and providing it to the trotter steps. It suffices to supply a symbolic representation of a SWAP gate by specifying the two nodes that should be swapped. In our case, this can be achieved using
\code{316}{321}
Now that we have a notion of Trotterisation/splitting Hamiltonians, we can utilise this concept in our first time-evolution algorithm.

\begin{exercise}{Splitting Hamiltonians}
    \begin{enumerate}
        \item Generalise the two splittings \eqref{eq:SuzTrot_decomp} and \eqref{eq:strang_splitting} to three matrices $A$, $B$ and $C$.
        \item Implement the Strang splitting \eqref{eq:strang_splitting} for the Hamiltonian $H_\txt{ann}$.
        \item On the same tree structure as we used for the Hamiltonian $H_\txt{ann}$ in \eqref{eq:almost_nn_ham} implement the Suzuki-Trotter and Strang splitting for
        \begin{equation}
            H_\txt{exc} = X_{01}X_{10} + X_{21}X_{10}.
        \end{equation}
    \end{enumerate}
\end{exercise}

\subsection{Time Evolving Block Decimation Algorithm}\label{sec:TEBD}
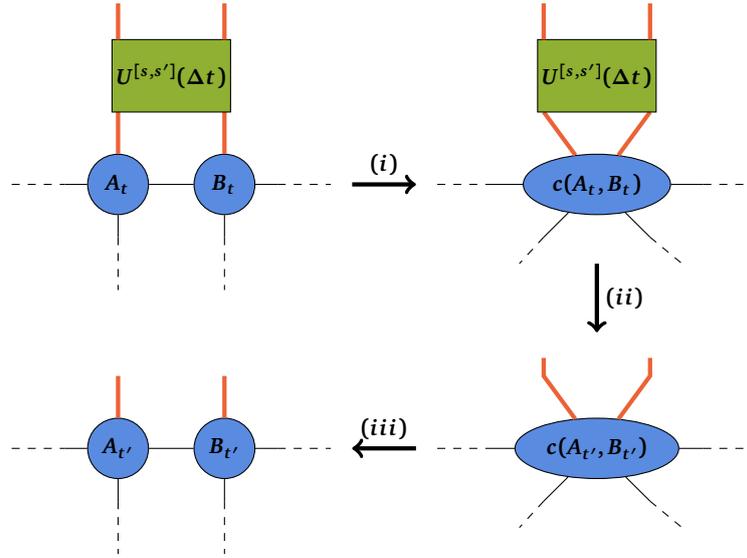
\begin{figure}
    \centering
    \begin{tikzpicture}[scale=0.8,every node/.style={scale=0.8}]
    \def\nodedist{1.75}
    \def\physdist{1.2}
    \def\minsize{1cm}

    \node[fill=mblue,draw,circle,minimum size=\minsize] (A) at (0,0) {$A_t$};
    \node[fill=mblue,draw,circle,minimum size=\minsize] (B) at (\nodedist,0){$B_t$};
    \draw[dashed] (-\nodedist,0) -- (-0.5*\nodedist,0);
    \draw (-0.5*\nodedist,0) -- (A) -- (B) -- (1.5*\nodedist,0);
    \draw[dashed] (1.5*\nodedist,0) -- (2*\nodedist,0);
    \draw (A) -- (0,-0.5*\nodedist);
    \draw[dashed] (0,-0.5*\nodedist) -- (0,-\nodedist);
    \draw (B) -- (\nodedist,-0.5*\nodedist);
    \draw[dashed] (\nodedist,-0.5*\nodedist) -- (\nodedist,-\nodedist);

    \draw[mred,ultra thick] (A) -- (0,\physdist);
    \draw[mred,ultra thick] (B) -- (\nodedist,\physdist);
    \filldraw[fill=mgreen] (-0.1,\physdist) rectangle (\nodedist+0.1,2*\physdist);
    \node at (0.5*\nodedist,1.5*\physdist){$U^{[s,s']} (\Delta t)$};
    \draw[mred,ultra thick] (0,2*\physdist) -- (0,2.5*\physdist);
    \draw[mred,ultra thick] (\nodedist,2*\physdist) -- (\nodedist,2.5*\physdist);

    \draw[ultra thick,-to] (2.2*\nodedist,0) -- (2.8*\nodedist,0) node[midway,above] {$(i)$};

    \begin{scope}[shift={(4*\nodedist,0)}]
        \node[fill=mblue,draw,ellipse,minimum size=\minsize] (A) at (0.5*\nodedist,0) {$c(A_t,B_t)$};
        \draw[dashed] (-\nodedist,0) -- (-0.5*\nodedist,0);
        \draw (-0.5*\nodedist,0) -- (A) -- (1.5*\nodedist,0);
        \draw[dashed] (1.5*\nodedist,0) -- (2*\nodedist,0);
        \draw (A) -- (0,-0.5*\nodedist);
        \draw[dashed] (0,-0.5*\nodedist) -- (-0.4,-0.75*\nodedist);
        \draw (A) -- (\nodedist,-0.5*\nodedist);
        \draw[dashed] (\nodedist,-0.5*\nodedist) -- (1.3*\nodedist,-0.75*\nodedist);
    
        \draw[mred,ultra thick] (A) -- (0,\physdist);
        \draw[mred,ultra thick] (A) -- (\nodedist,\physdist);
        \filldraw[fill=mgreen] (-0.1,\physdist) rectangle (\nodedist+0.1,2*\physdist);
        \node at (0.5*\nodedist,1.5*\physdist){$U^{[s,s']} (\Delta t)$};
        \draw[mred,ultra thick] (0,2*\physdist) -- (0,2.5*\physdist);
        \draw[mred,ultra thick] (\nodedist,2*\physdist) -- (\nodedist,2.5*\physdist);
    \end{scope}

    \draw[ultra thick,-to] (4.5*\nodedist,-0.75*\nodedist) -- (4.5*\nodedist,-1.4*\nodedist) node[midway,right] {$(ii)$};

    \begin{scope}[shift={(4*\nodedist,-2.5*\nodedist)}]
        \node[fill=mblue,draw,ellipse,minimum size=\minsize] (A) at (0.5*\nodedist,0) {$c(A_{t'},B_{t'})$};
        \draw[dashed] (-\nodedist,0) -- (-0.5*\nodedist,0);
        \draw (-0.5*\nodedist,0) -- (A) -- (1.5*\nodedist,0);
        \draw[dashed] (1.5*\nodedist,0) -- (2*\nodedist,0);
        \draw (A) -- (0,-0.5*\nodedist);
        \draw[dashed] (0,-0.5*\nodedist) -- (-0.4,-0.75*\nodedist);
        \draw (A) -- (\nodedist,-0.5*\nodedist);
        \draw[dashed] (\nodedist,-0.5*\nodedist) -- (1.3*\nodedist,-0.75*\nodedist);
    
        \draw[mred,ultra thick] (A) -- (0,\physdist) -- (0,\physdist+0.3);
        \draw[mred,ultra thick] (A) -- (\nodedist,\physdist) -- (\nodedist,\physdist+0.3);
    \end{scope}

    \draw[ultra thick,-to] (2.8*\nodedist,-2.5*\nodedist) -- (2.2*\nodedist,-2.5*\nodedist) node[midway,above] {$(iii)$};

    \begin{scope}[shift={(0,-2.5*\nodedist)}]
        \node[fill=mblue,draw,circle,minimum size=\minsize] (A) at (0,0) {$A_{t'}$};
        \node[fill=mblue,draw,circle,minimum size=\minsize] (B) at (\nodedist,0){$B_{t'}$};
        \draw[dashed] (-\nodedist,0) -- (-0.5*\nodedist,0);
        \draw (-0.5*\nodedist,0) -- (A) -- (B) -- (1.5*\nodedist,0);
        \draw[dashed] (1.5*\nodedist,0) -- (2*\nodedist,0);
        \draw (A) -- (0,-0.5*\nodedist);
        \draw[dashed] (0,-0.5*\nodedist) -- (0,-\nodedist);
        \draw (B) -- (\nodedist,-0.5*\nodedist);
        \draw[dashed] (\nodedist,-0.5*\nodedist) -- (\nodedist,-\nodedist);

        \draw[mred,ultra thick] (A) -- (0,\physdist);
        \draw[mred,ultra thick] (B) -- (\nodedist,\physdist);
    \end{scope}

\end{tikzpicture}
    \caption{A depiction of a TEBD step, if the associated unitary operator $U^{[s,s']}(\Delta t)$ is acting on two sites $s'$ and $s$. At time $t$, the sites have the tensors $A_t$ and $B_t$ associated with them, respectively. In step (i) the two tensors are contracted with each other into $c(A_t, B_t)$. In step (ii), the unitary operator is contracted with the open legs of the contracted site tensor. We use $t' = t + \Delta t$ to save space. In the final step (iii), the two sites are split back into two nodes using an SVD. During the SVD splitting, a truncation is performed if required. This leaves the final site tensors to be $A_{t'}$ and $B_{t'}$.}
    \label{fig:tebd_step}
\end{figure}

In this section we will see the first time evolution algorithm to simulate the dynamics of TTNS. We will also introduce the utility functions shared by all time-evolution algorithms. The TTNS time evolution algorithm of interest is called \emph{time-evolving block decimation} (TEBD). It was one of the first time evolution algorithms to emerge for MPS \cite{Verstraete2004, Vidal2004, Daley2004} and was quickly adapted to TTN \cite{Shi2006}. Due to its close relationship with the Trotterisation introduced in the previous subsection \ref{sec:trotterisation}, it is sometimes called Trotter gate time evolution. The TEBD algorithm takes a Trotterisation and applies each Trotter step one after the other to a given initial state to simulate a single time step. More specifically, a single site unitary operator $U^{[s]} (\Delta t)$ acting on the site $s$ is absorbed. On the other hand, if a two site operator $U^{[s,s']} (\Delta t)$ is applied, the two nodes $s$ and $s'$ are contracted into a two-site tensor $N^{[s,s']} (t)$. Therefore $s$ and $s'$ need to be neighbours. The two-site tensor $N^{[s,s']} (t)$ is contracted with $U^{[s,s']} (\Delta t)$ yielding $N^{[s,s']}(t+\Delta t)$. $N^{[s,s']}(t+\Delta t)$ is using an SVD to return to the original tree topology, yielding two updated node tensors $N^{[s]}(t+\Delta t)$ and $N^{[s']}(t+\Delta t)$. Figure~\ref{fig:tebd_step} shows a graphical depiction of this process.

Now we take a look at the {\ptn}-implementation of the TEBD time evolution. TEBD is implemented in the \texttt{TEBD} class. As any TTN time evolution algorithm in {\ptn} it inherits from the \texttt{TTNTimeEvolution} class and in turn from the \texttt{TimeEvolution} class. Starting at the top of the hierarchy, the \texttt{TimeEvolution} class provides the general methods to run a time evolution and deal with operator measurements and the measurement results. A time evolution always requires an initial state, a time step size and final time, and some operators to be evaluated. If the latter are supplied with keywords, the resulting expectation values can also be obtained using the keywords. Otherwise, one will need to manually keep track of the operators; all of their expectation values are returned in one large array. Any actual time evolution scheme must implement the \texttt{run\_one\_time\_step} and \texttt{evaluate\_operator} methods. For example, the latter is implemented in the \texttt{TTNTimeEvolution} class. As the state to be time-evolved is a TTNS, the evaluation of operator expectation values is done using the same procedures that were discussed in Sections~\ref{sec:TTNS} and~\ref{sec:TTNO}. Accordingly, the operators will be tensor products or TTNOs. Furthermore, we can decide if the bond dimension should be recorded during the simulation. Finally, the \texttt{TEBD} class implements the above application of all Trotter steps in a Trotterisation as the time stepping method. To do so, we need to supply the Trotterisation itself and the truncation parameters to be used during the SVD splitting. A TEBD algorithm can be created and run once all the parameters are specified.

\begin{figure}
     \centering
     \begin{subfigure}[b]{0.45\textwidth}
         \centering
         \includegraphics[width=\textwidth]{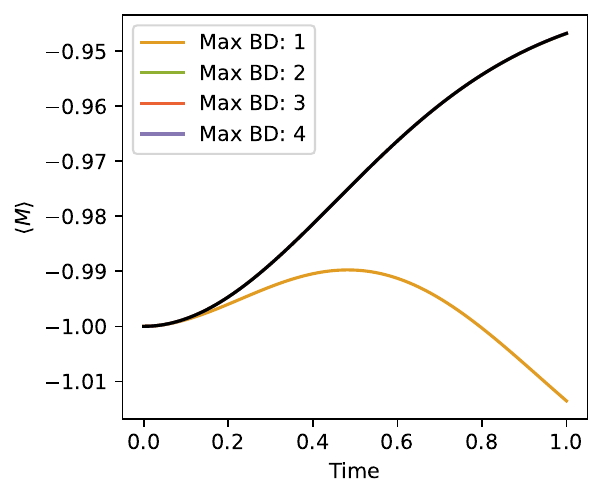}
         \caption{The dynamics of the total magnetisation $M$ expectation value. The exact solution is plotted in black.}
         \label{fig:simple_magn_magn_tebd}
     \end{subfigure}
     \hfill
     \begin{subfigure}[b]{0.45\textwidth}
         \centering
         \includegraphics[width=\textwidth]{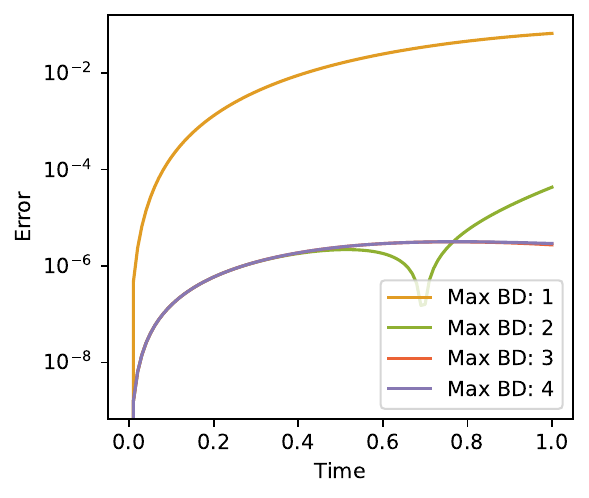}
         \caption{The error $\mathcal{E}$ of the total magnetisation obtained via TEBD simulations with respect to the exact solution.}
         \label{fig:simple_magn_error_tebd}
     \end{subfigure}
     \hfill
     \begin{subfigure}[b]{0.45\textwidth}
         \centering
         \includegraphics[width=\textwidth]{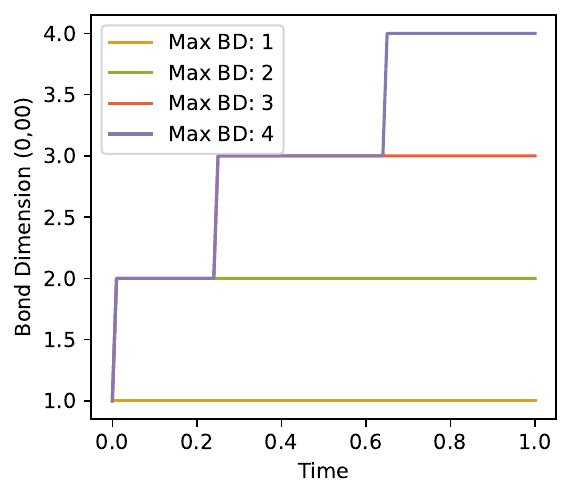}
         \caption{The change in bond dimension during TEBD simulation at the bond $(0,00)$.}
         \label{fig:simple_magn_bd_tebd}
     \end{subfigure}
        \caption{TEBD simulation results for the total magnetisation $M$ with initial state $\ket{\Psi_0}$ as defined in \eqref{eq:init_state} for different maximum bond dimensions.}
        \label{fig:simple_magn_tebd}
\end{figure}

We will demonstrate this with an example. Consider the transverse-field Ising (TFI) model
\begin{equation}\label{eq:transverse-field ising}
    H_\txt{TFI} = -J \sum_{\Delta (i,j) = 1} Z_i Z_j - g \sum_{i \in \mathcal{Q}[L]} X_i,
\end{equation}
where $J, g \in \mathbb{R}$ are coupling constants and $\mathcal{Q}[L]$ is the usual tree topology quantum system \eqref{eq:example_ttns} with all physical dimensions $d=2$. However, we introduced the parameter $L$. It denotes the length of each chain attached to the root node $0$. $X$ and $Z$ are the Pauli-operators. We want to simulate the time dependency of the total magnetisation
\begin{equation}\label{eq:total_magnetisation}
	M = \bigotimes_{i\in \mathcal{Q}[L]} Z_i.
\end{equation}
The initial state of the time evolution is the product state
\begin{equation}\label{eq:init_state}
    \ket{\Psi_0} = \bigotimes_{i\in \mathcal{Q}[L]} \ket{k^{\Delta (i,r)\mod 2}}_i,
\end{equation}
where $r$ denotes the root node. This is the product state where the root is in state $\ket{0}$; its neighbours are in the flipped state $\ket{1}$, and for every step away from the root, the state is flipped again until the end of the chains is reached. For $L=2$, we can construct the initial state with
\code{326}{343}
Furthermore, we choose $J=1$ and $g=0.1$. Considering a final time $T=1$ and a time step size $\Delta t = 0.01$, we can construct and run the desired TEBD time evolution using the code
\code{346}{372}
We can then extract the time steps, bond dimensions, and expectation values for $M$, as well as perform some sanity checks using the utility methods inherent to every time evolution object
\code{375}{382}
We run the above simulation for different maximum bond dimensions to show the effect of an ill-chosen truncation parameter. Additionally, the system is still small enough for us to run an exact time evolution. This allows us to compute the error of the TEBD algorithm. The results of the simulations can be found in Figure~\ref{fig:simple_magn_tebd}. The dynamics of the expectation value of the total magnetisation $M$ is shown in Figure~\ref{fig:simple_magn_magn_tebd}. The TEBD result approaches the exact solution already for a maximum bond dimension of $2$. The error is defined as
\begin{equation}\label{eq:error}
    \mathcal{E} (t) = \norm{ \bra{\Psi_\txt{exc} (t)} M \ket{\Psi_\txt{exc} (t)} - \bra{\Psi_\txt{TEBD} (t)} M \ket{\Psi_\txt{TEBD} (t)} },
\end{equation}
where $\ket{\Psi_\txt{exc} (t)}$ and $\ket{\Psi_\txt{TEBD} (t)}$ are the exact state obtained via state vector simulation and the state obtained via TEBD respectively. $\mathcal{E} (t)$ is shown in Figure~\ref{fig:simple_magn_error_tebd}. The error does not decrease significantly for an increase of the bond dimension above $2$. The error is around $\mathcal{E} (t) \sim 10^{-6}$ for a maximum bond dimension $\geq 2$, which is of the same order of magnitude as the error expected from the Trotterisation. Therefore, the main error source is not the truncation to a lower bond dimension but a too-large time step. However, the state will still occupy the maximum bond dimension allowed. This can be seen in Figure~\ref{fig:simple_magn_bd_tebd}, where the time dependence of the dimension of the bond connecting nodes $0$ and $00$ is shown. For our example, considering only this bond to see the full impact of the maximum bond dimension as a parameter suffices. Due to the symmetry under exchange, the other bonds of the root node would show the same behaviour. On the other hand, the dimension of the second bonds in the chains require a maximum dimension of $2$ anyways to be encoded exactly. This is reached as soon as the state is not a product state, resulting in rather irrelevant temporal behaviour of the bond dimension at the legs not connected to node $0$. As would be expected, we see that the bond dimension increases over time. However, not every time step leads to a direct increase in dimensionality. This shows that starting with a small bond dimension and slowly increasing it is reasonable and admits a reduction in computational resources required for the simulations.

However, as evidenced by our ability to compute an exact reference solution, the above simulation could be easily achieved without resorting to any tensor network method. To showcase the power of TTN methods, we will simulate an example that goes far beyond the capability of any state-vector simulation. The dimension of $\mathcal{Q}[L]$ is
\begin{equation}
    \txt{dim} \left( \mathcal{Q}[L] \right) = 2^{3L+1}.
\end{equation}
Due to this exponential scaling with respect to $L$, a state-vector approach will break down around $L=5$. On the other hand, the tensor network methods can simulate the dynamics for much larger systems.
\begin{figure}
    \centering
    \includegraphics[width=0.9\textwidth]{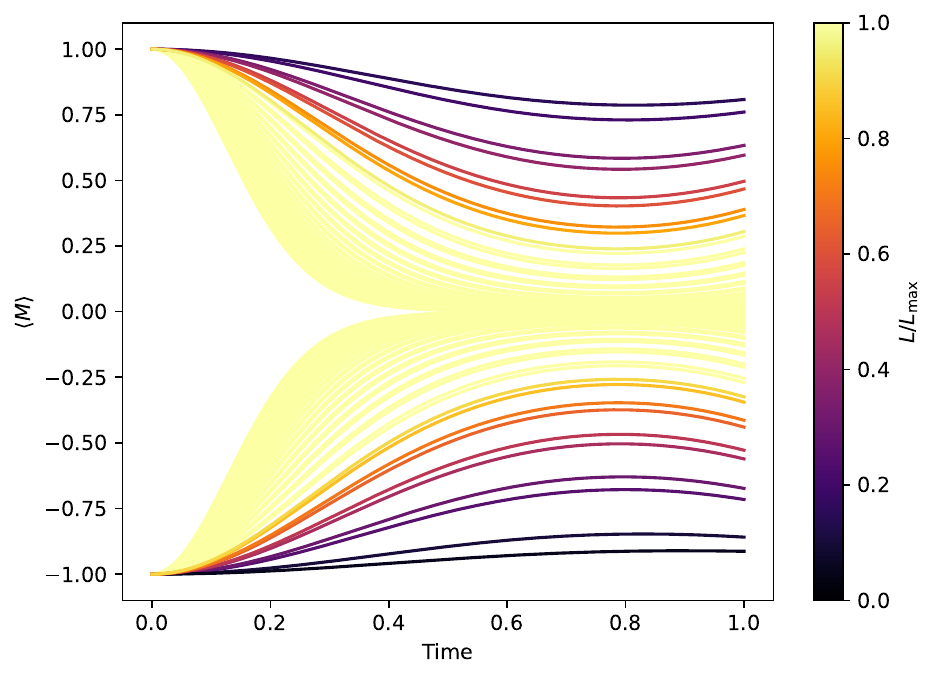}
    \caption{The dynamics of the total magnetisation expectation value $\langle M \rangle$ for different values of $L$. $L_\txt{max} = 500$ denotes the maximum length of the simulated chains.}
    \label{fig:tebd_dep_on_length}
\end{figure}
We demonstrate this ability of tensor network methods to perform many-body system simulations by repeating the above simulation for larger chain lengths $L$. This means we simulate the dynamics of the expectation values of the total magnetisation $M_{[L]} (t)$ as given in \eqref{eq:total_magnetisation} under the transverse-field Ising model $H_\txt{TFI} [L]$ \eqref{eq:transverse-field ising}, where $J=1$ and $g=0.1$, with the initial state $\ket{\Psi_0 [L]}$ as defined in \eqref{eq:init_state}. We perform simulations until $L=L_\txt{max} = 500$ and choose the maximum bond dimension for each run as $D_\txt{max} = \left\lceil L / 2 \right\rceil$. Therefore, the required memory scales linearly in $L$ rather than exponentially. The numerical results are given in Figure~\ref{fig:tebd_dep_on_length}. The plot has two noteworthy main characteristics. The first is that the results split into two separate sets. This happens as the magnetisation of the initial state depends on $L$
\begin{equation}
    \langle M_0 [L] \rangle = \bra{\Psi_0 [L]} M \ket{\Psi_0 [L]} = (-1)^{\, 3 \,\left( \left\lfloor \frac{L}{2} \right\rfloor +1 \right) }.
\end{equation}
Depending on $\langle M_0 [L] \rangle$, the magnetisation has to increase or decrease to reach zero. The second observation is that the approach of the magnetisation to zero happens more quickly for higher $L$. Truthfully, the transverse-field Ising model is not very complicated. However, it serves well as an example to showcase the abilities of tree tensor networks and how they can simulate large systems. In the next section, we will introduce the second class of time evolution methods implemented in {\ptn} that make good use of the TTNO introduced in Section~\ref{sec:TTNO}.

\begin{exercise}{Further exploration of the TFI model}
We applied the TEBD-TTN method to the TFI model in the above subsection. However, we only scratched the surface with our investigation. This exercise asks you to explore the TFI further.
\begin{enumerate}
    \item For the small system $L=2$, we only checked the behaviour for a differing maximum bond dimension. How does the error behave relative to the other truncation parameters $r_\txt{rel}$ and $r_\txt{tot}$?
    \item To simulate the large systems, we set the maximum bond dimension as $D_\txt{max}= \left\lceil L / 2 \right\rceil$. What happens if the bond dimension has a sublinear scaling $D_\txt{max}= \left\lceil L^{0.5} \right\rceil$? (Hint: Only use $L_\txt{max} = 100$ to require less simulation time.)
    \item In the above simulations, we always set $g=0.1$. How does the dynamic of $M$ change for different values of $g$? Choose $L=20$ and $g\in \{-0.5,-0.1,0,0.1,0.5 \}$.
    \item Find the value $L$ for which your machine lacks the resources to perform the exact state vector simulation. (\textbf{Caution}: Your machine might crash. Make sure all important documents are saved beforehand!)
\end{enumerate}
\end{exercise}

\subsection{Time Dependent Variational Principle for Trees}
The second TTN time evolution algorithm implemented in {\ptn} is the time-dependent variational principle (TDVP). The TDVP method is a general method to variationally time-evolve a parametrised set of quantum states \cite{Kramer1981} and is closely related to the Dirac-Frenkel variational method \cite{Broeckhove1988}. The TDVP was then adapted to matrix product states \cite{Haegeman2011, Haegeman2016}, where it turned out to have a similar structure to the matrix product density renormalisation group methods \cite{Haegeman2016}. The adaptation of TDVP to TTN was initially restricted to binary TTNSs \cite{Lubich2013} but then generalised to arbitrary tree topologies \cite{Bauernfeind2020}. The TDVP algorithm uses the fact that the set of TTNSs with a given tree topology $T = (V,E)$ and the same set of bond dimensions $\mathcal{D} = \{ D_e | e \in E \}$ forms a manifold $\mathcal{M}\left[T, \mathcal{D}\right]$. For a given state, the local tensors are variationally time-evolved by solving the projected Schrödinger equation
\begin{equation}\label{eq:proj_schroed_equation}
    \frac{d \ket{T}}{dt} = -i \mathcal{P}_{\mathcal{T}_{\ket{T}} \mathcal{M}\left[T,\mathcal{D}\right]} H\ket{T},
\end{equation}
where $\mathcal{P}_{\mathcal{T}_{\ket{T}} \mathcal{M}\left[T,\mathcal{D}\right]}$ is the projector onto the tangent space $\mathcal{T}_{\ket{T}} \mathcal{M}\left[T,\mathcal{D}\right]$ of the manifold $\mathcal{M}\left[T,\mathcal{D}\right]$ at state $\ket{T}$ and $H$ the system Hamiltonian. Note, that we need the Hamiltonian to be a TTNO with the same underlying tree structure $T$ of the TTNSs in the manifold $\mathcal{M}\left[T, \mathcal{D}\right]$. The exact form of the projector $\mathcal{P}_{\mathcal{T}_{\ket{T}} \mathcal{M}\left[T,\vec{D}\right]}$ allows a split of the projected Schrödinger equation \eqref{eq:proj_schroed_equation} into differential equations, that optimise only a few close sites, while contracting the remaining tree tensor network almost into an expectation value of the Hamiltonian \eqref{eq:ttno_ttns_contraction}. Details on the derivation and a detailed description of the projector can be found in \cite{Bauernfeind2020}. Using the usual TTNS \eqref{eq:example_ttns} we will now exemplify the two different TDVP classes implemented in {\ptn}.

\subsubsection{One-Site TDVP}\label{sec:1tdvp}
In the one-site TDVP or 1TDVP, the time update is always performed on the orthogonalisation centre $s$ of the TTNS. This update of site $s$ can be reduced to two steps. The first step is to evolve the tensor $N^{[s]}$ at site $s$ in the TTNS according to the equation
\begin{equation}\label{eq:1tdvp_site_deq}
    \frac{d N^{[s]}}{dt} = - i H_\txt{eff}^{[s]} \cdot N^{[s]},
\end{equation}
where $H_\txt{eff}^{[s]}$ is an effective site Hamiltonian, whose general exact form can be obtained from evaluating the projector $\mathcal{P}_{\mathcal{T}_{\ket{T}}}$ in \eqref{eq:proj_schroed_equation}, and $\cdot$ denotes the matrix multiplication of a vectorised  $ N^{[s]}$ with a matricised version of $H_\txt{eff}^{[s]}$. Assuming the TTNS structure \eqref{eq:example_ttns} and that site $0$ is the orthogonalisation center, we find:
\begin{equation}\label{eq:1tdvp_ste_ham_ex}
    \raisebox{2cm}{$H_\txt{eff}^{[0]}= \, \,$}
    \begin{tikzpicture}[scale=0.8,every node/.style={scale=0.6}]
        \pgfsetxvec{\pgfpoint{1cm}{0cm}}
        \pgfsetyvec{\pgfpoint{0.4cm}{0.3cm}}
        \pgfsetzvec{\pgfpoint{0cm}{1cm}}
        \def\nodedist{1.6}
        \def\disty{4}
        \def\distyhalf{\disty / 2}
        \def\minsize{1.1cm}
        \node[circle,minimum size=\minsize] (Rs) at (0,\disty,0){};
        \node[fill=mpurple,draw,minimum size=\minsize,
                isosceles triangle,
                isosceles triangle apex angle=70,
                rotate=0] (N00s) at (-\nodedist,\disty,0){$00^*$};
        \node[fill=mpurple,draw,minimum size=\minsize,
                isosceles triangle,
                isosceles triangle apex angle=70,
                rotate=0] (N01s) at (-2*\nodedist,\disty,0){$01^*$};
        \node[fill=mpurple,draw,minimum size=\minsize,
                isosceles triangle,
                isosceles triangle apex angle=70,
                rotate=90] (N10s) at (0,\disty,-\nodedist){};
        \node at (N10s){$10^*$};
        \node[fill=mpurple,draw,minimum size=\minsize,
                isosceles triangle,
                isosceles triangle apex angle=70,
                rotate=90] (N11s) at (0,\disty,-2*\nodedist){};
        \node at (N11s){$11^*$};
        \node[fill=mpurple,draw,minimum size=\minsize,
                isosceles triangle,
                isosceles triangle apex angle=70,
                rotate=180] (N20s) at (\nodedist,\disty,0){};
        \node at (N20s){$20^*$};
        \node[fill=mpurple,draw,minimum size=\minsize,
                isosceles triangle,
                isosceles triangle apex angle=70,
                rotate=180] (N21s) at (2*\nodedist,\disty,0){};
        \node at (N21s){$21^*$};
        \draw (N01s) -- (N00s);
        \draw[ultra thick] (N00s) -- (Rs);
        \draw[ultra thick] (Rs) -- (N20s);
        \draw (N20s) -- (N21s);
        \draw[ultra thick] (Rs) -- (N10s);
        \draw (N10s) -- (N11s);
        \node[fill=mgreen,draw,circle,minimum size=\minsize] (OR) at (0,0,0){$A_0$};
        \node[fill=mgreen,draw,circle,minimum size=\minsize] (O00) at (-\nodedist,0,0){$A_{00}$};
        \node[fill=mgreen,draw,circle,minimum size=\minsize] (O01) at (-2*\nodedist,0,0){$A_{01}$};
        \node[fill=mgreen,draw,circle,minimum size=\minsize] (O10) at (0,0,-\nodedist){$A_{10}$};
        \node[fill=mgreen,draw,circle,minimum size=\minsize] (O11) at (0,0,-2*\nodedist){$A_{11}$};
        \node[fill=mgreen,draw,circle,minimum size=\minsize] (O20) at (\nodedist,0,0){$A_{20}$};
        \node[fill=mgreen,draw,circle,minimum size=\minsize] (O21) at (2*\nodedist,0,0){$A_{21}$};
        \draw (O01) -- (O00) -- (OR) -- (O20) -- (O21);
        \draw (OR) -- (O10) -- (O11);
        \draw[mred,ultra thick] (Rs.center) -- (OR);
        \draw[mred] (N00s.center) -- (O00);
        \draw[mred] (N01s.center) -- (O01);
        \draw[mred] (N10s.center) -- (O10);
        \draw[mred] (N11s.center) -- (O11);
        \draw[mred] (N20s.center) -- (O20);
        \draw[mred] (N21s.center) -- (O21);
        \node[circle,minimum size=\minsize] (R) at (0,-\disty,0){};
        \node[fill=mblue,draw,minimum size=\minsize,
                isosceles triangle,
                isosceles triangle apex angle=70,
                rotate=0] (N00) at (-\nodedist,-\disty,0){$00$};
        \node[fill=mblue,draw,minimum size=\minsize,
                isosceles triangle,
                isosceles triangle apex angle=70,
                rotate=0] (N01) at (-2*\nodedist,-\disty,0){$01$};
        \node[fill=mblue,draw,minimum size=\minsize,
                isosceles triangle,
                isosceles triangle apex angle=70,
                rotate=90] (N10) at (0,-\disty,-\nodedist){};
        \node at (N10){$10$};
        \node[fill=mblue,draw,minimum size=\minsize,
                isosceles triangle,
                isosceles triangle apex angle=70,
                rotate=90] (N11) at (0,-\disty,-2*\nodedist){};
        \node at (N11){$11$};
        \node[fill=mblue,draw,minimum size=\minsize,
                isosceles triangle,
                isosceles triangle apex angle=70,
                rotate=180] (N20) at (\nodedist,-\disty,0){};
        \node at (N20){$20$};
        \node[fill=mblue,draw,minimum size=\minsize,
                isosceles triangle,
                isosceles triangle apex angle=70,
                rotate=180] (N21) at (2*\nodedist,-\disty,0){};
        \node at (N21){$21$};
        \draw (N01) -- (N00);
        \draw[ultra thick] (N00) -- (R);
        \draw[ultra thick] (R) -- (N20);
        \draw (N20) -- (N21);
        \draw[ultra thick] (R) -- (N10);
        \draw (N10) -- (N11);
        \draw[mred,ultra thick] (R) -- (OR.center);
        \draw[mred] (N00) -- (O00.center);
        \draw[mred] (N01) -- (O01.center);
        \draw[mred] (N10) -- (O10.center);
        \draw[mred] (N11) -- (O11.center);
        \draw[mred] (N20) -- (O20.center);
        \draw[mred] (N21) -- (O21.center);
        \node[anchor=north] at (R.west) {$i_0$};
        \node[anchor=north] at (R.east) {$i_2$};
        \node[anchor=west] at (R.south) {$i_1$};
        \node[anchor=south west] at (R) {$i_3$};
        \node[anchor=south] at (Rs.west) {$f_0$};
        \node[anchor=south] at (Rs.east) {$f_2$};
        \node[anchor=south west] at (Rs.south) {$f_1$};
        \node[anchor=south] at (Rs.center) {$f_3$};
    \end{tikzpicture}
    \raisebox{1cm}{
    \raisebox{1.2cm}{$= \, \,$}
    \begin{tikzpicture}
        \filldraw[fill=morange] (-0.2,0) rectangle (1.7,1);
        \foreach \i\j in {0/0,0.5/1,1/2,1.5/3}{
            \draw (\i,-0.5) -- (\i,0);
            \node[anchor=west] at (\i-0.1,-0.5){$i_{\j}$};
            \draw (\i,1) -- (\i,1.5);
            \node[anchor=west] at (\i-0.1,1.5){$f_{\j}$};
        }
        \node at (0.75,0.4) {$H_\txt{eff}^{[0]}$};
    \end{tikzpicture}.}
\end{equation}
The input legs $\left(i_0,i_1,i_2,i_3\right)$ and output legs $(f_0,f_1,f_2,f_3)$ of the matricised $H_\txt{eff}^{[0]}$ in \eqref{eq:1tdvp_site_deq} are the legs that would be contracted with $N^{[0]}$ and $\left(N^{[0]}\right)^*$ respectively. The effective one-site Hamiltonian generally has a form similar to the one above in \eqref{eq:1tdvp_ste_ham_ex}. To create it, the TTNS and TTNO tensors of all sites except the site $s$ that is to be updated are contracted. Then all virtual legs are contracted, including the virtual legs of the TTNO tensor of site $s$, but excluding the TTNS tensors of site $s$. The differential equation \eqref{eq:1tdvp_site_deq} is solved by
\begin{equation}\label{eq:1tdvp_site_update}
    N^{[s]} (t+ \Delta t) = e^{-i H_\txt{eff}^{[s]} \Delta t} N^{[s]} (t).
\end{equation}
Once node $N^{[s]}$ was updated, we split it using a QR-decomposition, where all but one leg of $N^{[s]}$ are associated with the resulting isometric $Q^{[s]}$-tensor. The leg associated with the $R^{[(s,s')]}$-tensor is the leg towards the next node $s'$ that is to be updated. Before proceeding with the next site update, the tensor $R^{[(s,s')}$ is updated according to
\begin{equation}\label{eq:1tdvp_link_deq}
    \frac{d R^{[(s,s')]}}{dt} = + i H_\txt{eff}^{[(s,s')]} \cdot R^{[(s,s')]},
\end{equation}
where we once more have an effective Hamiltonian $H_\txt{eff}^{[(s,s')]}$. Continuing on from the example in \eqref{eq:1tdvp_ste_ham_ex}, we find the effective Hamiltonian to be
\begin{equation}\label{eq:1tdvp_link_ham_ex}
    \raisebox{2cm}{$H_\txt{eff}^{[(0,20)]}= \, \,$}
    \begin{tikzpicture}[scale=0.8,every node/.style={scale=0.6}]
        \pgfsetxvec{\pgfpoint{1cm}{0cm}}
        \pgfsetyvec{\pgfpoint{0.4cm}{0.3cm}}
        \pgfsetzvec{\pgfpoint{0cm}{1cm}}
        \def\nodedist{1.6}
        \def\disty{4}
        \def\distyhalf{\disty / 2}
        \def\minsize{1.1cm}
        \def\rightoffset{1.2}
        \node[fill=mpurple,draw,minimum size=\minsize,
                isosceles triangle,
                isosceles triangle apex angle=70,
                rotate=0] (Rs) at (0,\disty,0){$0^*$};
        \node[fill=mpurple,draw,minimum size=\minsize,
                isosceles triangle,
                isosceles triangle apex angle=70,
                rotate=0] (N00s) at (-\nodedist,\disty,0){$00^*$};
        \node[fill=mpurple,draw,minimum size=\minsize,
                isosceles triangle,
                isosceles triangle apex angle=70,
                rotate=0] (N01s) at (-2*\nodedist,\disty,0){$01^*$};
        \node[fill=mpurple,draw,minimum size=\minsize,
                isosceles triangle,
                isosceles triangle apex angle=70,
                rotate=90] (N10s) at (0,\disty,-\nodedist){};
        \node at (N10s){$10^*$};
        \node[fill=mpurple,draw,minimum size=\minsize,
                isosceles triangle,
                isosceles triangle apex angle=70,
                rotate=90] (N11s) at (0,\disty,-2*\nodedist){};
        \node at (N11s){$11^*$};
        \node[fill=mpurple,draw,minimum size=\minsize,
                isosceles triangle,
                isosceles triangle apex angle=70,
                rotate=180] (N20s) at (\nodedist+\rightoffset,\disty,0){};
        \node at (N20s){$20^*$};
        \node[fill=mpurple,draw,minimum size=\minsize,
                isosceles triangle,
                isosceles triangle apex angle=70,
                rotate=180] (N21s) at (2*\nodedist+\rightoffset,\disty,0){};
        \node at (N21s){$21^*$};
        \draw[ultra thick] (Rs) -- (\rightoffset,\disty,0);
        \draw[ultra thick] (\nodedist,\disty,0) -- (N20s);
        \draw (N01s) -- (N00s) -- (Rs);
        \draw (N20s) -- (N21s);
        \draw (Rs) -- (N10s) -- (N11s);
        \node[fill=mgreen,draw,circle,minimum size=\minsize] (OR) at (0,0,0){$A_0$};
        \node[fill=mgreen,draw,circle,minimum size=\minsize] (O00) at (-\nodedist,0,0){$A_{00}$};
        \node[fill=mgreen,draw,circle,minimum size=\minsize] (O01) at (-2*\nodedist,0,0){$A_{01}$};
        \node[fill=mgreen,draw,circle,minimum size=\minsize] (O10) at (0,0,-\nodedist){$A_{10}$};
        \node[fill=mgreen,draw,circle,minimum size=\minsize] (O11) at (0,0,-2*\nodedist){$A_{11}$};
        \node[fill=mgreen,draw,circle,minimum size=\minsize] (O20) at (\nodedist+\rightoffset,0,0){$A_{20}$};
        \node[fill=mgreen,draw,circle,minimum size=\minsize] (O21) at (2*\nodedist+\rightoffset,0,0){$A_{21}$};
        \draw (O01) -- (O00) -- (OR) -- (O20) -- (O21);
        \draw (OR) -- (O10) -- (O11);
        \draw[mred] (Rs.center) -- (OR);
        \draw[mred] (N00s.center) -- (O00);
        \draw[mred] (N01s.center) -- (O01);
        \draw[mred] (N10s.center) -- (O10);
        \draw[mred] (N11s.center) -- (O11);
        \draw[mred] (N20s.center) -- (O20);
        \draw[mred] (N21s.center) -- (O21);
        \node[fill=mblue,draw,minimum size=\minsize,
                isosceles triangle,
                isosceles triangle apex angle=70,
                rotate=0] (R) at (0,-\disty,0){$0$};
        \node[fill=mblue,draw,minimum size=\minsize,
                isosceles triangle,
                isosceles triangle apex angle=70,
                rotate=0] (N00) at (-\nodedist,-\disty,0){$00$};
        \node[fill=mblue,draw,minimum size=\minsize,
                isosceles triangle,
                isosceles triangle apex angle=70,
                rotate=0] (N01) at (-2*\nodedist,-\disty,0){$01$};
        \node[fill=mblue,draw,minimum size=\minsize,
                isosceles triangle,
                isosceles triangle apex angle=70,
                rotate=90] (N10) at (0,-\disty,-\nodedist){};
        \node at (N10){$10$};
        \node[fill=mblue,draw,minimum size=\minsize,
                isosceles triangle,
                isosceles triangle apex angle=70,
                rotate=90] (N11) at (0,-\disty,-2*\nodedist){};
        \node at (N11){$11$};
        \node[fill=mblue,draw,minimum size=\minsize,
                isosceles triangle,
                isosceles triangle apex angle=70,
                rotate=180] (N20) at (\nodedist+\rightoffset,-\disty,0){};
        \node at (N20){$20$};
        \node[fill=mblue,draw,minimum size=\minsize,
                isosceles triangle,
                isosceles triangle apex angle=70,
                rotate=180] (N21) at (2*\nodedist+\rightoffset,-\disty,0){};
        \node at (N21){$21$};
        \draw (N01) -- (N00) -- (R);
        \draw[ultra thick] (R) -- (\rightoffset,-\disty,0);
        \draw[ultra thick] (\nodedist,-\disty,0) -- (N20);
        \draw (N20) -- (N21);
        \draw (R) -- (N10) -- (N11);
        \draw[mred] (R) -- (OR.center);
        \draw[mred] (N00) -- (O00.center);
        \draw[mred] (N01) -- (O01.center);
        \draw[mred] (N10) -- (O10.center);
        \draw[mred] (N11) -- (O11.center);
        \draw[mred] (N20) -- (O20.center);
        \draw[mred] (N21) -- (O21.center);
        \node[anchor=south] at (\rightoffset,-\disty,0){$i_0$};
        \node[anchor=south] at (\nodedist,-\disty,0){$i_1$};
        \node[anchor=south] at (\rightoffset,\disty,0){$f_0$};
        \node[anchor=south] at (\nodedist,\disty,0){$f_1$};
    \end{tikzpicture}
    \raisebox{1cm}{
    \raisebox{1.2cm}{$= \, \,$}
    \begin{tikzpicture}
        \filldraw[fill=morange] (-0.2,0) rectangle (1.7,1);
        \foreach \i\j in {0.4/0,1.1/1}{
            \draw (\i,-0.5) -- (\i,0);
            \node[anchor=west] at (\i-0.1,-0.5){$i_{\j}$};
            \draw (\i,1) -- (\i,1.5);
            \node[anchor=west] at (\i-0.1,1.5){$f_{\j}$};
        }
        \node at (0.75,0.4) {$H_\txt{eff}^{[(0,20)]}$};
    \end{tikzpicture}.}
\end{equation}
To solve \eqref{eq:1tdvp_link_deq} the effective Hamiltonian $H_\txt{eff}^{[(0,20)]}$ and the link tensor $R^{[(0,20)]}$ are matricised and vectorised analogously to the tensors $H_\txt{eff}^{[(0)]}$ and $N^{[0]}$ above. In general $H_\txt{eff}^{[(s,s')]}$ for a link $(s,s')$ is obtained by contracting the entire tensor network required for an expectation value of a TTNO with respect to a TTNS as discussed in Section~\ref{sec:TTNO} except for the two link tensors $R^{[(s,s')]}$ and $R^{[(s,s')]*}$. We solve the differential equation quation~\eqref{eq:1tdvp_link_deq} by updating the link tensor according to
\begin{equation}\label{eq:1tdvp_link_update}
    R^{[(s,s')]} (t + \Delta t) = e^{iH_\txt{eff}^{[(0,20)]} \Delta t} R^{[(s,s')]} (t).
\end{equation}
The updated link tensor $R^{[(s,s')]} (t + \Delta t)$ is then absorbed into the next site $s'$, making $s'$ the new orthogonalisation centre.

But how do we determine the next site $s'$? In principle, the order in which we run through the TTNS and update the sites is arbitrary as long as every site is updated once and the orthogonality centre is shifted accordingly to ensure it is the site that is updated. However, moving the orthogonality centre can lead to many unnecessary QR-decompositions. Therefore, we choose a path $\gamma$ through the TTNS that requires the least moves of the orthogonality centre. To achieve this, we choose the two leaves which are furthest apart as the start node $\ell_s$ and end node $\ell_e$. The nodes along the path $\gamma (\ell_s, \ell_e)$ are updated according to the scheme described above. Whenever there is a node $s \in \gamma (\ell_s, \ell_e)$ with neighbours $\nu \notin \gamma (\ell_s, \ell_e)$, all subtrees $\mathcal{S}_\nu^{[(n,s)]}$ originating from these neighbours are updated, before moving to the next site in $\gamma (\ell_s, \ell_e)$. Borrowing a simplified notation from \cite{Bauernfeind2020}, the entire procedure is exemplified for the usual example tree structure \eqref{eq:example_ttns} in Figure~\ref{fig:1tdvp}.
\begin{figure}
    \centering
    \begin{tikzpicture}[>=stealth]
    \def\nodedist{1}
    \def\minsize{0.25cm}
    \def\insep{1pt}
    \def\shiftx{5}
    \def\shifty{-4}
    \def\orthccolour{morange}
    \def\nupdtcolour{mblue}
    \def\updtcolour{mred}
    \def\rcolour{mgreen}

    \def\rdiamond(#1,#2){
        \node[fill=\rcolour,draw,
              minimum size=\minsize, inner sep=\insep,
              diamond] (R) at (#1,#2){};}
    \def\ortccircle(#1,#2,#3){
        \node[fill=\orthccolour,circle,draw,
              minimum size=\minsize,inner sep=\insep] (#3) at (#1,#2){};
    }
    \def\nupdtriangle(#1,#2,#3,#4){
        \node[fill=\nupdtcolour,draw,
              minimum size=\minsize, inner sep=\insep,
              isosceles triangle,
              isosceles triangle apex angle=70,
              rotate=#1] (#4) at (#2,#3){};
    }
    \def\updtriangle(#1,#2,#3,#4){
        \node[fill=\updtcolour,draw,
              minimum size=\minsize, inner sep=\insep,
              isosceles triangle,
              isosceles triangle apex angle=70,
              rotate=#1] (#4) at (#2,#3){};
    }
    \def\drawlines{
    \draw (N01) -- (N00) -- (N0) -- (N20) -- (N21);
    \draw (N0) -- (N10) -- (N11);
    }
    \def\drawarrow(#1,#2,#3,#4){
        \draw[ultra thick,->] (#1,#2) -- (#3,#4);
    }
    
    \ortccircle(-2*\nodedist,0,N01)
    \nupdtriangle(180,-1*\nodedist,0,N00)
    \nupdtriangle(180,0,0,N0)
    \nupdtriangle(90,0,-1*\nodedist,N10)
    \nupdtriangle(90,0,-2*\nodedist,N11)
    \nupdtriangle(180,1*\nodedist,0,N20)
    \nupdtriangle(180,2*\nodedist,0,N21)
    \drawlines

    \drawarrow(1.75*\nodedist,-1*\nodedist,
                \shiftx-1.75*\nodedist,-1*\nodedist)

    \begin{scope}[shift={(\shiftx,0)}]
        \updtriangle(0,-2*\nodedist,0,N01)
        \nupdtriangle(180,-1*\nodedist,0,N00)
        \nupdtriangle(180,0,0,N0)
        \nupdtriangle(90,0,-1*\nodedist,N10)
        \nupdtriangle(90,0,-2*\nodedist,N11)
        \nupdtriangle(180,1*\nodedist,0,N20)
        \nupdtriangle(180,2*\nodedist,0,N21)
        \drawlines
        \rdiamond(-1.5*\nodedist,0)
    \end{scope}

    \drawarrow(\shiftx+1.75*\nodedist,-1*\nodedist,
                2*\shiftx-1.75*\nodedist,-1*\nodedist)

    \begin{scope}[shift={(2*\shiftx,0)}]
        \updtriangle(0,-2*\nodedist,0,N01)
        \ortccircle(-1*\nodedist,0,N00)
        \nupdtriangle(180,0,0,N0)
        \nupdtriangle(90,0,-1*\nodedist,N10)
        \nupdtriangle(90,0,-2*\nodedist,N11)
        \nupdtriangle(180,1*\nodedist,0,N20)
        \nupdtriangle(180,2*\nodedist,0,N21)
        \drawlines
    \end{scope}

    \drawarrow(2*\shiftx,-2.4*\nodedist,
                2*\shiftx,\shifty+0.4*\nodedist)

    \begin{scope}[shift={(0,\shifty)}]
        \updtriangle(0,-2*\nodedist,0,N01)
        \updtriangle(0,-1*\nodedist,0,N00)
        \nupdtriangle(-90,0,0,N0)
        \nupdtriangle(-90,0,-1*\nodedist,N10)
        \updtriangle(90,0,-2*\nodedist,N11)
        \nupdtriangle(180,1*\nodedist,0,N20)
        \nupdtriangle(180,2*\nodedist,0,N21)
        \drawlines
        \rdiamond(0,-1.5*\nodedist)
    \end{scope}

    \drawarrow(2*\shiftx-1.75*\nodedist,-1*\nodedist+\shifty,
                \shiftx+1.75*\nodedist,-1*\nodedist+\shifty)

    \begin{scope}[shift={(\shiftx,\shifty)}]
        \updtriangle(0,-2*\nodedist,0,N01)
        \updtriangle(0,-1*\nodedist,0,N00)
        \nupdtriangle(-90,0,0,N0)
        \nupdtriangle(-90,0,-1*\nodedist,N10)
        \ortccircle(0,-2*\nodedist,N11)
        \nupdtriangle(180,1*\nodedist,0,N20)
        \nupdtriangle(180,2*\nodedist,0,N21)
        \drawlines
    \end{scope}

    \drawarrow(\shiftx-1.75*\nodedist,-1*\nodedist+\shifty,
                1.75*\nodedist,-1*\nodedist+\shifty)

    \begin{scope}[shift={(2*\shiftx,\shifty)}]
        \updtriangle(0,-2*\nodedist,0,N01)
        \updtriangle(0,-1*\nodedist,0,N00)
        \nupdtriangle(180,0,0,N0)
        \nupdtriangle(90,0,-1*\nodedist,N10)
        \nupdtriangle(90,0,-2*\nodedist,N11)
        \nupdtriangle(180,1*\nodedist,0,N20)
        \nupdtriangle(180,2*\nodedist,0,N21)
        \drawlines
        \rdiamond(-0.5*\nodedist,0)
    \end{scope}

    \drawarrow(0,\shifty-2.4*\nodedist,
                0,2*\shifty+0.4*\nodedist)

    \begin{scope}[shift={(0,2*\shifty)}]
        \updtriangle(0,-2*\nodedist,0,N01)
        \updtriangle(0,-1*\nodedist,0,N00)
        \nupdtriangle(-90,0,0,N0)
        \ortccircle(0,-1*\nodedist,N10)
        \updtriangle(90,0,-2*\nodedist,N11)
        \nupdtriangle(180,1*\nodedist,0,N20)
        \nupdtriangle(180,2*\nodedist,0,N21)
        \drawlines
    \end{scope}

    \drawarrow(1.75*\nodedist,-1*\nodedist+2*\shifty,
                \shiftx-1.75*\nodedist,-1*\nodedist+2*\shifty)

    \begin{scope}[shift={(\shiftx,2*\shifty)}]
        \updtriangle(0,-2*\nodedist,0,N01)
        \updtriangle(0,-1*\nodedist,0,N00)
        \nupdtriangle(-90,0,0,N0)
        \updtriangle(90,0,-1*\nodedist,N10)
        \updtriangle(90,0,-2*\nodedist,N11)
        \nupdtriangle(180,1*\nodedist,0,N20)
        \nupdtriangle(180,2*\nodedist,0,N21)
        \drawlines
        \rdiamond(0,-0.5*\nodedist)
    \end{scope}

    \drawarrow(\shiftx+1.75*\nodedist,-1*\nodedist+2*\shifty,
                2*\shiftx-1.75*\nodedist,-1*\nodedist+2*\shifty)

    \begin{scope}[shift={(2*\shiftx,2*\shifty)}]
        \updtriangle(0,-2*\nodedist,0,N01)
        \updtriangle(0,-1*\nodedist,0,N00)
        \ortccircle(0,0,N0)
        \updtriangle(90,0,-1*\nodedist,N10)
        \updtriangle(90,0,-2*\nodedist,N11)
        \nupdtriangle(180,1*\nodedist,0,N20)
        \nupdtriangle(180,2*\nodedist,0,N21)
        \drawlines
    \end{scope}

    \drawarrow(2*\shiftx,-2.4*\nodedist+2*\shifty,
                2*\shiftx,3*\shifty+0.4*\nodedist)

    \begin{scope}[shift={(2*\shiftx,3*\shifty)}]
        \updtriangle(0,-2*\nodedist,0,N01)
        \updtriangle(0,-1*\nodedist,0,N00)
        \updtriangle(0,0,0,N0)
        \updtriangle(90,0,-1*\nodedist,N10)
        \updtriangle(90,0,-2*\nodedist,N11)
        \nupdtriangle(180,1*\nodedist,0,N20)
        \nupdtriangle(180,2*\nodedist,0,N21)
        \drawlines
        \rdiamond(0.5,0)
    \end{scope}

    \drawarrow(2*\shiftx-1.75*\nodedist,-1*\nodedist+3*\shifty,
                \shiftx+1.75*\nodedist,-1*\nodedist+3*\shifty)

    \begin{scope}[shift={(\shiftx,3*\shifty)}]
        \updtriangle(0,-2*\nodedist,0,N01)
        \updtriangle(0,-1*\nodedist,0,N00)
        \updtriangle(0,0,0,N0)
        \updtriangle(90,0,-1*\nodedist,N10)
        \updtriangle(90,0,-2*\nodedist,N11)
        \ortccircle(1*\nodedist,0,N20)
        \nupdtriangle(180,2*\nodedist,0,N21)
        \drawlines
    \end{scope}

    \drawarrow(\shiftx-1.75*\nodedist,-1*\nodedist+3*\shifty,
                1.75*\nodedist,-1*\nodedist+3*\shifty)

    \begin{scope}[shift={(0,3*\shifty)}]
        \updtriangle(0,-2*\nodedist,0,N01)
        \updtriangle(0,-1*\nodedist,0,N00)
        \updtriangle(0,0,0,N0)
        \updtriangle(90,0,-1*\nodedist,N10)
        \updtriangle(90,0,-2*\nodedist,N11)
        \updtriangle(0,1*\nodedist,0,N20)
        \nupdtriangle(180,2*\nodedist,0,N21)
        \drawlines
        \rdiamond(1.5,0)
    \end{scope}

    \drawarrow(0.5*\nodedist,3*\shifty-2.4*\nodedist,
                \shiftx - 2.5 * \nodedist, 4*\shifty + 0.5*\nodedist)

    \begin{scope}[shift={(\shiftx,4*\shifty)}]
        \updtriangle(0,-2*\nodedist,0,N01)
        \updtriangle(0,-1*\nodedist,0,N00)
        \updtriangle(0,0,0,N0)
        \updtriangle(90,0,-1*\nodedist,N10)
        \updtriangle(90,0,-2*\nodedist,N11)
        \updtriangle(0,1*\nodedist,0,N20)
        \ortccircle(2*\nodedist,0,N21)
        \drawlines
        \rdiamond(1.5,0)
    \end{scope}
\end{tikzpicture}
    \caption{One time step performed with the one-site TDVP method on the tree structure \eqref{eq:example_ttns}. Orange circles denote site tensors that are updated according to \eqref{eq:1tdvp_site_update}. They are also the current orthogonalisation centre. The green diamonds are link tensors that are updated according to \eqref{eq:1tdvp_link_update} and originate from a QR-decomposition. The triangles point towards the orthogonalisation centre. The blue triangles represent sites that have not been updated, while the orange triangles represent updated sites.}
    \label{fig:1tdvp}
\end{figure}

As an example, assume once more the TFI model shown in \eqref{eq:transverse-field ising}, where $J=1$ and $g=0.1$. We want to find the dynamics of the total magnetisation $M$ \eqref{eq:total_magnetisation} for the initial state $\ket{\Psi_0}$ as defined in \eqref{eq:init_state} and a chain length $L=2$. TDVP methods can deal with multi-site interactions more easily than the TEBD method could. To showcase this, we include an additional four-site interaction around the root
\begin{equation}\label{eq:mod_tfi}
    H_\txt{mod} = H_\txt{TFI} + Z_{0} Z_{00} Z_{10} Z_{20}.
\end{equation}
Notably, one-site TDVP does not dynamically adapt the bond dimension during run-time. Therefore, we have do initialise the TTNS with the desired maximum bond dimension 
\code{388}{412}
The TTNO representing the modified TFI Hamiltonian $H_\txt{TFI}$ and the tensor product form of the magnetisation $M$ can be obtained as discussed in the previous sections \ref{sec:ham_and_sd} and \ref{sec:TTNS} respectively.
\code{414}{434}
Once a time step size and final evolution time are defined, we can initialise the one-site TDVP similarly to the TEDB algorithm shown in Section~\ref{sec:TEBD}.
\code{437}{444}
The class to run the TDVP method explained above is called \texttt{FirstOrderOneSiteTDVP}. Similar to the Trotter splittings explained in Section~\ref{sec:trotterisation}, we can define higher orders of TDVP. These orders are related to the order of solving the local partial differential equations. One \emph{sweep} through the system is the process of updating every node from the starting node $\ell_s$ to the end node $\ell_e$. The sweep above was always performed with $\Delta t$ and left $\ell_e$ as the orthogonality centre. Therefore, we need to recanonicalise the TTNS as a final step such that $\ell_s$ is the orthogonality centre. However, we could also perform updates on the way back through the TTNS. If we update on the forward sweep with $\Delta t /2$, we can perform all updates again in the exact opposite order with the same time step $\Delta t /2$. After one forward sweep and one backwards sweep, we end up with $\ell_s$ as the orthogonalisation centre and have updated the entire system by $\Delta t$. So in the example shown in Figure~\ref{fig:1tdvp}, we perform the backward sweep starting from the last picture and go opposite to the arrow direction. This is easily performed by initialising and running the \texttt{SecondOrderOneSiteTDVP} class.

We will leave any analysis of simulation results to the next subsection, where we compare it to the second TDVP method currently available in {\ptn}.

\subsubsection{Two-Site TDVP}\label{sec:2tdvp}
In the previous section, we introduced the one-site version of TDVP. We can define a similar algorithm, which updates a contraction of two sites and then evolves the second site backwards instead of updating a link. This method is called two-site TDVP or 2TDVP. To update a site $s$, which is again the orthogonality centre of the TTN, we contract its corresponding tensor $N^{[s]}$ with the tensor $N^{[s']}$ of a neighbouring site $s'$ into a two-site tensor $M^{[ss']}$. The two-site tensor is updated according to
\begin{equation}\label{eq:2tdvp_forward_deq}
    \frac{d M^{[ss']}}{dt} = - i H_\txt{eff}^{[ss']} \cdot M^{[ss']},
\end{equation}
where $H_\txt{eff}^{[ss']}$ is the matricised version of an effective Hamiltonian. $H_\txt{eff}^{[ss']}$ is similar to $H_\txt{eff}^{[s]}$ in \eqref{eq:1tdvp_site_update}, but instead of not contracting the TTNS tensor of the single site $s$, we do not contract the two-site tensor $M^{[ss']}$. So once again assuming the TTNS structure \eqref{eq:example_ttns} and that site $0$ is the orthogonalisation center, we find for $s'=20$
\begin{equation}\label{eq:2tdvp_twosite_ham_ex}
    \raisebox{2cm}{$H_\txt{eff}^{[0,20]}= \, \,$}
    \begin{tikzpicture}[scale=0.8,every node/.style={scale=0.6}]
        \pgfsetxvec{\pgfpoint{1cm}{0cm}}
        \pgfsetyvec{\pgfpoint{0.4cm}{0.3cm}}
        \pgfsetzvec{\pgfpoint{0cm}{1cm}}
        \def\nodedist{1.6}
        \def\disty{3.5}
        \def\distyhalf{\disty / 2}
        \def\minsize{1.1cm}
        \node[circle,minimum size=\minsize] (Rs) at (0,\disty,0){};
        \node[fill=mpurple,draw,minimum size=\minsize,
                isosceles triangle,
                isosceles triangle apex angle=70,
                rotate=0] (N00s) at (-\nodedist,\disty,0){$00^*$};
        \node[fill=mpurple,draw,minimum size=\minsize,
                isosceles triangle,
                isosceles triangle apex angle=70,
                rotate=0] (N01s) at (-2*\nodedist,\disty,0){$01^*$};
        \node[fill=mpurple,draw,minimum size=\minsize,
                isosceles triangle,
                isosceles triangle apex angle=70,
                rotate=90] (N10s) at (0,\disty,-\nodedist){};
        \node at (N10s){$10^*$};
        \node[fill=mpurple,draw,minimum size=\minsize,
                isosceles triangle,
                isosceles triangle apex angle=70,
                rotate=90] (N11s) at (0,\disty,-2*\nodedist){};
        \node at (N11s){$11^*$};
        \node[circle,minimum size=\minsize] (N20s) at (\nodedist,\disty,0){};
        \node[fill=mpurple,draw,minimum size=\minsize,
                isosceles triangle,
                isosceles triangle apex angle=70,
                rotate=180] (N21s) at (2*\nodedist,\disty,0){};
        \node at (N21s){$21^*$};
        \draw (N01s) -- (N00s);
        \draw[ultra thick] (N00s) -- (Rs);
        \draw[ultra thick] (N20s.east) -- (N21s);
        \draw[ultra thick] (Rs) -- (N10s);
        \draw (N10s) -- (N11s);
        \node[fill=mgreen,draw,circle,minimum size=\minsize] (OR) at (0,0,0){$A_0$};
        \node[fill=mgreen,draw,circle,minimum size=\minsize] (O00) at (-\nodedist,0,0){$A_{00}$};
        \node[fill=mgreen,draw,circle,minimum size=\minsize] (O01) at (-2*\nodedist,0,0){$A_{01}$};
        \node[fill=mgreen,draw,circle,minimum size=\minsize] (O10) at (0,0,-\nodedist){$A_{10}$};
        \node[fill=mgreen,draw,circle,minimum size=\minsize] (O11) at (0,0,-2*\nodedist){$A_{11}$};
        \node[fill=mgreen,draw,circle,minimum size=\minsize] (O20) at (\nodedist,0,0){$A_{20}$};
        \node[fill=mgreen,draw,circle,minimum size=\minsize] (O21) at (2*\nodedist,0,0){$A_{21}$};
        \draw (O01) -- (O00) -- (OR) -- (O20) -- (O21);
        \draw (OR) -- (O10) -- (O11);
        \draw[mred,ultra thick] (Rs.center) -- (OR);
        \draw[mred] (N00s.center) -- (O00);
        \draw[mred] (N01s.center) -- (O01);
        \draw[mred] (N10s.center) -- (O10);
        \draw[mred] (N11s.center) -- (O11);
        \draw[mred,ultra thick] (N20s.center) -- (O20);
        \draw[mred] (N21s.center) -- (O21);
        \node[circle,minimum size=\minsize] (R) at (0,-\disty,0){};
        \node[fill=mblue,draw,minimum size=\minsize,
                isosceles triangle,
                isosceles triangle apex angle=70,
                rotate=0] (N00) at (-\nodedist,-\disty,0){$00$};
        \node[fill=mblue,draw,minimum size=\minsize,
                isosceles triangle,
                isosceles triangle apex angle=70,
                rotate=0] (N01) at (-2*\nodedist,-\disty,0){$01$};
        \node[fill=mblue,draw,minimum size=\minsize,
                isosceles triangle,
                isosceles triangle apex angle=70,
                rotate=90] (N10) at (0,-\disty,-\nodedist){};
        \node at (N10){$10$};
        \node[fill=mblue,draw,minimum size=\minsize,
                isosceles triangle,
                isosceles triangle apex angle=70,
                rotate=90] (N11) at (0,-\disty,-2*\nodedist){};
        \node at (N11){$11$};
        \node[circle, minimum size=\minsize] (N20) at (\nodedist,-\disty,0){};
        \node[fill=mblue,draw,minimum size=\minsize,
                isosceles triangle,
                isosceles triangle apex angle=70,
                rotate=180] (N21) at (2*\nodedist,-\disty,0){};
        \node at (N21){$21$};
        \draw (N01) -- (N00);
        \draw[ultra thick] (N00) -- (R);
        \draw[ultra thick] (N20.east) -- (N21);
        \draw[ultra thick] (R) -- (N10);
        \draw (N10) -- (N11);
        \draw[mred,ultra thick] (R) -- (OR.center);
        \draw[mred] (N00) -- (O00.center);
        \draw[mred] (N01) -- (O01.center);
        \draw[mred] (N10) -- (O10.center);
        \draw[mred] (N11) -- (O11.center);
        \draw[mred,ultra thick] (N20) -- (O20.center);
        \draw[mred] (N21) -- (O21.center);
        \node[anchor=north] at (R.west) {$i_0$};
        \node[anchor=north] at (N20.east) {$i_2$};
        \node[anchor=west] at (R.south) {$i_1$};
        \node[anchor=south west] at (R) {$i_3$};
        \node[anchor=south west] at (N20) {$i_4$};
        \node[anchor=south] at (Rs.west) {$f_0$};
        \node[anchor=south] at (N20s.east) {$f_2$};
        \node[anchor=south west] at (Rs.south) {$f_1$};
        \node[anchor=south] at (Rs.center) {$f_3$};
        \node[anchor=south east] at (N20s) {$f_4$};
    \end{tikzpicture}
    \raisebox{1cm}{
    \raisebox{1.2cm}{$= \, \,$}
    \begin{tikzpicture}
        \filldraw[fill=morange] (-0.2,0) rectangle (2.2,1);
        \foreach \i\j in {0/0,0.5/1,1/2,1.5/3,2/4}{
            \draw (\i,-0.5) -- (\i,0);
            \node[anchor=west] at (\i-0.1,-0.5){$i_{\j}$};
            \draw (\i,1) -- (\i,1.5);
            \node[anchor=west] at (\i-0.1,1.5){$f_{\j}$};
        }
        \node at (0.75,0.4) {$H_\txt{eff}^{[0]}$};
    \end{tikzpicture}.}
\end{equation}
To solve \eqref{eq:2tdvp_forward_deq}, we evolve $M^{[ss']}$ with
\begin{equation}\label{eq:2tdvp_forwards_update}
    M^{[ss']} (t+ \Delta t) = e^{-i H_\txt{eff}^{[ss']} \Delta t} M^{[ss']} (t).
\end{equation}
Once updated, we split the two-site tensor $M^{[ss']}$ via a tensor decomposition into two single-site tensors $N^{[s']}$ and $N^{[s']}$. If the SVD decomposition is used, we can truncate the bond dimension during this step. This dynamic bond dimension adaptation significantly improves 2TVDP compared to 1TDVP, where a bond dimension has to be chosen at the start. As mentioned before, we then evolve the second site tensor backwards in time according to
\begin{equation}\label{eq:2tdvp_backwards_deq}
    \frac{d N^{[s']}}{dt} = i H_\txt{eff}^{[s']} \cdot N^{[s']},
\end{equation}
where the effective Hamiltonian $H_\txt{eff}^{[s']}$, is the exact same as in the 1TDVP-site-update \eqref{eq:1tdvp_site_deq}. Accordingly, the dynamics are solved by performing the tensor update
\begin{equation}\label{eq:2tdvp_backwards_update}
    N^{[s']} (t+ \Delta t) = e^{-i H_\txt{eff}^{[s']} \Delta t} N^{[s'} (t).
\end{equation}
The order in which we update the sites, is the exact same as explained for the 1TDVP case. A graphical depiction of a full forwards sweep with the 2TDVP method through the example tree structure \eqref{eq:example_ttns} can be found in Figure~\ref{fig:2tdvp_example}.\\

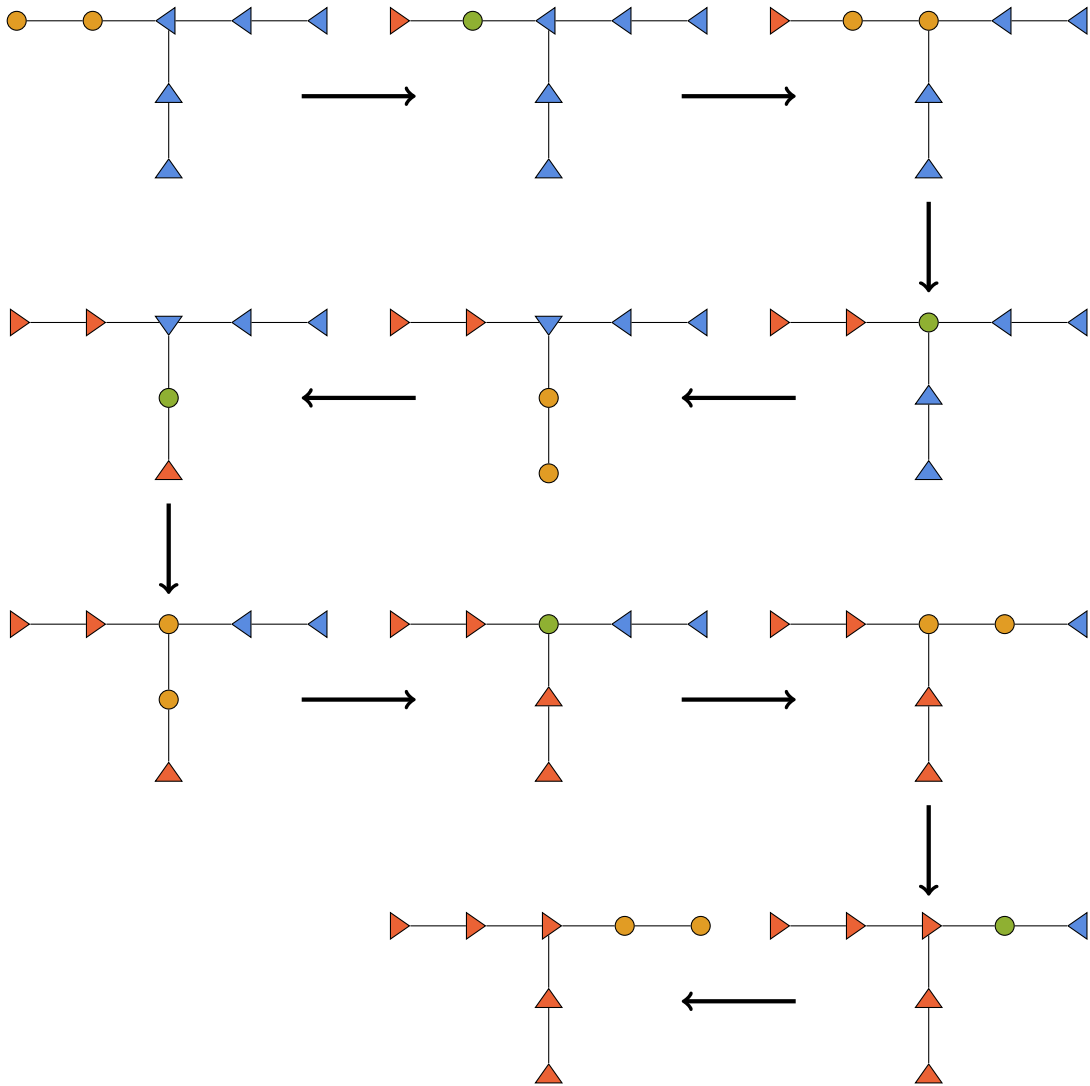
\begin{figure}
    \centering
    \begin{tikzpicture}
    \def\nodedist{1}
    \def\minsize{0.25cm}
    \def\insep{1pt}
    \def\shiftx{5}
    \def\shifty{-4}
    \def\orthccolour{morange}
    \def\nupdtcolour{mblue}
    \def\updtcolour{mred}
    \def\rcolour{mgreen}

    \def\rdiamond(#1,#2){
        \node[fill=\rcolour,draw,
              minimum size=\minsize, inner sep=\insep,
              diamond] (R) at (#1,#2){};}
    \def\ortccircle(#1,#2,#3){
        \node[fill=\orthccolour,circle,draw,
              minimum size=\minsize,inner sep=\insep] (#3) at (#1,#2){};
    }
    \def\backwcircle(#1,#2,#3){
        \node[fill=\rcolour,circle,draw,
              minimum size=\minsize,inner sep=\insep] (#3) at (#1,#2){};
    }
    \def\nupdtriangle(#1,#2,#3,#4){
        \node[fill=\nupdtcolour,draw,
              minimum size=\minsize, inner sep=\insep,
              isosceles triangle,
              isosceles triangle apex angle=70,
              rotate=#1] (#4) at (#2,#3){};
    }
    \def\updtriangle(#1,#2,#3,#4){
        \node[fill=\updtcolour,draw,
              minimum size=\minsize, inner sep=\insep,
              isosceles triangle,
              isosceles triangle apex angle=70,
              rotate=#1] (#4) at (#2,#3){};
    }
    \def\drawlines{
    \draw (N01) -- (N00) -- (N0) -- (N20) -- (N21);
    \draw (N0) -- (N10) -- (N11);
    }
    \def\drawarrow(#1,#2,#3,#4){
        \draw[ultra thick,-to] (#1,#2) -- (#3,#4);
    }
    
    \ortccircle(-2*\nodedist,0,N01)
    \ortccircle(-1*\nodedist,0,N00)
    \nupdtriangle(180,0,0,N0)
    \nupdtriangle(90,0,-1*\nodedist,N10)
    \nupdtriangle(90,0,-2*\nodedist,N11)
    \nupdtriangle(180,1*\nodedist,0,N20)
    \nupdtriangle(180,2*\nodedist,0,N21)
    \drawlines

    \drawarrow(1.75*\nodedist,-1*\nodedist,
                \shiftx-1.75*\nodedist,-1*\nodedist)

    \begin{scope}[shift={(\shiftx,0)}]
        \updtriangle(0,-2*\nodedist,0,N01)
        \backwcircle(-1*\nodedist,0,N00)
        \nupdtriangle(180,0,0,N0)
        \nupdtriangle(90,0,-1*\nodedist,N10)
        \nupdtriangle(90,0,-2*\nodedist,N11)
        \nupdtriangle(180,1*\nodedist,0,N20)
        \nupdtriangle(180,2*\nodedist,0,N21)
        \drawlines
    \end{scope}

    \drawarrow(\shiftx+1.75*\nodedist,-1*\nodedist,
                2*\shiftx-1.75*\nodedist,-1*\nodedist)

    \begin{scope}[shift={(2*\shiftx,0)}]
        \updtriangle(0,-2*\nodedist,0,N01)
        \ortccircle(-1*\nodedist,0,N00)
        \ortccircle(0,0,N0)
        \nupdtriangle(90,0,-1*\nodedist,N10)
        \nupdtriangle(90,0,-2*\nodedist,N11)
        \nupdtriangle(180,1*\nodedist,0,N20)
        \nupdtriangle(180,2*\nodedist,0,N21)
        \drawlines
    \end{scope}

    \drawarrow(2*\shiftx,-2.4*\nodedist,
                2*\shiftx,\shifty+0.4*\nodedist)

    \begin{scope}[shift={(0,\shifty)}]
        \updtriangle(0,-2*\nodedist,0,N01)
        \updtriangle(0,-1*\nodedist,0,N00)
        \nupdtriangle(-90,0,0,N0)
        \backwcircle(0,-1*\nodedist,N10)
        \updtriangle(90,0,-2*\nodedist,N11)
        \nupdtriangle(180,1*\nodedist,0,N20)
        \nupdtriangle(180,2*\nodedist,0,N21)
        \drawlines
    \end{scope}

    \drawarrow(2*\shiftx-1.75*\nodedist,-1*\nodedist+\shifty,
                \shiftx+1.75*\nodedist,-1*\nodedist+\shifty)

    \begin{scope}[shift={(\shiftx,\shifty)}]
        \updtriangle(0,-2*\nodedist,0,N01)
        \updtriangle(0,-1*\nodedist,0,N00)
        \nupdtriangle(-90,0,0,N0)
        \ortccircle(0,-1*\nodedist,N10)
        \ortccircle(0,-2*\nodedist,N11)
        \nupdtriangle(180,1*\nodedist,0,N20)
        \nupdtriangle(180,2*\nodedist,0,N21)
        \drawlines
    \end{scope}

    \drawarrow(\shiftx-1.75*\nodedist,-1*\nodedist+\shifty,
                1.75*\nodedist,-1*\nodedist+\shifty)

    \begin{scope}[shift={(2*\shiftx,\shifty)}]
        \updtriangle(0,-2*\nodedist,0,N01)
        \updtriangle(0,-1*\nodedist,0,N00)
        \backwcircle(0,0,N0)
        \nupdtriangle(90,0,-1*\nodedist,N10)
        \nupdtriangle(90,0,-2*\nodedist,N11)
        \nupdtriangle(180,1*\nodedist,0,N20)
        \nupdtriangle(180,2*\nodedist,0,N21)
        \drawlines
    \end{scope}

    \drawarrow(0,\shifty-2.4*\nodedist,
                0,2*\shifty+0.4*\nodedist)

    \begin{scope}[shift={(0,2*\shifty)}]
        \updtriangle(0,-2*\nodedist,0,N01)
        \updtriangle(0,-1*\nodedist,0,N00)
        \ortccircle(0,0,N0)
        \ortccircle(0,-1*\nodedist,N10)
        \updtriangle(90,0,-2*\nodedist,N11)
        \nupdtriangle(180,1*\nodedist,0,N20)
        \nupdtriangle(180,2*\nodedist,0,N21)
        \drawlines
    \end{scope}

    \drawarrow(1.75*\nodedist,-1*\nodedist+2*\shifty,
                \shiftx-1.75*\nodedist,-1*\nodedist+2*\shifty)

    \begin{scope}[shift={(\shiftx,2*\shifty)}]
        \updtriangle(0,-2*\nodedist,0,N01)
        \updtriangle(0,-1*\nodedist,0,N00)
        \backwcircle(0,0,N0)
        \updtriangle(90,0,-1*\nodedist,N10)
        \updtriangle(90,0,-2*\nodedist,N11)
        \nupdtriangle(180,1*\nodedist,0,N20)
        \nupdtriangle(180,2*\nodedist,0,N21)
        \drawlines
    \end{scope}

    \drawarrow(\shiftx+1.75*\nodedist,-1*\nodedist+2*\shifty,
                2*\shiftx-1.75*\nodedist,-1*\nodedist+2*\shifty)

    \begin{scope}[shift={(2*\shiftx,2*\shifty)}]
        \updtriangle(0,-2*\nodedist,0,N01)
        \updtriangle(0,-1*\nodedist,0,N00)
        \ortccircle(0,0,N0)
        \updtriangle(90,0,-1*\nodedist,N10)
        \updtriangle(90,0,-2*\nodedist,N11)
        \ortccircle(1*\nodedist,0,N20)
        \nupdtriangle(180,2*\nodedist,0,N21)
        \drawlines
    \end{scope}

    \drawarrow(2*\shiftx,-2.4*\nodedist+2*\shifty,
                2*\shiftx,3*\shifty+0.4*\nodedist)

    \begin{scope}[shift={(2*\shiftx,3*\shifty)}]
        \updtriangle(0,-2*\nodedist,0,N01)
        \updtriangle(0,-1*\nodedist,0,N00)
        \updtriangle(0,0,0,N0)
        \updtriangle(90,0,-1*\nodedist,N10)
        \updtriangle(90,0,-2*\nodedist,N11)
        \backwcircle(1*\nodedist,0,N20)
        \nupdtriangle(180,2*\nodedist,0,N21)
        \drawlines
    \end{scope}

    \drawarrow(2*\shiftx-1.75*\nodedist,-1*\nodedist+3*\shifty,
                \shiftx+1.75*\nodedist,-1*\nodedist+3*\shifty)

    \begin{scope}[shift={(\shiftx,3*\shifty)}]
        \updtriangle(0,-2*\nodedist,0,N01)
        \updtriangle(0,-1*\nodedist,0,N00)
        \updtriangle(0,0,0,N0)
        \updtriangle(90,0,-1*\nodedist,N10)
        \updtriangle(90,0,-2*\nodedist,N11)
        \ortccircle(1*\nodedist,0,N20)
        \ortccircle(2*\nodedist,0,N21)
        \drawlines
    \end{scope}
\end{tikzpicture}
    \caption{One forwards sweep through the tree structure \eqref{eq:example_ttns} performed with the two-site TDVP method. Orange circles denote the two sites whose tensors are contracted into one and updated according to \eqref{eq:2tdvp_forwards_update}. A green circle denotes site tensors that have evolved backwards in time according to \eqref{eq:2tdvp_backwards_update}. The blue triangles represent sites that have not been updated, while the orange triangles represent updated sites.}
    \label{fig:2tdvp_example}
\end{figure}

\begin{figure}
     \centering
     \begin{subfigure}[b]{0.45\textwidth}
         \centering
         \includegraphics[width=\textwidth]{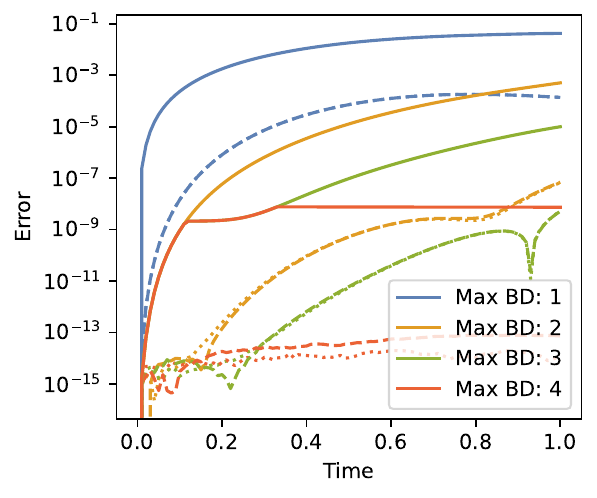}
         \caption{The error $\mathcal{E}$ of the total magnetisation obtained via TDVP simulations with respect to the exact solution.}
         \label{fig:simple_magn_error_tdvp}
     \end{subfigure}
     \hfill
     \begin{subfigure}[b]{0.45\textwidth}
         \centering
         \includegraphics[width=\textwidth]{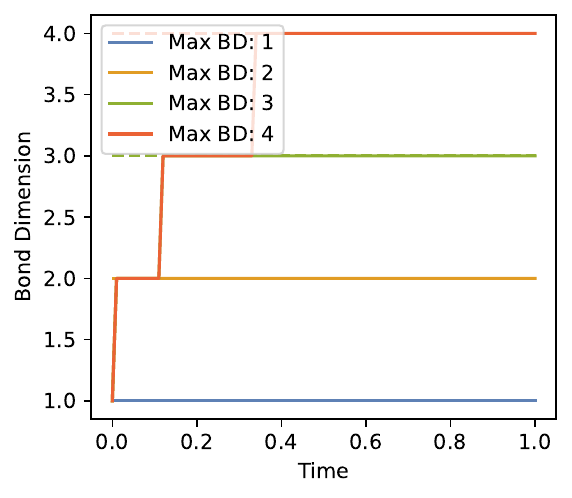}
         \caption{The change in bond dimension during TDVP simulation at the bond $(0,00)$.}
         \label{fig:simple_magn_bd_tdvp}
     \end{subfigure}
        \caption{TDVP simulation results for the total magnetisation $M$ with initial state $\ket{\Psi_0}$ as defined in \eqref{eq:init_state} for different maximum bond dimensions under the time evolution governed by the modified TFI model \eqref{eq:mod_tfi}. The dotted lines are the results of the first-order single-site TDVP, the dashed lines are the results of the second-order single-site TDVP, and the full lines are the results of the two-site TDVP algorithm.}
        \label{fig:simple_magn_tdvp}
\end{figure}
We will now explore the simulation of the modified TFI model \eqref{eq:mod_tfi}. In Figure~\ref{fig:simple_magn_tdvp}, we plotted the results of the simulations for $L=2$. In Figure~\ref{fig:simple_magn_error_tdvp}, the error $\mathcal{E}(t)$ as defined in \eqref{eq:error} is plotted for the different TDVP methods for varying maximum bond dimensions. The larger error in the two-site TDVP originates from the additional truncation during each time step. We can also see that the first and second-order one-site TDVP perform very similarly. In Figure~\ref{fig:simple_magn_bd_tdvp}, the bond dimension at the bond $(0,00)$ is plotted. As Section~\ref{sec:1tdvp} mentions, the bond dimensions during the one-site TDVP are fixed to the bond dimension of the initial state. The initial state can be padded with zeros to increase its bond dimension, above what is required to represent it exactly. Therefore, the bond dimension is constant in Figure~\ref{fig:simple_magn_bd_tdvp}. On the other hand, the two-site TDVP can adapt the bond dimensions dynamically due to the truncation step. It increases step by step until the required maximum bond dimension is reached.
\begin{figure}
     \centering
     \begin{subfigure}[b]{0.45\textwidth}
         \centering
         \includegraphics[width=\textwidth]{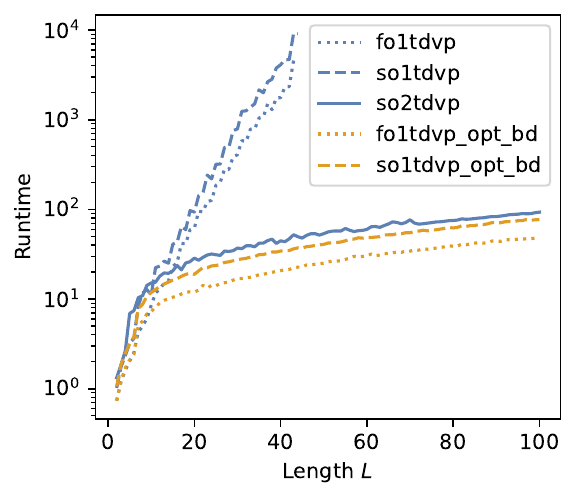}
         \caption{The computational runtime for the different TDVP algorithms. The blue graphs denote the standard methods, while the orange graphs denote the one-site TDVP methods with an adapted bond dimension.}
         \label{fig:tdvp_runtime_dep_length}
     \end{subfigure}
     \hfill
     \begin{subfigure}[b]{0.45\textwidth}
         \centering
         \includegraphics[width=\textwidth]{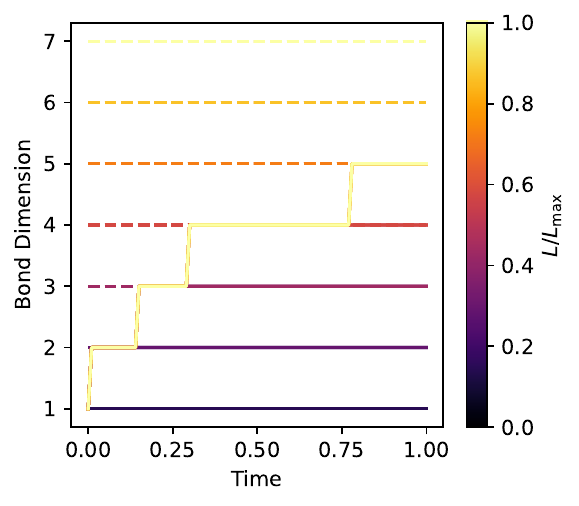}
         \caption{The change in bond dimension during TDVP simulation at the bond $(0,00)$ over time for different lengths $L$. The colour of the graph depends on the length.}
         \label{fig:tdvp_bd_dep_length}
     \end{subfigure}
        \caption{TDVP simulation results for initial state $\ket{\Psi_0}$ as defined in \eqref{eq:init_state} for different lengths $L$. The modified TFI model \eqref{eq:mod_tfi} governs the time evolution. The dotted lines are the results of the first-order single-site TDVP, the dashed lines are the results of the second-order single-site TDVP, and the full lines are the results of the two-site TDVP algorithm.}
        \label{fig:tdvp_dep_on_length}
\end{figure}
We can see the significance of this dynamic increase in Figure~\ref{fig:tdvp_dep_on_length}. The results were once more obtained by running the time evolution governed by the modified TFI \eqref{eq:mod_tfi}. However, we ran the simulation for different chain lengths $L$. In the first instance, we set the maximum bond dimension for all TDVP methods to $D_\txt{max} = \left\lceil L / 2 \right\rceil$. Then we measured the runtime of each algorithm. As we have to fix the bond dimensions for the one-site TDVP methods at the beginning, their runtime scales badly with the system size compared to the two-site TDVP. It also requires too much memory during the time evolution to run on a common machine for $L \gtrsim 45$. It turns out that for the modified TFI, a bond dimension of $5$ is sufficient to represent the state at time $t=1$. This can be seen in Figure~\ref{fig:tdvp_bd_dep_length}. Even though the length $L$ is increased until $14$, as evidenced by the constant graphs of the one-site TDVP methods, the two-site TDVP only increase the bond dimension to $5$. In the second step, we used the bond dimension generated by the two-site TDVP at the final time as the maximum bond dimension of the one-site TDVP algorithms. The run time is plotted in Figure~\ref{fig:tdvp_runtime_dep_length} as the orange graphs. It turns out that, for equal bond dimensions, the one-site TDVP methods run faster than the two-site TDVP. This is due to the time steps being faster, a QR-decomposition of one site is significantly faster than an SVD for a two-site tensor. Therefore, running a two-site TDVP method can be reasonable until a maximum desired bond dimension or even convergence is reached before switching to a one-site algorithm. Even though these simulations go far beyond the capabilities of state-vector simulations, we can still improve on them. We will discuss some of these improvements in the next section.

\section{Further Developements}\label{sec:developements}
{\ptn} already has a plethora of utilities and can be used to implement any method using tree tensor networks. However, the current focus of the library is on the simulation of the time evolution of quantum systems. A different popular tensor network method is the density matrix renormalisation group algorithm (DMRG) \cite{Schollwock2011}. It is a variational algorithm used for ground state search. Since the DMRG was generalised to TTN \cite{Nakatani2013}, the DMRG algorithm could be directly included in {\ptn}. 

As we have seen in Section~\ref{sec:2tdvp}, the one-site TDVP method is faster for the same bond dimension than the two-site TDVP. However, for one-site TDVP, one has to set the bond dimension when starting the time evolution. Since there is no general way to a priory know the best bond dimension, there will be either too large of an error or an unnecessary use of computational resources. Therefore, adding a dynamic bond adaptation to {\ptn} for one-site TDVP would be advantageous. For matrix product states, such dynamic bond adaptation methods are already established \cite{Yang2020, Dunnett2021, Li2022}. These methods only take the directly attached nodes of the bond into account or require the existence of a canonical form. Therefore, they should be straightforward to be generalised to tree structures.

It would also be beneficial to support more subclasses of tree structures directly. While {\ptn} already implements the matrix product \cite{Schollwock2011} and fork tree structures \cite{Bauernfeind2017, Bauernfeind2018}, other common tree structures such as three-legged TTN \cite{Gunst2018} and binary TTN \cite{Tagliacozzo2009} are still missing. To get the most efficiency out of these structures, it could be necessary to treat nodes of trivial physical dimension in the TTNS algorithms separately. Currently, the trivial dimension has to be added explicitly.

Another way {\ptn} could be made more efficient is by including quantum numbers leading to sparse tensors \cite{Singh2010, Singh2011, Singh2012}. This could also extend to include fermionic tensors and fermionic tensor networks that require special treatment due to their anti-commutation relations \cite{Mortier2024}. In both cases, we can resort to existing tensor network libraries \cite{Yosprakob2023, Wu2024, Aolomonik2014, chemtensor} whose tensor classes would replace the NumPy arrays as the data structure representing tensors in {\ptn}.

So, while there are still a lot of possible extensions for {\ptn}, it already has sufficient content to serve as a useful library for tree tensor network simulations.

\paragraph{Funding information}
The research is part of the Munich Quantum Valley, which is supported by the Bavarian state government with funds from the Hightech Agenda Bayern Plus. The research is also supported by the Bavarian Ministry of Economic Affairs, Regional Development and Energy via the project BayQS with funds from the Hightech Agenda Bayern.





\bibliography{references.bib}

\begin{thebibliography}{100}
\providecommand{\url}[1]{\texttt{#1}}
\providecommand{\urlprefix}{URL }
\expandafter\ifx\csname urlstyle\endcsname\relax
  \providecommand{\doi}[1]{doi:\discretionary{}{}{}#1}\else
  \providecommand{\doi}{doi:\discretionary{}{}{}\begingroup
  \urlstyle{rm}\Url}\fi
\providecommand{\eprint}[2][]{\url{#2}}

\bibitem{Szalay2015}
S.~Szalay, M.~Pfeffer, V.~Murg, G.~Barcza, F.~Verstraete, R.~Schneider and
  O.~Legeza,
\newblock \emph{Tensor product methods and entanglement optimization for ab
  initio quantum chemistry},
\newblock Int. J. Quantum Chem. \textbf{115}(19) (2015),
\newblock \doi{10.1002/qua.24898}.

\bibitem{Wood2015}
C.~J. Wood, J.~D. Biamonte and D.~G. Cory,
\newblock \emph{Tensor networks and graphical calculus for open quantum
  systems},
\newblock arXiv (arXiv:1111.6950) (2015),
\newblock \doi{10.48550/arXiv.1111.6950}.

\bibitem{Jaschke2018}
D.~Jaschke, S.~Montangero and L.~D. Carr,
\newblock \emph{One-dimensional many-body entangled open quantum systems with
  tensor network methods},
\newblock Quantum Sci. Technol. \textbf{4}(1) (2018),
\newblock \doi{10.1088/2058-9565/aae724}.

\bibitem{Strathearn2018}
A.~Strathearn, P.~Kirton, D.~Kilda, J.~Keeling and B.~W. Lovett,
\newblock \emph{Efficient non-markovian quantum dynamics using time-evolving
  matrix product operators},
\newblock Nat. Commun. \textbf{9}(1) (2018),
\newblock \doi{10.1038/s41467-018-05617-3}.

\bibitem{Schollwock2011}
U.~Schollw\"ock,
\newblock \emph{The density-matrix renormalization group in the age of matrix
  product states},
\newblock Annals of Physics \textbf{326}(1) (2011),
\newblock \doi{10.1016/j.aop.2010.09.012}.

\bibitem{Banuls2023}
M.~C. Ba\~nuls,
\newblock \emph{Tensor network algorithms: {A} route map},
\newblock Annu. Rev. Condens. Matter Phys. \textbf{14}(1) (2023),
\newblock \doi{10.1146/annurev-conmatphys-040721-022705}.

\bibitem{Jahn2021}
A.~Jahn and J.~Eisert,
\newblock \emph{Holographic tensor network models and quantum error correction:
  {A} topical review},
\newblock Quantum Sci. Technol. \textbf{6}(3) (2021),
\newblock \doi{10.1088/2058-9565/ac0293}.

\bibitem{Melnikov2023}
A.~A. Melnikov, A.~A. Termanova, S.~V. Dolgov, F.~Neukart and M.~R.
  Perelshtein,
\newblock \emph{Quantum state preparation using tensor networks},
\newblock Quantum Sci. Technol. \textbf{8}(3) (2023),
\newblock \doi{10.1088/2058-9565/acd9e7}.

\bibitem{Patra2024}
S.~Patra, S.~S. Jahromi, S.~Singh and R.~Or\'us,
\newblock \emph{Efficient tensor network simulation of ibm{\textquoteright}s
  largest quantum processors},
\newblock Phys. Rev. Research \textbf{6}(1) (2024),
\newblock \doi{10.1103/PhysRevResearch.6.013326}.

\bibitem{Rieser2023}
H.-M. Rieser, F.~K\"oster and A.~P. Raulf,
\newblock \emph{Tensor networks for quantum machine learning},
\newblock Proc. Math. Phys. Eng. Sci. \textbf{479}(2275) (2023),
\newblock \doi{10.1098/rspa.2023.0218}.

\bibitem{Ji2019}
Y.~Ji, Q.~Wang, X.~Li and J.~Liu,
\newblock \emph{A survey on tensor techniques and applications in machine
  learning},
\newblock IEEE Access \textbf{7}, 162950{\textendash}162990 (2019),
\newblock \doi{10.1109/ACCESS.2019.2949814}.

\bibitem{Panagakis2021}
Y.~Panagakis, J.~Kossaifi, G.~G. Chrysos, J.~Oldfield, M.~A. Nicolaou,
  A.~Anandkumar and S.~Zafeiriou,
\newblock \emph{Tensor methods in computer vision and deep learning},
\newblock Proc. IEEE \textbf{109}(5) (2021),
\newblock \doi{10.1109/JPROC.2021.3074329}.

\bibitem{Sengupta2022}
R.~Sengupta, S.~Adhikary, I.~Oseledets and J.~Biamonte,
\newblock \emph{Tensor networks in machine learning},
\newblock JEMS (126) (2022),
\newblock \doi{10.4171/mag/101}.

\bibitem{Stoudenmire2016}
E.~Stoudenmire and D.~J. Schwab,
\newblock \emph{Supervised learning with tensor networks},
\newblock In D.~Lee, M.~Sugiyama, U.~Luxburg, I.~Guyon and R.~Garnett, eds.,
  \emph{Advances in neural information processing systems}, vol.~29 (2016).

\bibitem{AboKhamis2016}
M.~Abo~Khamis, H.~Q. Ngo and A.~Rudra,
\newblock \emph{Faq: Questions asked frequently},
\newblock In \emph{Proceedings of the 35th ACM SIGMOD-SIGACT-SIGAI Symposium on
  Principles of Database Systems}. San Francisco California USA,
\newblock ISBN 978-1-4503-4191-2,
\newblock \doi{10.1145/2902251.2902280} (2016).

\bibitem{Dudek2020}
J.~M. Dudek, L.~Due\~nas Osorio and M.~Y. Vardi,
\newblock \emph{Efficient contraction of large tensor networks for weighted
  model counting through graph decompositions},
\newblock arXiv (arXiv:1908.04381) (2020),
\newblock \doi{10.48550/arXiv.1908.04381}.

\bibitem{Stoian2023}
M.~Stoian, R.~Milbradt and C.~B. Mendl,
\newblock \emph{On the optimal linear contraction order of tree tensor
  networks, and beyond},
\newblock arXiv (arXiv:2209.12332) (2023),
\newblock \doi{10.48550/arXiv.2209.12332}.

\bibitem{Silvi2019}
P.~Silvi, F.~Tschirsich, M.~Gerster, J.~J\"unemann, D.~Jaschke, M.~Rizzi and
  S.~Montangero,
\newblock \emph{The tensor networks anthology: {S}imulation techniques for
  many-body quantum lattice systems},
\newblock SciPost Phys. Lecture Notes  (2019),
\newblock \doi{10.21468/SciPostPhysLectNotes.8}.

\bibitem{Ran2020Book}
S.-J. Ran, E.~Tirrito, C.~Peng, X.~Chen, L.~Tagliacozzo, G.~Su and
  M.~Lewenstein,
\newblock \emph{Tensor Network Contractions: {M}ethods and Applications to
  Quantum Many-Body Systems}, vol. 964 of \emph{Lecture Notes in Physics},
\newblock Cham,
\newblock ISBN 978-3-030-34488-7,
\newblock \doi{10.1007/978-3-030-34489-4} (2020).

\bibitem{Montangero2018}
S.~Montangero,
\newblock \emph{Introduction to Tensor Network Methods: {N}umerical simulations
  of low-dimensional many-body quantum systems},
\newblock Cham,
\newblock ISBN 978-3-030-01408-7,
\newblock \doi{10.1007/978-3-030-01409-4} (2018).

\bibitem{Evenbly2022}
G.~Evenbly,
\newblock \emph{A practical guide to the numerical implementation of tensor
  networks {I}: {C}ontractions, decompositions and gauge freedom}
  (arXiv:2202.02138) (2022),
\newblock \doi{10.48550/arXiv.2202.02138}.

\bibitem{Biamonte2017}
J.~Biamonte and V.~Bergholm,
\newblock \emph{Tensor networks in a nutshell},
\newblock arXiv (arXiv:1708.00006) (2017),
\newblock \doi{10.48550/arXiv.1708.00006}.

\bibitem{Biamonte2020}
J.~Biamonte,
\newblock \emph{Lectures on quantum tensor networks},
\newblock arXiv (arXiv:1912.10049) (2020),
\newblock \doi{10.48550/arXiv.1912.10049}.

\bibitem{Bridgeman2017}
J.~C. Bridgeman and C.~T. Chubb,
\newblock \emph{Hand-waving and interpretive dance: {A}n introductory course on
  tensor networks},
\newblock Journal of Physics A: Mathematical and Theoretical \textbf{50}(22)
  (2017),
\newblock \doi{10.1088/1751-8121/aa6dc3}.

\bibitem{Orus2019}
R.~Or\'us,
\newblock \emph{Tensor networks for complex quantum systems},
\newblock Nature Reviews Physics \textbf{1}(9) (2019),
\newblock \doi{10.1038/s42254-019-0086-7}.

\bibitem{Ren2022}
J.~Ren, W.~Li, T.~Jiang, Y.~Wang and Z.~Shuai,
\newblock \emph{Time-dependent density matrix renormalization group method for
  quantum dynamics in complex systems},
\newblock Wiley Interdiscip. Rev. Comput. Mol. Sci. \textbf{12}(6), e1614
  (2022),
\newblock \doi{10.1002/wcms.1614}.

\bibitem{Cirac2021}
J.~I. Cirac, D.~P\'erez-Garc{\'\i}a, N.~Schuch and F.~Verstraete,
\newblock \emph{Matrix product states and projected entangled pair states:
  Concepts, symmetries, theorems},
\newblock Rev. Mod. Phys. \textbf{93}(4) (2021),
\newblock \doi{10.1103/RevModPhys.93.045003}.

\bibitem{Verstraete2004PEPS}
F.~Verstraete and J.~I. Cirac,
\newblock \emph{Renormalization algorithms for quantum-many body systems in two
  and higher dimensions},
\newblock arXiv (arXiv:cond-mat/0407066) (2004),
\newblock \doi{10.48550/arXiv.cond-mat/0407066}.

\bibitem{Verstraete2006}
F.~Verstraete, M.~M. Wolf, D.~Perez-Garcia and J.~I. Cirac,
\newblock \emph{Criticality, the area law, and the computational power of
  projected entangled pair states},
\newblock Phys. Rev. Lett. \textbf{96}(22) (2006),
\newblock \doi{10.1103/PhysRevLett.96.220601}.

\bibitem{Schuch2007}
N.~Schuch, M.~M. Wolf, F.~Verstraete and J.~I. Cirac,
\newblock \emph{Computational complexity of projected entangled pair states},
\newblock Phys. Rev. Lett. \textbf{98}(14) (2007),
\newblock \doi{10.1103/PhysRevLett.98.140506}.

\bibitem{Lubasch2014}
M.~Lubasch, J.~I. Cirac and M.-C. Ba\~nuls,
\newblock \emph{Unifying projected entangled pair state contractions},
\newblock New J. Phys. \textbf{16}(3) (2014),
\newblock \doi{10.1088/1367-2630/16/3/033014}.

\bibitem{Shi2006}
Y.-Y. Shi, L.-M. Duan and G.~Vidal,
\newblock \emph{Classical simulation of quantum many-body systems with a tree
  tensor network},
\newblock Phys. Rev. A \textbf{74}(2) (2006),
\newblock \doi{10.1103/PhysRevA.74.022320}.

\bibitem{pytreenet}
R.~M. Milbradt, Q.~Huang and C.~B. Mendl,
\newblock \emph{{PyTreeNet-Repository}},
\newblock \eprint{https://github.com/Drachier/PyTreeNet}.

\bibitem{Bauernfeind2020}
D.~Bauernfeind and M.~Aichhorn,
\newblock \emph{Time dependent variational principle for tree tensor networks},
\newblock SciPost Phys. \textbf{8}(2) (2020),
\newblock \doi{10.21468/SciPostPhys.8.2.024}.

\bibitem{Frowis2010}
F.~Fr\"owis, V.~Nebendahl and W.~D\"ur,
\newblock \emph{Tensor operators: {C}onstructions and applications for
  long-range interaction systems},
\newblock Phys. Rev. A \textbf{81}(6) (2010),
\newblock \doi{10.1103/PhysRevA.81.062337}.

\bibitem{Bauernfeind2017}
D.~Bauernfeind, M.~Zingl, R.~Triebl, M.~Aichhorn and H.~G. Evertz,
\newblock \emph{Fork tensor-product states: {E}fficient multiorbital real-time
  {DMFT} solver},
\newblock Phys. Rev. X \textbf{7}(3) (2017),
\newblock \doi{10.1103/PhysRevX.7.031013}.

\bibitem{Murg2010}
V.~Murg, F.~Verstraete, O.~Legeza and R.~M. Noack,
\newblock \emph{Simulating strongly correlated quantum systems with tree tensor
  networks},
\newblock Phys. Rev. B \textbf{82}(20) (2010),
\newblock \doi{10.1103/PhysRevB.82.205105}.

\bibitem{Okunishi2023}
K.~Okunishi, H.~Ueda and T.~Nishino,
\newblock \emph{Entanglement bipartitioning and tree tensor networks},
\newblock Prog. Theor. Exp. Phys. \textbf{2023}(2) (2023),
\newblock \doi{10.1093/ptep/ptad018}.

\bibitem{Nakatani2013}
N.~Nakatani and G.~K.-L. Chan,
\newblock \emph{Efficient tree tensor network states ({TTNS}) for quantum
  chemistry: {G}eneralizations of the density matrix renormalization group
  algorithm},
\newblock J. Chem. Phys. \textbf{138}(13) (2013),
\newblock \doi{10.1063/1.4798639}.

\bibitem{Larsson2019}
H.~R. Larsson,
\newblock \emph{Computing vibrational eigenstates with tree tensor network
  states {(TTNS)}},
\newblock J. Chem. Phys. \textbf{151}(20) (2019),
\newblock \doi{10.1063/1.5130390}.

\bibitem{Gunst2018}
K.~Gunst, F.~Verstraete, S.~Wouters, O.~Legeza and D.~Van~Neck,
\newblock \emph{{T3NS: T}hree-legged tree tensor network states},
\newblock J. Chem. Theory Comput. \textbf{14}(4) (2018),
\newblock \doi{10.1021/acs.jctc.8b00098}.

\bibitem{Murg2015}
V.~Murg, F.~Verstraete, R.~Schneider, P.~R. Nagy and O.~Legeza,
\newblock \emph{Tree tensor network state with variable tensor order: {A}n
  efficient multireference method for strongly correlated systems},
\newblock J. Chem. Theory Comput. \textbf{11}(3) (2015),
\newblock \doi{10.1021/ct501187j}.

\bibitem{Verstraete2004}
F.~Verstraete, J.~J. Garc{\'\i}a-Ripoll and J.~I. Cirac,
\newblock \emph{Matrix product density operators: {S}imulation of
  finite-temperature and dissipative systems},
\newblock Phys. Rev. Lett. \textbf{93}(20) (2004),
\newblock \doi{10.1103/PhysRevLett.93.207204}.

\bibitem{Vidal2004}
G.~Vidal,
\newblock \emph{Efficient simulation of one-dimensional quantum many-body
  systems},
\newblock Phys. Rev. Lett. \textbf{93}(4) (2004),
\newblock \doi{10.1103/PhysRevLett.93.040502}.

\bibitem{Daley2004}
A.~J. Daley, C.~Kollath, U.~Schollw\"ock and G.~Vidal,
\newblock \emph{Time-dependent density-matrix renormalization-group using
  adaptive effective {H}ilbert spaces},
\newblock J. Stat. Mech.: Theor. Exp. \textbf{2004}(04) (2004),
\newblock \doi{10.1088/1742-5468/2004/04/P04005}.

\bibitem{Haegeman2011}
J.~Haegeman, J.~I. Cirac, T.~J. Osborne, I.~Pi\v{z}orn, H.~Verschelde and
  F.~Verstraete,
\newblock \emph{Time-dependent variational principle for quantum lattices},
\newblock Phys. Rev. Lett. \textbf{107}(7) (2011),
\newblock \doi{10.1103/PhysRevLett.107.070601}.

\bibitem{Haegeman2016}
J.~Haegeman, C.~Lubich, I.~Oseledets, B.~Vandereycken and F.~Verstraete,
\newblock \emph{Unifying time evolution and optimization with matrix product
  states},
\newblock Phys. Rev. B \textbf{94}(16) (2016),
\newblock \doi{10.1103/PhysRevB.94.165116}.

\bibitem{Harris2020}
C.~R. Harris, K.~J. Millman, S.~J. van~der Walt, R.~Gommers, P.~Virtanen,
  D.~Cournapeau, E.~Wieser, J.~Taylor, S.~Berg, N.~J. Smith, R.~Kern, M.~Picus
  \emph{et~al.},
\newblock \emph{Array programming with {NumPy}},
\newblock Nature \textbf{585}(7825) (2020),
\newblock \doi{10.1038/s41586-020-2649-2}.

\bibitem{Liu2023}
J.-G. Liu, X.~Gao, M.~Cain, M.~D. Lukin and S.-T. Wang,
\newblock \emph{Computing solution space properties of combinatorial
  optimization problems via generic tensor networks},
\newblock SIAM J. Sci. Comp. \textbf{45}(3) (2023),
\newblock \doi{10.1137/22M1501787}.

\bibitem{Hu2022}
M.~Hu and J.~Tura,
\newblock \emph{Tropical contraction of tensor networks as a {B}ell inequality
  optimization toolset},
\newblock arXiv (arXiv:2208.02798) (2022),
\newblock \doi{10.48550/arXiv.2208.02798}.

\bibitem{Verstraete2010}
F.~Verstraete and J.~I. Cirac,
\newblock \emph{Continuous matrix product states for quantum fields},
\newblock Phys. Rev. Lett. \textbf{104}(19) (2010),
\newblock \doi{10.1103/PhysRevLett.104.190405}.

\bibitem{Jennings2015}
D.~Jennings, C.~Brockt, J.~Haegeman, T.~J. Osborne and F.~Verstraete,
\newblock \emph{Continuum tensor network field states, path integral
  representations and spatial symmetries},
\newblock New J. Phys. \textbf{17}(6) (2015),
\newblock \doi{10.1088/1367-2630/17/6/063039}.

\bibitem{Tilloy2019}
A.~Tilloy and J.~I. Cirac,
\newblock \emph{Continuous tensor network states for quantum fields},
\newblock Phys. Rev. X \textbf{9}(2) (2019),
\newblock \doi{10.1103/PhysRevX.9.021040}.

\bibitem{Lu2023}
J.~Lu,
\newblock \emph{Matrix decomposition and applications},
\newblock arXiv (arXiv:2201.00145) (2023),
\newblock \doi{10.48550/arXiv.2201.00145}.

\bibitem{Eckart1936}
C.~Eckart and G.~Young,
\newblock \emph{The approximation of one matrix by another of lower rank},
\newblock Psychometrika \textbf{1}(3) (1936),
\newblock \doi{10.1007/BF02288367}.

\bibitem{Mirsky1960}
L.~Mirsky,
\newblock \emph{Symmetric gauge functions and unitarily invariant norms},
\newblock Q. J. Math. \textbf{11}(1), 50{\textendash}59 (1960),
\newblock \doi{10.1093/qmath/11.1.50}.

\bibitem{Krumke2005}
S.~O. Krumke and H.~Noltemeier,
\newblock \emph{6. {B}\"aume, {W}\"alder und {M}atroide}, p. 99–144,
\newblock Vieweg+Teubner Verlag,
\newblock ISBN 978-3-322-92112-3,
\newblock \doi{10.1007/978-3-322-92112-3_6} (2015).

\bibitem{Catarina2023}
G.~Catarina and B.~Murta,
\newblock \emph{Density-matrix renormalization group: {A} pedagogical
  introduction},
\newblock Eur. Phys. J. B \textbf{96}(8) (2023),
\newblock \doi{10.1140/epjb/s10051-023-00575-2}.

\bibitem{Schollwock2005}
U.~Schollw\"ock,
\newblock \emph{The density-matrix renormalization group},
\newblock Rev. Mod. Phys. \textbf{77}(1) (2005),
\newblock \doi{10.1103/RevModPhys.77.259}.

\bibitem{Hikihara2023}
T.~Hikihara, H.~Ueda, K.~Okunishi, K.~Harada and T.~Nishino,
\newblock \emph{Automatic structural optimization of tree tensor networks},
\newblock Physical Review Research \textbf{5}(1) (2023),
\newblock \doi{10.1103/PhysRevResearch.5.013031}.

\bibitem{Hikihara2024}
T.~Hikihara, H.~Ueda, K.~Okunishi, K.~Harada and T.~Nishino,
\newblock \emph{Visualization of entanglement geometry by structural
  optimization of tree tensor network},
\newblock arxiv (arXiv:2401.16000) (2024),
\newblock \doi{10.48550/arXiv.2401.16000}.

\bibitem{Bleh2012}
D.~Bleh, T.~Calarco and S.~Montangero,
\newblock \emph{Quantum game of life},
\newblock EPL \textbf{97}(2) (2012),
\newblock \doi{10.1209/0295-5075/97/20012}.

\bibitem{Ney2022}
P.-M. Ney, S.~Notarnicola, S.~Montangero and G.~Morigi,
\newblock \emph{Entanglement in the quantum game of life},
\newblock Phys. Rev. A \textbf{105}(1) (2022),
\newblock \doi{10.1103/PhysRevA.105.012416}.

\bibitem{McCulloch2007}
I.~P. McCulloch,
\newblock \emph{From density-matrix renormalization group to matrix product
  states},
\newblock J. Stat. Mech.: Theor. Exp. \textbf{2007}(10) (2007),
\newblock \doi{10.1088/1742-5468/2007/10/P10014}.

\bibitem{Crosswhite2008}
G.~M. Crosswhite and D.~Bacon,
\newblock \emph{Finite automata for caching in matrix product algorithms},
\newblock Phys. Rev. A \textbf{78}(1) (2008),
\newblock \doi{10.1103/PhysRevA.78.012356}.

\bibitem{Pirvu2010}
B.~Pirvu, V.~Murg, J.~I. Cirac and F.~Verstraete,
\newblock \emph{Matrix product operator representations},
\newblock New J. Phys. \textbf{12}(2) (2010),
\newblock \doi{10.1088/1367-2630/12/2/025012}.

\bibitem{Keller2015}
S.~Keller, M.~Dolfi, M.~Troyer and M.~Reiher,
\newblock \emph{An efficient matrix product operator representation of the
  quantum chemical hamiltonian},
\newblock J. Chem. Phys. \textbf{143}(24) (2015),
\newblock \doi{10.1063/1.4939000}.

\bibitem{Chan2016}
G.~K.-L. Chan, A.~Keselman, N.~Nakatani, Z.~Li and S.~R. White,
\newblock \emph{Matrix product operators, matrix product states, and ab initio
  density matrix renormalization group algorithms},
\newblock J. Chem. Phys. \textbf{145}(1) (2016),
\newblock \doi{10.1063/1.4955108}.

\bibitem{Hubig2017}
C.~Hubig, I.~P. McCulloch and U.~Schollw\"ock,
\newblock \emph{Generic construction of efficient matrix product operators},
\newblock Phys. Rev. B \textbf{95}(3) (2017),
\newblock \doi{10.1103/PhysRevB.95.035129}.

\bibitem{Ren2020}
J.~Ren, W.~Li, T.~Jiang and Z.~Shuai,
\newblock \emph{A general automatic method for optimal construction of matrix
  product operators using bipartite graph theory},
\newblock J. Chem. Phys. \textbf{153}(8) (2020),
\newblock \doi{10.1063/5.0018149}.

\bibitem{Wall2020}
M.~L. Wall,
\newblock \emph{Matrix product operator representation of polynomial
  interactions},
\newblock Journal of Physics A: Mathematical and Theoretical \textbf{53}(21)
  (2020),
\newblock \doi{10.1088/1751-8121/ab8675}.

\bibitem{Nusseler2021}
A.~N\"u{\ss}eler, I.~Dhand, S.~F. Huelga and M.~B. Plenio,
\newblock \emph{Efficient construction of matrix-product representations of
  many-body gaussian states},
\newblock Phys. Rev. A \textbf{104}(1) (2021),
\newblock \doi{10.1103/PhysRevA.104.012415}.

\bibitem{Cakir_to_be_publ}
H.~\c{C}akır, R.~M. Milbradt and C.~B. Mendl,
\newblock \emph{(publication in preparation)}.

\bibitem{Milbradt2024}
R.~M. Milbradt, Q.~Huang and C.~B. Mendl,
\newblock \emph{State diagrams to determine tree tensor network operators},
\newblock arXiv (arXiv:2311.13433) (2024),
\newblock \doi{10.48550/arXiv.2311.13433}.

\bibitem{Cakir2024}
H.~\c{C}akır,
\newblock \emph{Optimal Construction of Matrix Product Operators and Tree
  Tensor Network Operators},
\newblock Master's thesis, Technical University of Munich (2024),
  \eprint{https://mediatum.ub.tum.de/1740000}.

\bibitem{Mitra2018}
A.~Mitra,
\newblock \emph{Quantum quench dynamics},
\newblock Annu. Rev. Condens. Matter Phys. \textbf{9}(1) (2018),
\newblock \doi{10.1146/annurev-conmatphys-031016-025451}.

\bibitem{Smith2019}
A.~Smith, M.~S. Kim, F.~Pollmann and J.~Knolle,
\newblock \emph{Simulating quantum many-body dynamics on a current digital
  quantum computer},
\newblock npj Quantum Inf. \textbf{5}(1) (2019),
\newblock \doi{10.1038/s41534-019-0217-0}.

\bibitem{Zhou2020}
Y.~Zhou, E.~M. Stoudenmire and X.~Waintal,
\newblock \emph{What limits the simulation of quantum computers?},
\newblock Phys. Rev. X \textbf{10}(4) (2020),
\newblock \doi{10.1103/PhysRevX.10.041038}.

\bibitem{Pan2021}
F.~Pan and P.~Zhang,
\newblock \emph{Simulating the {S}ycamore quantum supremacy circuits},
\newblock arXiv (arXiv:2103.03074) (2021),
\newblock \doi{10.48550/arXiv.2103.03074}.

\bibitem{Trotter1959}
H.~F. Trotter,
\newblock \emph{On the product of semi-groups of operators},
\newblock Proc. Amer. Math. Soc. \textbf{10}(4), 545{\textendash}551 (1959),
\newblock \doi{10.1090/S0002-9939-1959-0108732-6}.

\bibitem{Suzuki1976}
M.~Suzuki,
\newblock \emph{Generalized {T}rotter{\textquoteright}s formula and systematic
  approximants of exponential operators and inner derivations with applications
  to many-body problems},
\newblock Commun. Math. Phys. \textbf{51}(2) (1976),
\newblock \doi{10.1007/BF01609348}.

\bibitem{Strang1968}
G.~Strang,
\newblock \emph{On the construction and comparison of difference schemes},
\newblock SIAM J. Numer. Anal. \textbf{5}(3) (1968),
\newblock \doi{10.1137/0705041}.

\bibitem{Ostmeyer2023}
J.~Ostmeyer,
\newblock \emph{Optimised trotter decompositions for classical and quantum
  computing},
\newblock Journal of Physics A: Mathematical and Theoretical \textbf{56}(28)
  (2023),
\newblock \doi{10.1088/1751-8121/acde7a}.

\bibitem{Stoudenmire2010}
E.~M. Stoudenmire and S.~R. White,
\newblock \emph{Minimally entangled typical thermal state algorithms},
\newblock New J. Phys. \textbf{12}(5) (2010),
\newblock \doi{10.1088/1367-2630/12/5/055026}.

\bibitem{Kramer1981}
P.~Kramer and M.~Saraceno,
\newblock \emph{Geometry of the Time-Dependent Variational Principle in Quantum
  Mechanics}, vol. 140 of \emph{Lecture Notes in Physics},
\newblock Springer, Berlin, Heidelberg,
\newblock ISBN 978-3-540-10579-4,
\newblock \doi{10.1007/3-540-10579-4} (1981).

\bibitem{Broeckhove1988}
J.~Broeckhove, L.~Lathouwers, E.~Kesteloot and P.~Van~Leuven,
\newblock \emph{On the equivalence of time-dependent variational principles},
\newblock Chem. Phys. Lett. \textbf{149}(5) (1988),
\newblock \doi{10.1016/0009-2614(88)80380-4}.

\bibitem{Lubich2013}
C.~Lubich, T.~Rohwedder, R.~Schneider and B.~Vandereycken,
\newblock \emph{Dynamical approximation by hierarchical {T}ucker and
  tensor-train tensors},
\newblock SIAM Journal on Matrix Analysis and Applications \textbf{34}(2)
  (2013),
\newblock \doi{10.1137/120885723}.

\bibitem{Yang2020}
M.~Yang and S.~R. White,
\newblock \emph{Time-dependent variational principle with ancillary {K}rylov
  subspace},
\newblock Phys. Rev. B \textbf{102}(9) (2020),
\newblock \doi{10.1103/PhysRevB.102.094315}.

\bibitem{Dunnett2021}
A.~J. Dunnett and A.~W. Chin,
\newblock \emph{Efficient bond-adaptive approach for finite-temperature open
  quantum dynamics using the one-site time-dependent variational principle for
  matrix product states},
\newblock Phys. Rev. B \textbf{104}(21) (2021),
\newblock \doi{10.1103/PhysRevB.104.214302}.

\bibitem{Li2022}
J.-W. Li, A.~Gleis and J.~von Delft,
\newblock \emph{Time-dependent variational principle with controlled bond
  expansion for matrix product states},
\newblock Phys. Rev. Lett. \textbf{133}(2), 026401 (2024),
\newblock \doi{10.1103/PhysRevLett.133.026401}.

\bibitem{Bauernfeind2018}
D.~Bauernfeind,
\newblock \emph{Fork tensor product states: {E}fficient multi-orbital impurity
  solver for dynamical mean field theory},
\newblock Ph.D. thesis, Technische Universit\"at Graz, Graz (2018),
  \eprint{https://permalink.obvsg.at/tug/AC15076600}.

\bibitem{Tagliacozzo2009}
L.~Tagliacozzo, G.~Evenbly and G.~Vidal,
\newblock \emph{Simulation of two-dimensional quantum systems using a tree
  tensor network that exploits the entropic area law},
\newblock Phys. Rev. B \textbf{80}(23) (2009),
\newblock \doi{10.1103/PhysRevB.80.235127}.

\bibitem{Singh2010}
S.~Singh, R.~N.~C. Pfeifer and G.~Vidal,
\newblock \emph{Tensor network decompositions in the presence of a global
  symmetry},
\newblock Phys. Rev. A \textbf{82}(5) (2010),
\newblock \doi{10.1103/PhysRevA.82.050301}.

\bibitem{Singh2011}
S.~Singh, R.~N.~C. Pfeifer and G.~Vidal,
\newblock \emph{Tensor network states and algorithms in the presence of a
  global {U}(1) symmetry},
\newblock Phys. Rev. B \textbf{83}(11) (2011),
\newblock \doi{10.1103/PhysRevB.83.115125}.

\bibitem{Singh2012}
S.~Singh and G.~Vidal,
\newblock \emph{Tensor network states and algorithms in the presence of a
  global {SU}(2) symmetry},
\newblock Phys. Rev. B \textbf{86}(19) (2012),
\newblock \doi{10.1103/PhysRevB.86.195114}.

\bibitem{Mortier2024}
Q.~Mortier, L.~Devos, L.~Burgelman, B.~Vanhecke, N.~Bultinck, F.~Verstraete,
  J.~Haegeman and L.~Vanderstraeten,
\newblock \emph{Fermionic tensor network methods},
\newblock arXiv (arXiv:2404.14611) (2024),
\newblock \doi{10.48550/arXiv.2404.14611}.

\bibitem{Yosprakob2023}
A.~Yosprakob,
\newblock \emph{{GrassmannTN: A Python package for Grassmann tensor network
  computations}},
\newblock SciPost Physics Codebases  (2023),
\newblock \doi{10.21468/SciPostPhysCodeb.20}.

\bibitem{Wu2024}
K.-H. Wu, C.-T. Lin, K.~Hsu, H.-T. Hung, M.~Schneider, C.-M. Chung, Y.-J. Kao
  and P.~Chen,
\newblock \emph{The {C}ytnx library for tensor networks},
\newblock arXiv (arXiv:2401.01921) (2024),
\newblock \doi{10.48550/arXiv.2401.01921}.

\bibitem{Aolomonik2014}
E.~Solomonik, D.~Matthews, J.~R. Hammond, J.~F. Stanton and J.~Demmel,
\newblock \emph{A massively parallel tensor contraction framework for
  coupled-cluster computations},
\newblock J. Parallel Distrib. Comput. \textbf{74}(12), 3176 (2014),
\newblock \doi{10.1016/j.jpdc.2014.06.002.}

\bibitem{chemtensor}
C.~B. Mendl,
\newblock \emph{{ChemTensor}},
\newblock \eprint{https://github.com/qc-tum/chemtensor}.

\end{thebibliography}


\end{document}